\documentclass[10pt,journal,compsoc]{IEEEtran}

\ifCLASSOPTIONcompsoc
  \usepackage[nocompress]{cite}
\else
  \usepackage{cite}
\fi
\usepackage{booktabs} 

\usepackage[utf8]{inputenc} 
\usepackage[T1]{fontenc}
\usepackage{microtype}

\usepackage{alltt}
\usepackage{amsfonts}
\usepackage{amsmath}
\usepackage{array}
\usepackage{booktabs}
\usepackage{boxhandler}
\usepackage{colortbl}
\usepackage{enumerate}
\usepackage{enumitem}
\usepackage{graphicx}
\usepackage{latexsym}
\usepackage{listings}
\usepackage{makecell}
\usepackage{multirow}
\usepackage{proof}
\usepackage{stackengine}
\usepackage{url}
\usepackage{subcaption}
\usepackage{tcolorbox}
\usepackage{mdframed}
\usepackage{textcomp}
\usepackage{url}
\usepackage{xcolor}
\usepackage{xspace}
\usepackage{tabularx}
\usepackage{tikz} 
\usepackage{soul}

\newcommand{\ie}{\textit{i.e., }}
\newcommand{\aka}{\textit{aka }}

\newcommand{\our}{{\em Cerebro}\xspace}
\newcommand{\dt}{{\em Decision Trees}\xspace}
\newcommand{\rnd}{{\em Random}\xspace}

\newcommand{\changing}[1]{{#1}}
\newcommand{\minrev}[1]{{#1}}

\definecolor{pblue}{rgb}{0.13,0.13,1}
\definecolor{pgreen}{rgb}{0,0.5,0}
\definecolor{pred}{rgb}{0.9,0,0}
\definecolor{pgrey}{rgb}{0.46,0.45,0.48}

\usepackage{algorithm,algorithmicx,algpseudocode}

\usepackage{listings}
 \lstdefinelanguage{PrettyJava}
{ 
  language=Java,
  numbers=left,escapeinside={(*@}{@*)},
  showspaces=false,
  showtabs=false,
  breaklines=true,
  showstringspaces=false,
  breakatwhitespace=true,
  commentstyle=\color{pgreen},
  keywordstyle=\color{pblue},
  stringstyle=\color{pred},
  basicstyle=\fontsize{6.5}{6.6} \ttfamily,
  moredelim=[is][\bfseries\textcolor{pgreen}]{/+}{+/}
}
\let\origthelstnumber\thelstnumber
\makeatletter
\newcommand*\Suppressnumber{%
  \lst@AddToHook{OnNewLine}{%
    \let\thelstnumber\relax%
     \advance\c@lstnumber-\@ne\relax%
    }%
}

\newcommand*\Reactivatenumber[1]{%
  \setcounter{lstnumber}{\numexpr#1-1\relax}
  \lst@AddToHook{OnNewLine}{%
   \let\thelstnumber\origthelstnumber%
   \refstepcounter{lstnumber}
  }%
}
\makeatother

\newcommand{\mutant}[1]{
\begin{lstlisting}[language=PrettyJava]
#1
\end{lstlisting}
}

\newcommand{\ourtitle}{Cerebro: Static Subsuming Mutant Selection}

\begin{document}
\title{\ourtitle}

\author{Aayush~Garg,
        Milos~Ojdanic,
        Renzo~Degiovanni,\\
        Thierry~Titcheu~Chekam,
        Mike~Papadakis
        and~Yves~Le~Traon
\IEEEcompsocitemizethanks{\IEEEcompsocthanksitem  A. Garg, M. Ojdanic, R. Degiovanni, M. Papadakis and Y. Le Traon are with the University of Luxembourg, Luxembourg.\protect\\
\IEEEcompsocthanksitem T. Checkam is with SES, Luxembourg.}
}

\markboth{\ourtitle}%
{Garg \MakeLowercase{\textit{et al.}}: \ourtitle}

\IEEEtitleabstractindextext{%
\begin{abstract}
Mutation testing research has indicated that a major part of its application cost is due to the large number of low utility mutants that it introduces. Although previous research has identified this issue, no previous study has proposed any \changing{effective} solution to the problem. Thus, it remains unclear how to mutate and test a given piece of code in a \changing{best effort way, i.e., achieving a good trade-off between invested effort and test effectiveness}. To achieve this, we propose \our, a machine learning approach that \emph{statically} selects subsuming mutants, i.e., the set of mutants that resides on the top of the subsumption hierarchy, based on the mutants’ surrounding code context. We evaluate \our using 48 and 10 programs written in C and Java, respectively, and demonstrate that it preserves the mutation testing benefits while limiting application cost, i.e., reduces all cost application factors such as equivalent mutants, mutant executions, and the mutants requiring analysis. We demonstrate that \our has strong inter-project prediction ability, which is significantly higher than two baseline methods, i.e., supervised learning on features proposed by state-of-the-art, and random mutant selection. More importantly, our results show that \our's selected mutants lead to strong tests that are respectively capable of killing 2 times higher than the number of subsuming mutants killed by the baselines when selecting the same number of mutants. At the same time, \our reduces the cost-related factors, as it selects, on average, \changing{68\%} fewer equivalent mutants, while requiring \changing{90\%} fewer test executions than the baselines.
\end{abstract}

}

\maketitle

\IEEEraisesectionheading{
\section{Introduction}
\label{sec:introduction}
}

Research and practice with mutation testing has shown that it can effectively guide developers in improving their test suite strengths~\cite{ChekamPTH17, Ammann_2014}, and can be used to reliably compare test techniques~\cite{Andrews+2006, PapadakisSYB18}. A key issue though, is that it is expensive, as a large number of mutants are involved, the majority of which are of low utility, i.e., they do not contribute to the testing process~\cite{JiaH09, 5693206, Ammann_2014}. This means that mutation testers should filter their mutant sets using manual analysis to identify equivalent mutants
~\cite{BuddA82}, and perform numerous test executions to discard mutants that do not provide testing value, i.e., mutants  that are detected by the tests designed to detect other mutants~\cite{JiaH09, 5693206, Ammann_2014}. 

Working with large real-world systems makes the problem almost intractable due to the vast numbers of mutants involved.  Test execution overheads alone can limit the scalability of the technique. For instance, in our experiments, we needed around 48 hours to execute the mutants for a single component of the systems we examined. At the same time the manual effort required by testers is escalated with larger programs as the number of mutants grows proportionally to program size. 

To reduce application cost, it is imperative to limit the number of mutants to those that are actually useful, prior to any manual mutant analysis  or test execution. 
Thus, we need to identify which mutants are killable in order to limit the manual effort involved in their identification, and also to identify the mutants that are subsuming (disjoint)\footnote{The term disjoint mutants refers to \changing{a minimal} subset of mutants that need to be killed in order to reciprocally kill the original set~\cite{5693206, PapadakisCT18}.}, in order to reduce unnecessary computations, and to provide accurate adequacy measurements~\cite{PapadakisHHJT16}. 

This problem is known as  the mutant selection problem~\cite{Papadakis+2019} and has been studied in the form of selective mutation~\cite{OffuttLRUZ96, ZhangGMK13}, i.e., restricting the number of transformations to be used, with limited success~\cite{Kurtz+2016, ChekamPBTS20}. Though, the key issue with mutant selection is the simple 
syntactic-based nature of the \changing{selection process.} \changing{The issue is that mutants are introduced everywhere with respect to simple language operators, e.g., by replacing an operator with another, that completely ignore the program and particular location semantics.} This operator matching mutant selection has the unfortunate effect of introducing mutants independent of their context and program semantics. 

We propose \our\footnote{Cerebro is a fictional device appearing in Marvel comics used by the X-Men to detect human mutants. More details in \url{https://en.wikipedia.org/wiki/Cerebro}.}, a machine learning technique that learns to identify interesting mutants given their context. In particular we learn the associations between mutants and their surrounding code. Our learning scope is a relatively small area around the mutation point that differentiates locally, the mutants that are useful from those that are not. This allows mutating the program elements to fit best to their context, instead of mutating entire codebases with every possible transformation, enabling inter-project predictions. 

\our operates at lexical level, with a simple code preprocessing. In particular, a mutant and its surrounding code is represented as a vector of tokens where all literals and identifiers, i.e., user defined variables, types, and method calls, are replaced with \changing{predefined, hence predictable,} identifier names. This allows restricting the related vocabulary and learning scope to a relatively small fixed size of tokens around the mutation points. Learning is performed using a powerful and language-agnostic machine translation technique\changing{~\cite{britz_arxiv_2017}} that we train on related code fragments and their labels. 


We consider useful, the subset of mutants that resides on top of the subsumption hierarchy and subsumes the others~\cite{Kurtz_2014}, \aka \emph{subsuming mutants}\changing{~\cite{JiaH09}}, for the set of all possible mutant instances produced by a given set of mutation operators. Mutant $M_1$ subsumes mutant $M_2$ if \changing{every test case detecting $M_1$} also detects $M_2$. This implies that the tests detecting the subsuming mutant will also detect the subsumed ones thereby making subsumed mutants redundant. 

We implemented \our and evaluated its ability to predict (inter-project predictions) subsuming mutants on a large set of programs, composed of 48 C programs (CoreUtils) and 10 Java projects (Apache Commons, Joda-Time, and Jsoup). Our results demonstrate that \our significantly outperforms both, random mutant selection and a supervised machine learning approach (used by previous research) on both, C and Java benchmarks. 

In particular, our results show that \our significantly outperforms the baselines. In Java projects, \our obtained \changing{2.81} 
times higher MCC\footnote{The \emph{Matthews Correlation Coefficient}~(MCC)~\cite{MATTHEWS1975442} is a reliable metric of the quality of prediction models~\cite{6824804}, relevant when the classes are of very different sizes, e.g. in case of C programs, 10.2\% subsuming mutants (positives) over 89.8\% non-subsuming mutants (negatives).}
 values,  an improvement of \changing{82\%} 
 in F-measure, \changing{68.88\%} 
 in Precision, and \changing{85.71\%} 
 in Recall over the state-of-the-art supervised machine learning. In C programs, \our obtained \changing{2.76} 
 times higher MCC values, \changing{3.72} 
 times higher precision, \changing{and} slightly \changing{increased} Recall value (\changing{4\% higher}). 
 The improvement measured in F-measure is approximately \changing{65\%}.

To \changing{put the predictions into a context and} understand its influence on mutation testing, we also validated \our~ \changing{in a controlled simulation of} the envisioned use case. In particular, we simulate a scenario where testers are guided by mutation testing,  i.e., they design test cases based on mutants. Therefore, fewer mutants imply less effort, while stronger mutants imply stronger tests. Our analysis shows that \our achieved more than twice the subsuming mutation scores\footnote{\emph{Subsuming mutation score} (MS*) is the ratio of the killed and the total number of subsuming mutants.} in both, C and Java programs that we use. At the same time \our required significantly less effort in terms of both, \changing{analyzed} equivalent mutants and test executions. 
In C programs, \changing{3.70\%} 
of the mutants analyzed by \our are equivalent, while \changing{55.56\% and 53.33\%} 
analyzed by random mutant selection and supervised learning, respectively are equivalent; \our also required 
\changing{91\% fewer} test executions than random selection and supervised learning, respectively. In Java programs, \our required the analysis of 
\changing{41\% and 36\% fewer} equivalent mutants, and 
\changing{92\% and 87\% fewer} test executions than random mutant selection and supervised learning, respectively.

%
%
%
%
%

%
All-in-all our paper makes the following contributions:
\begin{enumerate}
    \item We present \our, a powerful static subsuming mutant selection technique.
    
    \item \changing{We provide} evidence suggesting that \our successfully predicts subsuming mutants with \changing{0.85} 
    Precision, \changing{0.33} 
    Recall and \changing{0.46} 
    MCC.  
    
	\item We \changing{show} that \our significantly outperforms the current state-of-the-art, i.e., random mutant selection and previously proposed machine learning technique, by revealing 2 times the subsuming mutants, while analyzing 
	\changing{64\% to 67\% fewer} equivalent mutants and requiring 
	\changing{89\% to 92\% fewer} test executions.  
\end{enumerate}

The remainder of the paper is organized as follows. Section \ref{sec:background} introduces preliminary concepts necessary in subsequent sections. 
\changing{Section~\ref{sec:use-case} describes the envisioned use case for \our and elaborates on a particular motivating example.} 
Section \ref{sec:approach} describes the approach in detail. Section \ref{sec:research-question} introduces the research questions and Section \ref{sec:exp-setup} details the experimental setup. The results of our experimental evaluation are summarized in Section~\ref{sec:exp-results}. 
We discuss threats to validity in Section \ref{sec:threats-validity}. 
\changing{In Section~\ref{sec:discussion} we also discuss the impact of the abstraction process and mutants' context size on \our's prediction performance}.  
Finally, we discuss related work in Section~\ref{sec:related-work}, and present our conclusion \changing{and future work} in Section~\ref{sec:conclusion}.

\section{Background}
\label{sec:background}



\subsection{Subsuming Mutants}
\label{subsec:subsuming-mutants}


Mutation is a test adequacy criterion in which test requirements are represented by mutants \changing{that are} obtained by performing slight syntactic modifications to the original program. 
Then, the tester needs to design test cases in order to \emph{kill} the mutants, \ie to distinguish  the observable behavior between the mutant and the original program. Some mutants cannot be killed as they are functionally \emph{equivalent} to the original program. 
Hence, the quality of a test suite is measured by the mutation (adequacy) score, a percentage metric obtained by the ratio of killed mutants over the total number of \changing{(non-equivalent)} generated mutants.

Mutation testing is a promising, empirically validated software testing technique that hasn't achieved its full potential yet~\cite{Papadakis+2019}. 
It is often considered as computationally expensive, mainly due to the large number of  mutants that it introduces, which require analysis and execution with the related test suites. 
One may notice that the number of mutants is disproportionate with the number of test cases to kill them, since one test case can kill several mutants at the same time. Thus, the effort put \changing{into} analyzing and executing mutants that do not help \changing{to improve} test suites is wasted. 
Hence, it is desirable to analyze only the mutants that add value, i.e., subsuming mutants~\cite{JiaH09, 5693206, Kurtz_2014,Ammann_2014}. 

Intuitively, subsuming mutants are the minimum subset of all mutants that when killed, by any possible test suite, results in killing the entire set of \changing{killable} mutants.
\changing{
Given two mutants $M_1$ and $M_2$, it is said that $M_1$ subsumes $M_2$ if every test suite $T$ killing $M_1$ also kills $M_2$. 
Unfortunately, identifying subsuming mutants is undecidable as it is not possible to know a mutant's behavior under every possible input. Thus, researchers typically approximate them through test suites \cite{JiaH09, Ammann_2014, PapadakisHHJT16, Kurtz+2016, PapadakisCT18}.}

More precisely, let $M_1$, $M_2$ and $T$ be two mutants and a test suite, respectively, where $T_1 \subseteq T$ and $T_2 \subseteq T$ are the set of tests from $T$ that kill mutants $M_1$ and $M_2$, respectively, and $T_1 \neq \emptyset$ and $T_2 \neq \emptyset$, indicating that both $M_1$ and $M_2$ are killable mutants. 
We will say that mutant $M_1$ subsumes mutant $M_2$, if and only if, $T_1 \subseteq T_2$. 
In case $T_1=T_2$, we say that mutants $M_1$ and $M_2$ are indistinguishable \changing{for $T$}. The set of mutants which are both killable, and subsumed only by indistinguishable mutants are called \emph{subsuming mutants}. 

\changing{For example, if we have a mutant set of 3 mutants ($M_1$, $M_2$, and $M_3$) and a test set $T = \{t_1, t_2, t_3\}$, where $M_1$ is killed by $T_1 = \{t_1\}$; $M_2$ is killed by $T_2 = \{t_1, t_2\}$; and $M_3$ is killed by $T_3=\{t_3\}$.
We can notice that every time that we run a test ($t_1$) to kill mutant $M_1$ we will also kill mutant $M_2$. However, the opposite does not hold. 
Thus, we have two subsuming mutants, i.e., $M_1$ and $M_3$. 
}




\changing{
Subsuming mutation score (MS*) is the ratio between killed subsuming mutants over the total number of subsuming mutants~\cite{PapadakisHHJT16}. 
Subsuming mutation score has been proposed~\cite{Ammann_2014,PapadakisHHJT16,5693206} as a reliable metric to evaluate the effectiveness of testing techniques as it does not consider the presence of subsumed mutants. Subsumed mutants can artificially inflate the mutation score of a testing technique and can mislead its apparent ability to detect faults.
For instance, following our previous example, a test suite $\{t_1$, $t_2\}$ kills 66.7\% of all the mutants (i.e., $M_1$ and $M_2$), but 50\% of the subsuming ones ($M_3$ is not killed). 
}

Interestingly, killing subsuming mutants leads to the killing of all killable mutants, thus, testers needs to focus mutation analysis on subsuming mutants. 
\changing{
The problem though, is that one needs to know the subsumption relations between mutants in advance, before starting to analyze the mutants and designing tests. 
}
To deal with this issue, we \changing{introduce \emph{Cerebro}, a \emph{static} technique that predicts subsuming mutants \changing{without requiring any dynamic analysis}, with the aim to help testers decide on which mutants to use when performing mutation-guided test generation \cite{PapadakisM10b,FraserZeller2010}}. 


\subsection{Machine Translation}
\label{subsec:machine-translation}
Machine Translation can be considered as a transformation function $\mathit{transform(X) = Y}$, where the input $\mathit{X = \{x_1, x_2, \ldots, x_n\}}$ is a set of \emph{entities} that represents a component to be transformed, to produce the output $\mathit{Y = \{y_1, y_2, \ldots, y_n\}}$, which is a set of entities that represent a transformed (desired) component.
In the training phase, the transformation function learns on the example pairs $\mathit{(X,Y)}$ available in the training dataset.
In our context, $X$ contains the source code with an annotation that indicates the location and type of the mutation operator applied, and $Y$ contains the same information, plus a label that indicates whether the mutant is subsuming or not.

The transformation function is trained to append the label to a given mutant by training the function on the example pairs (Code+MutationAnnotation, Code+MutationAnnotation+Label), 
where Code+MutationAnnotation represents the source code with an annotation in the statement to indicate the mutation operator type applied. 
This learned transformation is used as our prediction model for predicting subsuming mutants. 
Among the several machine translation algorithms that have been suggested over the past years, we use the  RNN Encoder-Decoder which is established and is used by many recent studies~\cite{tufano_icsme_2019,tufano_tosem_2019,sutskever_arxiv_2014}.

\subsection{RNN Encoder-Decoder architecture}
\label{subsec:encoder-decoder-architecture}
The RNN Encoder-Decoder machine translation is composed of two major components: an RNN Encoder to encode a sequence of terms $x$ into a vector representation, and an RNN Decoder to decode the representation into another sequence of terms $y$.
The model learns a conditional distribution over an (output) sequence conditioned on another (input) sequence of terms: $P(y_1; \ldots; y_m|x_1; \ldots; x_n)$, where $n$ and $m$ may differ. 
For example, given an input sequence $x$ = $Sequence_{in}$ = $(x_1; \ldots; x_n)$ and a target sequence $y$ = $Sequence_{out}$ =  $(y_1; \ldots; y_m)$, the model is trained to learn the conditional distribution: $P(Sequence_{out}|Sequence_{in}) = P(y_1; \ldots; y_m|x_1; \ldots; x_n)$, where $x_i$ and $y_j$ are space-separated tokens.
A bi-directional RNN Encoder~\cite{britz_arxiv_2017} (formed by a backward RNN and a forward RNN) is considered the most efficient to create representations as it takes into account both past and future inputs while reading a sequence~\cite{bahdanau_arxiv_2014}.


\changing{
\section{Use Case Scenario and Motivation}
\label{sec:use-case}

\subsection{Use Case Scenario}
\label{sec:use-case-scenario}
Figure~\ref{fig:mutation-testing} shows an overview of how the testing process is performed when it is guided by mutation. We adapted this figure from the one published in \cite[Figure 5.2]{AmmannOffutt2008}. 
Given a program $P$ as input, the mutation testing process starts by creating a set $M$ of mutants forming the test requirements. Test requirements are satisfied when tests kill the mutants. 
Since the number of mutants are excessive and form the key cost factor of mutation testing \cite{Papadakis+2019}, testers select a subset $M'$ of mutants from $M$ to focus on their analysis. Then, testers pick a mutant $m \in M'$ and design a test $t$ capable of killing $m$ or \minrev{judge} it as equivalent and discard it. The process is \minrev{repeated} until the design of test is capable of killing a predefined ratio of mutants (threshold). Finally, the designed test suite $T$ is used to check the correctness of program $P$ (w.r.t. test suite $T$). \minrev{If} test \minrev{suite} $T$ \minrev{detects} some bug in program $P$, then $P$ has to be fixed and the same mutation testing procedure can again be employed. 

\begin{figure}[t]
\begin{center}
\includegraphics[width=0.5\textwidth]{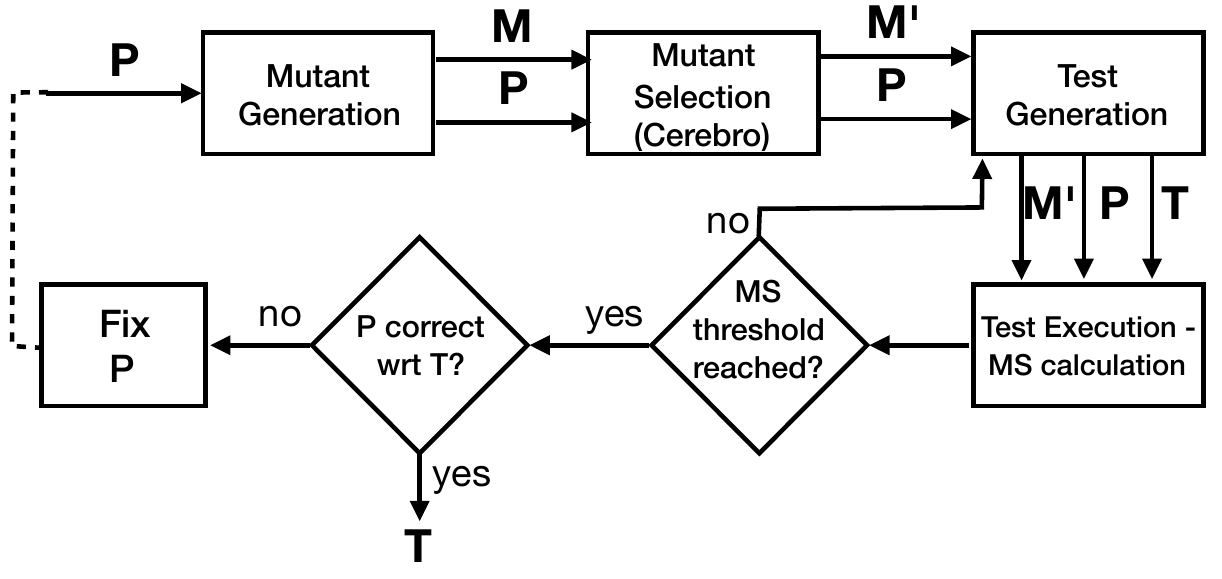}
\caption{\changing{\our Mutation Testing process. Given a program $P$ and a mutant set $M$, \our selects from $M$ a subset of mutants $M'$ to be used for test generation. $M'$ is then used to in Test generation, test execution and mutation score calculation steps.}}
\label{fig:mutation-testing}
\vspace{-1em}
\end{center}
\end{figure}

It is worth mentioning that there are two major cost factors in mutation testing, these are the equivalent and subsumed mutants. This is because they introduce overheads both during test generation and test execution, leading to minor test effectiveness improvements. Therefore,  to reduce mutation testing effort while preserving its effectiveness, it is essential to focus on subsuming mutants. 
}

\begin{figure*}[htp]
\begin{subfigure}{0.39\textwidth}
\begin{lstlisting}[language=PrettyJava]
int max(int a, int b, int c){
  if (a >= b && a >= c)(*@\Suppressnumber@*) //M0: (a < b && a >= c)
                        //M1: (a >= b && a > c)
                        //M2: (a >= b || a >= c)
                        //M3: (true && a >= c) (*@\Reactivatenumber{3}@*) 
    return a; //M4: return b;
  else if (b >= a && b >= c)(*@\Suppressnumber@*)//M5: (b < a && b >= c) 
                        //M6: (b >= a && b > c) 
                        //M7: (b >= a || b >= c) 
                        //M8: (false && b >= c) (*@\Reactivatenumber{5}@*) 
    return b; //M9: return a;
  else
    return c; //M10: return 0;
}
\end{lstlisting}
\caption{\changing{Code and mutants for function \texttt{max}.}}
\label{fig:max-function}
\end{subfigure}
\begin{subfigure}{0.24\textwidth}
\begin{center}
\includegraphics[width=\textwidth]{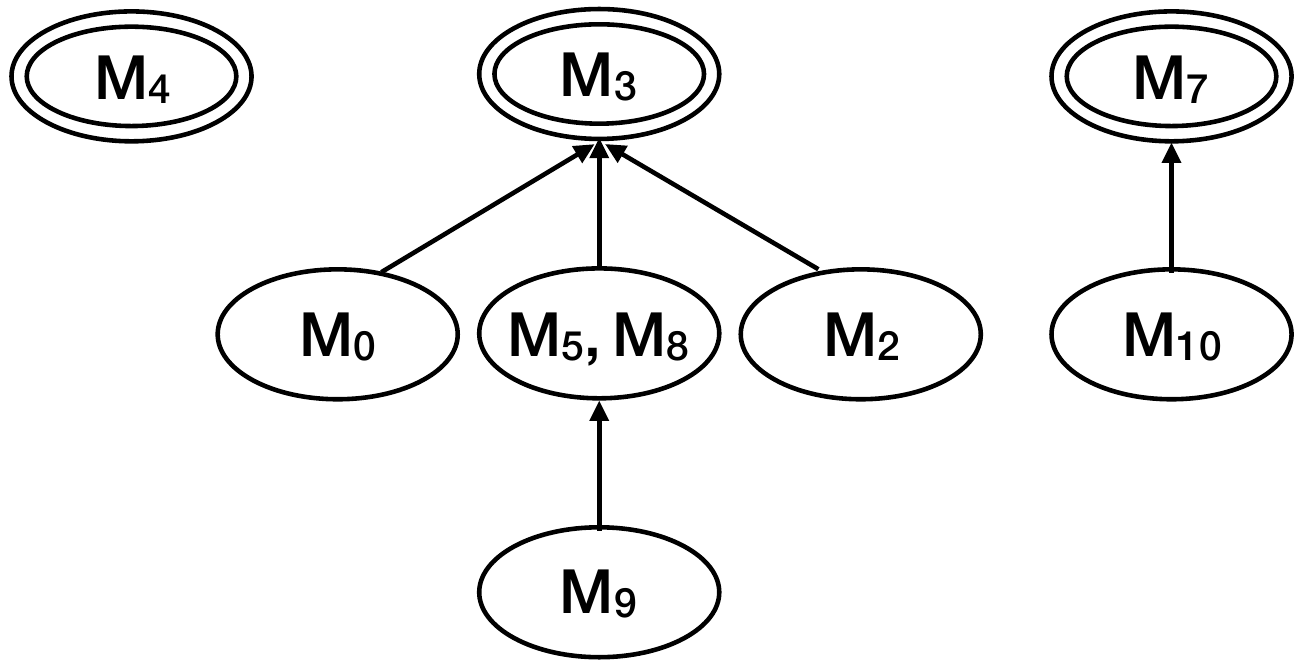}
\vspace{1em}
\caption{\changing{Subsuming Mutants Graph for function \texttt{max}.}}
\label{fig:max-subsuming-mutants}
\end{center}
\end{subfigure}
\hfill
\begin{subfigure}{0.35\textwidth}
\begin{center}
\includegraphics[width=\textwidth]{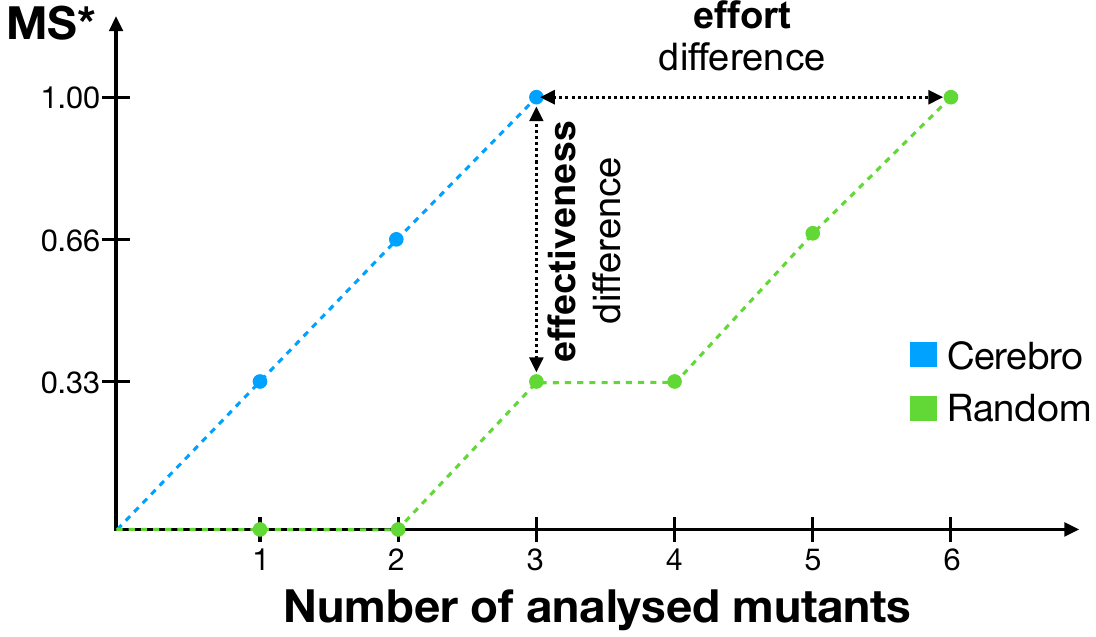}
\caption{\changing{Our motivating example shows that mutants selected by \our lead to stronger test suites than those designed to kill \minrev{randomly} selected mutants, when \minrev{equal} number of mutants is \minrev{analyzed}.}}
\label{fig:motivating_example_scenario}
\end{center}
\end{subfigure}
\caption{\changing{The example shows that by \minrev{analyzing} only the three subsuming mutants $M_3$, $M_4$ and $M_7$ is enough for covering all 9 killable mutants. Particularly, mutants $M_1$ and $M_6$ are equivalents.}}
\label{fig:motivating-example}
\end{figure*}

\changing{
Hence, we develop \our, a machine learning technique that learns from \minrev{mutants'} surrounding context to predict which mutants are subsuming.  
Given the input program, $P$ and the set $M$ of mutants, \our selects a subset $M'$ of mutants that is probably subsuming (predicted as subsuming by \our), to be used for mutation testing (to guide \minrev{testers} and evaluate test effectiveness). Based on $M'$, testers \minrev{and/}or automatic test generation techniques can focus on \minrev{the} few strong \minrev{mutants} and design effective test cases. 

\subsection{Motivating Example}
\label{sec:motivating-example}

Let us consider the code snippet of function \texttt{max} of Figure~\ref{fig:max-function}, which takes three integers as input and returns the maximum number among them. Also, consider (for simplicity) that we have the 11 mutants shown in the figure. For instance, mutant $M_0$ mutates sub-expression \texttt{a >= b} of line 2 into \texttt{a < b}. Similar mutations on relational operations were applied to produce mutants $M_1$, $M_3$, $M_5$, $M_6$ and $M_8$. 
Mutants $M_2$ and $M_7$ replace the conjunction (\texttt{\&\&}) by the disjunction (\texttt{||}). While mutants $M_4$, $M_9$ and $M_{10}$ replace the returned variable name by other variable name or constant ($M_{10}$ replaces variable name \texttt{c} by constant \texttt{0}).

For the sake of the thorough demonstration, we observed scenarios under the following testing conditions: 
A test case invoking \texttt{max(1,2,0)} and expecting \texttt{2} as a result, kills mutant $M_3$, as well as, mutants $M_0$, $ M_2$, $M_5$, $M_8$, and $M_9$. 
But tests invoking \texttt{max(2,0,1)}, \texttt{max(1,0,2)}, and \texttt{max(0,2,1)} will kill mutants $M_0$, $ M_2$, $M_5$, $M_8$, and $M_9$, except $M_3$.
Figure~\ref{fig:max-subsuming-mutants} shows a graph representation of the subsumption relation between the 9 killable mutants. Moreover, Figure~\ref{fig:max-subsuming-mutants} shows that $M_3$ subsumes $M_0$, $M_5$, $M_8$ and $M_2$. Particularly notice that mutants $M_5$ and $M_8$ are indistinguishable, since they are killed by the same tests, and subsume mutant $M_9$. Although, mutants $M_1$ and $M_6$ are equivalent. 
  
In summary, mutants $M_3$, $M_4$ and $M_7$ are subsuming, indicating that in order to kill every killable mutant it is sufficient to kill only these 3 subsuming mutants. 

\our will take as input the program \texttt{max} and the set of mutants, and it will point to those that are most likely subsuming. In an ideal scenario, \our would point only to $M_3$, $M_4$ and $M_7$, but it is possible, as in every machine learning based technique, that it does some mistakes, i.e., incorrect predictions of subsuming mutants, pointing to some non-subsuming (subsumed or equivalent mutants) as subsuming. 

For instance, consider the case in which \our predicts $M_3$ and $M_4$ and $M_{10}$ as subsuming mutants. 
Therefore, a tester will incrementally design test cases to kill all the predicted mutants.  
\minrev{Assume} that the tester starts by analyzing mutant $M_3$ and designs a test to kill it, e.g., by invoking \texttt{max(1,2,0)}. This test does not kill the rest of the selected mutants. 
\minrev{The} tester then proceeds to analyze the surviving mutant $M_4$, for which \minrev{he/she} designs a test \minrev{that invokes} \texttt{max(2,0,1)} to kill it. 
Finally, the tester designs a test by invoking \texttt{max(0,1,2)}, \minrev{which} kills mutant $M_{10}$ and also (non selected) subsuming mutant $M_7$. 
Notice that this test suite designed to kill all mutants selected by \our progressively increments the MS*:  first test kills subsuming mutant $M_3$ leading to a MS* of 33.33\%; second test kills subsuming mutant $M_4$, obtaining 66.66\% of MS*; and finally, third test kills collaterally subsuming mutant $M_7$ leading to a MS* of 100\%.  

Consider a scenario in which mutants are selected \emph{randomly}. For instance, assume that  $M_9$ is the first one to be selected for analysis for which a test case invoking \texttt{max(0,2,1)} is designed to kill it. This test  collaterally kills mutants $M_5$ and $M_8$, but it does not kill any subsuming mutant. 
Then, assume that equivalent mutant $M_1$ is randomly selected, adding no value to the testing process, but requiring analysis anyway. 
Afterwards mutant $M_0$ is randomly selected for which a test case invoking \texttt{max(2,0,1)} is designed to kill it, that fortunately also kills subsuming mutant $M_4$. 
Then, mutant $M_2$ is randomly selected for which the tester designs a test  to kill it by invoking \texttt{max(1,0,2)}. This test also kills mutant $M_{10}$, but no subsuming mutant is killed. 
After that, tester randomly selects mutant $M_3$ for analysis and designs a test by invoking \texttt{max(1,2,0)} to kill \minrev{it}. This test kills subsuming mutant $M_3$ and also mutant $M_2$. Finally, mutant $M_4$ is randomly selected for which the tester designs a test to kill it, by invoking \texttt{max(2,0,2)}. 
Hence, \minrev{all subsuming mutants are killed.}

In this particular scenario we can observe that MS* remains \minrev{at} 0\% after \minrev{analyzing the} first 2 mutants randomly selected, and reaches a MS* of 33.33\% after \minrev{analyzing} the third randomly selected mutant. 
The analysis of the fourth selected mutant (non-subsuming) did not add value (MS* remains the same). Finally, fifth and sixth \minrev{analyzed} mutants were subsuming, leading to a test suite that obtains \minrev{MS* of} 100\% after \minrev{analyzing} 6 mutants. 

Figure~\ref{fig:motivating_example_scenario} depicts the progress of MS* obtained by the test suites when guided by \our and random mutant selection in the previously described scenarios. Through this example we demonstrate a case where two approaches \minrev{analyze} the same number of mutants (same effort) with \our having higher effectiveness (MS*) than the random mutant selection baseline. At the same time, in order to reach the same MS* as \our, random mutant selection needs more effort, i.e., it will require the analysis of many more mutants than \our (in the example random baseline \minrev{analyzed} two times more mutants than \our).




There are several points we want to highlight about the particular scenarios just described. 
First, it is essential to notice that mutants selected by \our will be as close as possible to subsuming in the subsumption relation. Killing these (almost subsuming) mutants can help \minrev{in} killing subsuming mutants predicted as non-subsuming by \our, for instance, the test that kills subsumed mutant $M_{10}$, also kills subsuming mutant $M_7$ that was incorrectly predicted as non-subsuming by \our. 
Second, it is also important to notice that \our selects the least possible number of equivalent mutants, saving the time of analysis to the tester (in the example, \our did not predict any equivalent mutant as subsuming). 
Third, notice that the prediction performance obtained by \our does not necessarily reflect its effectiveness in practice, since mutant kills are not independent \minrev{of} one another. While \our reached 66.66\% of Precision and 66.66\% of Recall in the example, in practice, the test suite designed to kill all selected mutants obtains 100\% of subsuming mutation score (MS*). 
And fourth, it is worth to study the \minrev{trade-off} between the effectiveness and  effort of the different mutant selection techniques.  
We consider all these points in our empirical evaluation to assess the prediction performance, effectiveness\minrev{,} and effort required by \our and the related mutant selection techniques.
}

\section{Approach}
\label{sec:approach}

The main objective of \our is to automatically learn the silent features/patterns of the context surrounding subsuming mutants without requiring any features definition and/or selection by human intervention, that we can use later to predict if mutants on an unseen source code are likely to be subsuming or not. 
Thus, we train a machine translator (viz. an encoder-decoder model) to identify subsuming mutants, by feeding it with source code where the statement (to mutate) is annotated with the mutant type and its label (subsuming or not). 
Machine translators have been successfully used to translate text from one language to another, as they automatically recognize
(i) the features of the language (to be translated) and (ii) the required translation (to the desired language). 
In our case, it is used to automatically identify the features of subsuming mutants without any investment of time and/or resources to define features. 

After training, one can input to the translator, an unseen mutant (source code where the statement to mutate is annotated with the mutation annotation).
The translator will append the label to the mutant given as input, to predict whether it is subsuming or not. 

\begin{figure*}[htp]
\begin{center}
\includegraphics[width=\textwidth]{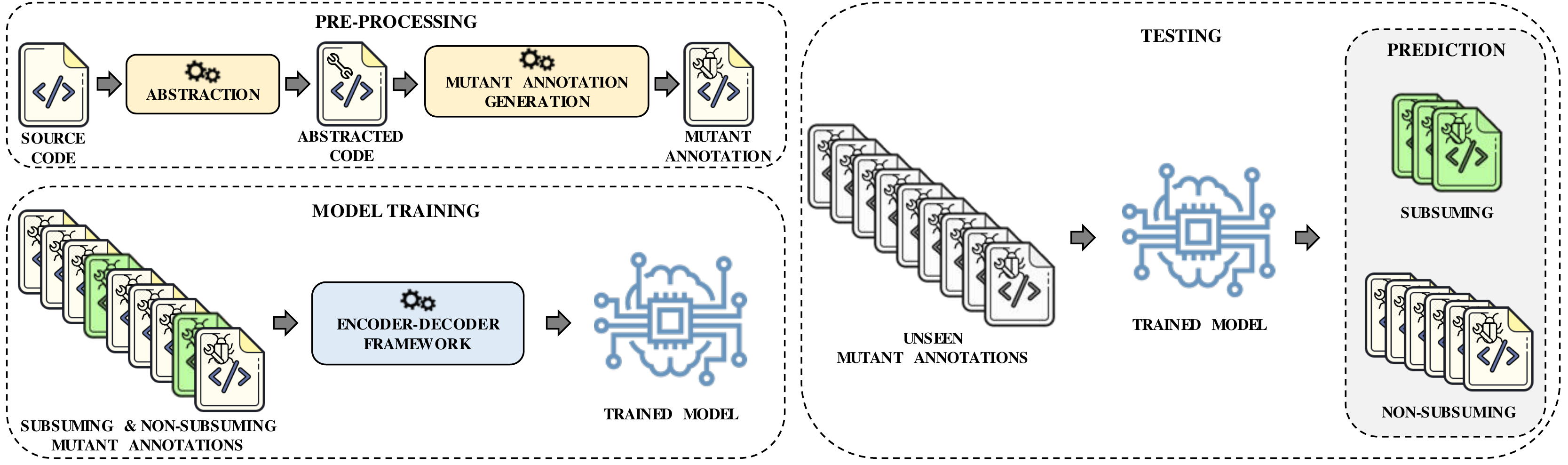}
\vspace{-1.5em}
\caption{Implementation: Source code is abstracted and attached with mutation annotation to produce mutant annotations. Model is trained on mutant annotations to further append the label (subsuming/non-subsuming).
Trained model is provided with an unseen mutant annotation to append the label. The appended label acts as the prediction for the unseen mutant annotation.}
\label{fig:1}
\vspace{-1em}
\end{center}
\end{figure*}

Figure~\ref{fig:1} shows an overview of the implementation. 
For training, \our takes a set of mutants and their corresponding label. 
In each mutant source code, the statement (to mutate) is annotated with the mutation annotation, and the model learns the label to be appended to this annotation, that indicates whether the mutant is subsuming or non-subsuming. 
We can summarize \our's pre-processing, training and testing steps as follows:

\begin{enumerate}
\item \emph{Abstraction:} Producing abstracted code of the actual source code by removing irrelevant information (e.g. comments) and replacing user-defined identifiers and literals (e.g. variable names) by predictable tokens; 

\item \emph{Pairs Generation:} Generating the pairs (input-expected~output) to be used for training, by adding the corresponding label into the mutation annotations; 

\item \emph{Training:} Training the machine translator to learn which label is to be appended to the mutation annotations;

\item \emph{Testing:} Utilizing the trained translator to predict and append labels to the mutation annotations present in unseen mutant source code.
\end{enumerate}

In the remainder of this section we describe each of the aforementioned phases of our approach, in detail.

\subsection{Abstracting the Irrelevant Information}

A major challenge in dealing with raw source code is the huge vocabulary created by the abundance of identifiers and literals used in the code.
On such a large scale, vocabulary may hinder the goal of learning features surrounding the subsuming mutants. 
Thus, to reduce vocabulary size, we abstract source code by replacing user-defined entities with re-usable identifiers.

\begin{figure*}[htp]
\vspace{-0.5em}
\begin{subfigure}{0.25\textwidth}
    \begin{lstlisting}[language=PrettyJava]
...
public String getOptionValue (
 final Option option ) {
 if ( option == null ) {
  return null ;
 }
 final String [] values = 
  getOptionValues ( option ) ;
 return ( values == null ) ? 
  null : values [ 0 ] ;
}
...
\end{lstlisting}
\vspace{-1em}
    \caption{Actual Source Code}
    \label{fig:2a}
  \end{subfigure}
\hfill
\begin{subfigure}{0.25\textwidth}
    \begin{lstlisting}[language=PrettyJava]
...
public String fn_3 (
 final tp_1 vr_3 ) {
 if ( vr_3 == null ) {
  return null ;
 }
 final String [] vr_5 = 
  fn_4 ( vr_3 ) ;
 return ( vr_5 == null ) ?
  null : vr_5 [ 0 ] ;
}
...
\end{lstlisting}
\vspace{-1em}
    \caption{Abstracted Code}
    \label{fig:2b}
  \end{subfigure}
\hfill
\begin{subfigure}{0.40\textwidth}
\begin{tikzpicture}
\node[anchor=south west,inner sep=0] at (0,0) {
\begin{lstlisting}[language=PrettyJava]
...
public String fn_3 (
 final tp_1 vr_3 ) {
 if ( vr_3 == null ) { 
  return null ; /+MST[ReturnValsMutator]MSP[  ]+/
 }
 final String [] vr_5 = 
  fn_4 ( vr_3 ) ;
 return ( vr_5 == null ) ?
  null : vr_5 [ 0 ] ; 
}
...
\end{lstlisting}
};

\draw[blue,thick,rounded corners] (5.78,1.65) rectangle (6.05,1.90);
\node (name) at (5.5,0.9) {\emph{Label}};
\node (label) at (5.5,0.6) {\textbf{S / N}};
\path[->] (5.9,1.64) edge[blue,-latex, line width=0.2ex] (name);
\end{tikzpicture}

    \caption{Mutant Annotation}
    \label{fig:2c}
  \end{subfigure}
% ...
\caption{Abstraction: Actual Source Code (\ref{fig:2a}) is abstracted by replacing user-defined entities (Function names, Type names, Variable names) with tokens (fn\_num, tp\_num, vr\_num) to achieve the Abstracted Code (\ref{fig:2b}). Mutant annotation (\ref{fig:2c}) is generated by adding the Mutation annotation with its corresponding label, \ie Subsuming (S) or Non-Subsuming (N). The trained model is used for prediction of unseen mutant annotations.}
\label{fig:abstraction}
\vspace{-1em}
\end{figure*}

Figure \ref{fig:abstraction} shows an actual code snippet (Figure~\ref{fig:2a}) converted into its abstract representation (Figure~\ref{fig:2b}). 
The purpose of this abstraction is to replace any reference to user-defined entities (function names, types, goto labels, variable names and string literals) by \changing{identifiers} that can be reused across source code file, hence reducing the vocabulary size. 
Thus, our abstraction approach first detects user-defined entities before replacing them with unique identifiers (new IDs). 

New IDs follow the regular expression \texttt{(fn|tp|lb|vr|lr)\_(num)$^+$}, where \texttt{num} stands for numbers $1,2,3, \ldots$ assigned in a sequential and positional fashion based on the occurrence of that entity.
All the user-defined \emph{Function} names, \emph{Type} names, \emph{Variable} names, \emph{Labels}, and \emph{String Literals} are replaced with \texttt{fn\_num}, \texttt{tp\_num}, \texttt{lb\_num}, \texttt{vr\_num}, and \texttt{lr\_num}, respectively.
Thus, the first function name found receives the ID \texttt{fn\_1}, the second receives the ID \texttt{fn\_2}, and so on.
If any of these entities appear multiple times in a source code file, it is replaced with the same ID.

Additionally, \changing{we remove code comments and add mutation annotations} to encode the mutation operator and the corresponding label (to be learned by the translator).
Our mutation annotations have the general shape ``\texttt{MST[}+MutationOperator+\texttt{]MSP[]}'', where MST and MSP denote mutation annotation start and stop, respectively, and MutationOperator indicates the applied mutation operation (in green in Figure~\ref{fig:2c}). 
Between the last brackets \texttt{[]}, our trained model adds one of the labels \texttt{S} or \texttt{N}, indicating that the mutant obtained by applying the mutation operation, is predicted as subsuming or non-subsuming, respectively.

\subsection{Pairs Generation}
\label{subsec:PairsGeneration}
The mutation operation (\texttt{ReturnValsMutator}\footnote{\url{https://pitest.org/quickstart/mutators/#RETURN_VALS}}) shown in Figure~\ref{fig:2c} represents a mutant in which the sentence  \texttt{return null} is replaced by \texttt{throw new java.lang.RuntimeException()} exception. 
Notice that this mutant is labeled as subsuming in our dataset, since there is only one test that can kill it, when the input \texttt{option} is null. Hence for training we consider  \texttt{S} as the label to be learned by the translator to predict this mutant as subsuming. 

To do so, we train in pairs (MutantAnnotation, MutantAnnotation+Label), where the first component is the annotated code shown in Figure~\ref{fig:2c}, and the second component is the same code with the predicted label, i.e., \texttt{MST[ReturnValsMutator]MSP[S]} in our case, to indicate that the mutant is subsuming.
The resulting text is arranged in a single sentence to represent a sequence of space-separated entities (the representation supported by the machine translator). 
The only difference between the input sequence given to the translator and the expected output sequence produced by it, is the predicted label \texttt{S} or \texttt{N}. \changing{Using these sequences,} we intend to capture as much code as possible around the mutant without incurring the exponential increase in training time. 

\subsection{Building the Machine Translator}

To build our machine translator, we train an encoder-decoder model that can transform an input sequence to a desired output sequence. 
In our representation, a sequence consists of tokens separated by spaces that ends with a newline character. 
Thus, we train the encoder-decoder by feeding it with pairs of sequences, produced in the previous step. 
The translator learns to replicate the abstracted source code with the mutation annotation and to append the label (\emph{S}/\emph{N}) that will be used as a prediction for the mutant.

\changing{We found that training the translator on sequences of maximum \emph{100} tokens in length is computationally feasible, but expensive (740 training hours required on a Tesla V100 GPU). Hence, we also experiment with sequences of \emph{50} tokens in length and demonstrate that the computation cost of training the translator can be further contained (360 training hours required). We name \our trained on sequences of 100 tokens in length as \emph{\our-100}. Following our naming convention, we name \our trained on sequences of 50 tokens in length as \emph{\our-50}.} 

\subsection{Predicting from appended labels}
\label{subsec:PredictingFromAppendedLabels}
To predict whether or not a certain mutation at a particular position in an unseen code is subsuming, we abstract the unseen code followed by sequence generation which results in abstracted code sequence attached with mutation annotation as depicted in Figure \ref{fig:1}. 
We feed this sequence into the trained machine translator to yield an output sequence with an appended label.
The appended label acts as a prediction (subsuming/non-subsuming) for this specific mutation. 
If the translator produces an output sequence with a change other than appending the predicted label, the input sequence is predicted as non-subsuming, by default. 
In our experiments reported in Section~\ref{sec:exp-results}, this happened in 4.2\% and 0.1\% of the sequences for C and Java programs, respectively.

\section{Research Questions}
\label{sec:research-question}

We start by checking the prediction ability of \our and ask: 
\begin{enumerate}
\item [\textbf{RQ1}] \emph{Prediction Performance:} How effective is \our in predicting subsuming mutants?
\end{enumerate}

We leverage two datasets, made of C and Java programs, 
for which extensive mutation analysis has been performed to identify subsuming mutants. 
We reimplemented 2 techniques that we use as baselines in our analysis. 
The first baseline is a \emph{Random} mutant sampling, while the second is a supervised machine learning method based on manually designed features that were used by previous work \cite{ChekamPBTS20} (e.g., data flow, control flow, etc.). These features are used to train a binary classifier in order to predict whether a mutant is subsuming or not. Further details about the baselines can be found in Section~\ref{subsec:baselines}.

After analyzing the predictions, we turn our attention to the envisioned application scenario; measuring test effectiveness of the predicted mutants. It is important to check the application case because a) predictions may select weak mutants \cite{ChekamPBTS20} (weak subsuming mutants result in lower test effectiveness \changing{than} the strong ones), b) selected mutants may not be diverse as they may include mutually subsuming mutants \cite{Kurtz_2014}, and c) tester benefits are unclear. Thus, we ask:

\begin{enumerate}
\item [\textbf{RQ2}] \emph{Effectiveness Evaluation:} How does \our compare with the baselines in terms of subsuming mutation score? 
\end{enumerate}

We perform a simulation of a mutation testing scenario where a tester analyzes the selected mutants in order to generate tests~\cite{Andrews+2006,Kurtz+2016, ChekamPBTS20}. For test effectiveness, we measure the subsuming mutation score (MS*) achieved by the tests that kill the selected mutants. In essence, we evaluate the guidance offered by the mutants when testers design tests to kill the selected mutants. \changing{It is worth noticing that in this part of the experiment} we control the number of mutants, i.e., all  techniques analyze the same number of mutants. Such simulation is typical in mutation testing literature \cite{Andrews+2006,Kurtz+2016,ChekamPBTS20} and aims at quantifying the benefit of an approach over the other.

Complementary to the previous question, we compare the effort required by each technique to obtain the same level of test effectiveness. Hence, we first investigate the human effort measured in terms of the number of mutants analyzed by the tester, to reach the same subsuming mutation score using \our and the baselines.  
Hence, we ask:

\begin{enumerate}
\item [\textbf{RQ3}] \emph{Manual Effort:} \changing{How many} mutants require manual analysis in order to reach a given level of subsuming mutation score? 
\end{enumerate}

We perform a similar simulation of a testing scenario in which we measure how many mutants the tester needs to analyze (generate a test case to kill or \changing{judge} equivalence), until he/she obtains a determined subsuming mutation score. 
This allows us to quantify the human effort required by each approach to obtain the same benefit. 

Related to the previous question, we also investigate the number of test executions necessary to reach the same subsuming mutation score, by following the incremental process of mutation analysis, \ie a tester picking a mutant and analyzing it. If the picked mutant is killable, he/she generates a test case that kills it, and then checks if the remaining alive (not analyzed and not killed) mutants are \changing{collaterally} killed by the same test (by executing the generated test on alive mutants). The killed mutants are removed from the set of alive mutants. \changing{Then, we ask:}
\begin{enumerate}
\item [\textbf{RQ4}] \emph{Computational Effort:} \changing{How many} \minrev{test} executions are required in order to reach a given level of subsuming mutation score? 
\end{enumerate}

We perform a simulation as before, but in this case, every time that a test is generated, we count the number of test executions and measure the attained subsuming mutation score, until we reach a given subsuming mutation score.

\section{Experimental Setup}
\label{sec:exp-setup}

\subsection{Benchmarks and Ground Truth}
\label{subsec:benchmarks}
In order to show that our approach is language agnostic, we make our evaluation on a set of C and Java programs.

\emph{C-Benchmark:}
\changing{To perform our study that requires strong test suites, we used an independently built dataset from related work~\cite{Chekam+2021}. It includes} C programs from the GNU Coreutils\footnote{https://www.gnu.org/software/coreutils/}, that consist of file, text and shell utility programs widely used in Unix systems.  The data-set is composed of 48 GNU Coreutils (v8.22) programs~\aka subjects \changing{(mentioned in Table~\ref{tab:java_benchmark_ver_info})}, each packaged with an accompanying system test suite, generated by developers. The size of these programs ranges from 1,000 to 14,000 lines of code (LOC), with a median size of 3,500 LOC. 
For each subject, the data-set includes a mutant-test killing matrix that records, for each mutant, a set of tests that kill it. 

\changing{The mutant-test killing matrices were obtained by generating mutants using the \emph{Mart} mutant generation tool~\cite{Chekam+2019} and executing them against large test pools. The test pools were built by considering developer tests and adding automatically generated tests using a 24 hours run of KLEE~\cite{Cadar+2008}. Additionally, mutation-based test suites were automatically generated using 128 different configurations of \emph{SEMu}~\cite{Chekam+2021}, each running for 2 hours, and an additional `seeded` test generation of KLEE. To reduce the total execution cost, for each program, the 3 functions that were covered by the largest number of developer tests were selected for mutation analysis, \ie mutants were generated only for these functions.}

We use these mutant-test killing matrices to compute the mutant subsumption, following the definition given in Section~\ref{subsec:subsuming-mutants}, and label each mutant as either subsuming or non-subsuming. To make the problem as balanced as possible (to assist in machine learning), we mark as subsuming all mutants in the top of the hierarchies, including mutually subsumed mutants.

Needless to say, it is possible to have some noise in our labeling process in the sense that mutants labeled as subsuming may be non-subsuming. \changing{The data-set reduced this noise by augmenting the test suites with} multiple large and diverse test suites generated by different state-of-the-art tools. Please refer to the threat in Section~\ref{sec:threats-validity} for a related discussion. 

\emph{Java-Benchmark:} For Java we select a set of well-tested open source projects from GitHub. We select projects from the Apache Commons Proper\footnote{https://commons.apache.org} repository of reusable Java components, Joda-Time\footnote{https://github.com/JodaOrg/joda-time/} - a date and time library, and Jsoup\footnote{https://github.com/jhy/jsoup} - an HTML manipulation library. The set counts 10 projects: \texttt{commons-cli}, \texttt{commons-codec}, \texttt{commons-collections}, \texttt{commons-csv}, \texttt{commons-io}, \texttt{commons-lang}, \texttt{commons-net}, \texttt{commons-text}, \texttt{jsoup}, \texttt{joda-time}. 
These projects contain up to 284 classes. \changing{Table~\ref{tab:java_benchmark_ver_info} reports the version/commit of each project we used for our study.} 
Following a similar procedure done for C in~\cite{Chekam+2021}, we also build test pools by using developer tests and adding automatically generated tests by running EvoSuite\cite{FraserZeller2010} for each project with the default running time, but with multiple coverage metrics\footnote{LINE:BRANCH:MUTATION:OUTPUT:METHOD:CBRANCH}. The mutant-test killing matrices were obtained using Pitest~\cite{Coles+2016}. For each project, we run the mutants on the test pools for 48 hours. To reduce execution time, we select the classes processed during that time lapse.

Table~\ref{tab:Subjects_info} records the total number of mutants, number (and percentage) of killable and subsuming mutants, and number of test cases conforming to the mutant-test killing matrices. Please note that the difference on the ratio of subsuming mutants with previous research \cite{PapadakisHHJT16, Ammann_2014,Kurtz_2014} is due to the inclusion of all mutually subsuming mutants. As already explained, we include all subsuming mutants to avoid misleading our learner. 

\begin{table}[!t]
  \caption{\changing{Benchmark}}
   \vspace{-1.0em}
  \label{tab:java_benchmark_ver_info}
  \resizebox{.48\textwidth}{!}{
  \begin{tabular}{| l | l | l |}
  \hline
  \rule{0pt}{3ex}\textbf{\changing{Project}} & \textbf{\changing{Web URL}} &  \textbf{\changing{\thead{Version /\\Commit}}} \\ \hline 
  \multicolumn{3}{|c|}{\rule{0pt}{3ex} \textbf{\changing{C}}} \\ \hline 
    \rule{0pt}{3ex} \changing{base64, }\changing{basename, } &  & \\
    \rule{0pt}{0ex} \changing{chcon, }\changing{chgrp, } &  & \\
    \rule{0pt}{0ex} \changing{chmod, }\changing{chown, } &  & \\
    \rule{0pt}{0ex} \changing{chroot, }\changing{cksum, } &  & \\
    \rule{0pt}{0ex} \changing{comm, }\changing{date, } &  & \\
    \rule{0pt}{0ex} \changing{df, }\changing{dirname, } &  & \\
    \rule{0pt}{0ex} \changing{echo, }\changing{expr, } &  & \\
    \rule{0pt}{0ex} \changing{factor, }\changing{false, } &  & \\
    \rule{0pt}{0ex} \changing{groups, }\changing{join, } &  & \\
    \rule{0pt}{0ex} \changing{link, }\changing{logname, } &  & \\
    \rule{0pt}{0ex} \changing{ls, }\changing{md5sum, } &  & \\
    \rule{0pt}{0ex} \changing{mkdir, }\changing{mkfifo, } & \changing{https://github.com/coreutils/coreutils.git} & \changing{v8.22} \\
    \rule{0pt}{0ex} \changing{mknod, }\changing{mktemp, } &  & \\
    \rule{0pt}{0ex} \changing{nproc, }\changing{numfmt, } &  & \\
    \rule{0pt}{0ex} \changing{pathchk, }\changing{printf, } &  & \\
    \rule{0pt}{0ex} \changing{pwd, }\changing{realpath, } &  & \\
    \rule{0pt}{0ex} \changing{rmdir, }\changing{sha256sum, } &  & \\
    \rule{0pt}{0ex} \changing{sha512sum, }\changing{sleep, } &  & \\
    \rule{0pt}{0ex} \changing{stdbuf, }\changing{sum, } &  & \\
    \rule{0pt}{0ex} \changing{sync, }\changing{tee, } &  & \\
    \rule{0pt}{0ex} \changing{touch, }\changing{truncate, } &  & \\
    \rule{0pt}{0ex} \changing{tty, }\changing{uname, } &  & \\
    \rule{0pt}{0ex} \changing{uptime, }\changing{users, } &  & \\
    \rule{0pt}{0ex} \changing{wc, }\changing{whoami~\cite{Chekam+2021}} &  & \\ \hline
  \multicolumn{3}{|c|}{\rule{0pt}{3ex} \textbf{\changing{Java}}} \\ \hline 
  \rule{0pt}{3ex} \changing{commons-cli} & \changing{https://github.com/apache/commons-cli.git} & \changing{6490067} \\ \hline 
  \rule{0pt}{3ex} \changing{commons-collections} & \changing{https://github.com/apache/commons-collections.git} & \changing{d6eeceb} \\ \hline 
  \rule{0pt}{3ex} \changing{commons-text} & \changing{https://github.com/apache/commons-text.git} & \changing{26a308f} \\ \hline 
  \rule{0pt}{3ex} \changing{commons-csv} & \changing{https://github.com/apache/commons-csv.git} & \changing{865872e} \\ \hline 
  \rule{0pt}{3ex} \changing{commons-lang} & \changing{https://github.com/apache/commons-lang.git} & \changing{2c0429a} \\ \hline 
  \rule{0pt}{3ex} \changing{commons-io} & \changing{https://github.com/apache/commons-io.git} & \changing{c126bdd} \\ \hline 
  \rule{0pt}{3ex} \changing{commons-net} & \changing{https://github.com/apache/commons-net.git} & \changing{33df028} \\ \hline 
  \rule{0pt}{3ex} \changing{commons-codec} & \changing{https://github.com/apache/commons-codec.git} & \changing{475910a} \\ \hline 
  \rule{0pt}{3ex} \changing{jsoup} & \changing{https://github.com/jhy/jsoup.git} & \changing{528ba55} \\ \hline 
  \rule{0pt}{3ex} \changing{joda-time} & \changing{https://github.com/JodaOrg/joda-time.git} & \changing{767c94e} \\ \hline
  \end{tabular}
  }
 \vspace{-0.5em}
\end{table}

\begin{table}[!t]
  \caption{Test Subjects}
   \vspace{-1.0em}
  \label{tab:Subjects_info}
  \resizebox{.48\textwidth}{!}{
  \begin{tabular}{| l | r | r | r | r | r |}
  \hline
  \rule{0pt}{3ex}\textbf{Language} & \textbf{\#Programs} &  \textbf{\#Mutants} & \textbf{\#Killed} &  \textbf{\#Subsuming} & \textbf{\#Testcases} \\ \hline 
  \rule{0pt}{3ex}C~\cite{Chekam+2021} & 48 & 71,850 & 49,530 (68.9\%) & 7,358 (10.2\%) & 136,412 \\ \hline 
  \rule{0pt}{3ex}Java & 10 & 153,823 &  124,064 (80.6\%) & 41,219 (26.8\%) & 21,878 \\ \hline 
  \end{tabular}
  }
 \vspace{-0.5em}
\end{table}

\changing{
\subsection{Equivalent Mutants}
\label{subsec:problem-of-mutant-equivalence}
Early research on mutation testing has demonstrated that deciding whether a mutant is equivalent is an undecidable problem~\cite{BuddA82}. Mutation testing may produce a mutant that is syntactically different from the original, yet semantically identical, \aka equivalent mutant~\cite{Kintis+2018}. Undecidability of equivalences means that it is impossible to automatically discard them all. As a result, the tester may never know whether he or she has failed to find a killing test case because the mutant is particularly hard to kill, yet remains killable (a ‘stubborn’ mutant~\cite{yao2014study}), or whether failure to find a killing test case derives from the fact that the mutant is equivalent. The best \minrev{options we have are} effective algorithms that can remove most equivalent mutants, e.g., in C data-set~\cite{Chekam+2021} authors applied TCE (Trivial Compiler Equivalence) \cite{Kintis+2018, HaririSFMM19} to filter out equivalent and duplicated mutants. Interestingly, early research on mutation testing \cite{Acree80} has shown that humans also \minrev{make} many mistakes (approximately 20\%) when judging mutants (as being equivalent or not). This means that it is unrealistic to expect that automated tools (or testers, in case of manual test case design) kill all killable mutants. 

To make a fair approximation of killable mutants we used state-of-the-art test generation tools (KLEE\cite{Cadar+2008}, SEMu\cite{Chekam+2021}, and EvoSuite\cite{FraserZeller2010}), together with mature developer test suites to identify killable mutants. For the remaining live mutants (i.e., mutants that are killed neither by developers written nor automatically generated test suites) we assumed that live mutants are equivalent. Although, this assumption may have some impact on our results (refer to Section \ref{sec:noise-equivalent-mutants} for an analysis of the impact of this assumption), it allows quantifying the effort involved by testers in analyzing low utility mutants when using the current state-of-the-art advances. Moreover, since \our performs machine learning, it learns from the employed data. This means that the availability of clean data, with a clear signal to learn, will allow \our make better predictions, thereby potentially improving its performance.}

\subsection{Baselines}
\label{subsec:baselines}
We consider 2 baselines. The first one is the \emph{Random} mutant sampling that samples uniformly from the entire set of mutants. The second baseline is a Decision Tree classification based on the features proposed by related work \cite{ChekamPBTS20, JustKA17}.

Previous works showed a strong connection between mutant utility and surrounding code (utility captured through CFG, data flows, AST, etc. features). Thus, we use the mutant features to predict subsuming mutants in both C and Java. Features belong to 4 categories: Mutant Type related features, Control-Flow graph related features, Control and Data dependency related features, and AST related features. In total we used the 28 features, used by the related work~\cite{ChekamPBTS20}, for the C programs, and implemented 16  of those features for Java\footnote{statementComplexity, expressionComplexity, MutantType, BlockDepth, CfgDepth, CfgPredNum, CfgSuccNum, NumInBlock, NumOutDataDeps, NumInDataDeps, NumOutCtrlDeps, NumInCtrlDeps, AstNodeParentType, NumberOfAstParents, AstNodeType, NumberOfAstChildren}. We excluded features  such as AstChildHasIdentifier and AstChildHasLiteral that we found unfeasible to implement in the employed tools, i.e., Pitest works at byte-code level making it difficult to identify the original source code expression. Nevertheless, the excluded features were approximated by mutant type.

After extracting the features, following the related work~\cite{ChekamPBTS20}, we trained a stochastic gradient boosted Decision Tree model by using the same configuration as the related work~\cite{ChekamPBTS20}. We followed the same validation setup for \our. 

%


\subsection{Implementation and Model Configuration}
We rely on the \emph{srcML} tool~\cite{7816536} to convert source code into an XML format to tag literals, keywords, identifiers, comments, and our mutation annotations. This helps in separating user-defined identifiers and string literals (the largest part of the vocabulary) from language keywords as srcML supports C, Java and other languages. 
Then, we implement the ID replacement to generate the abstracted code. 

We follow the sequence pair generation procedure mentioned in Section~\ref{subsec:PairsGeneration} to generate sequences from the abstracted code.
These sequences serve as training input for our encoder-decoder model, which we build using \emph{tf-seq2seq}~\cite{tensorflow2015-whitepaper}, a general-purpose encoder-decoder framework. 
Following previous works~\cite{tufano_icsme_2019,tufano_tosem_2019}, we configure our model with bidirectional encoder. 
%
We use a Gated Recurrent Units (GRU) network~\cite{Cho+2014} to act as the Recurrent Neural Network (RNN) cell, which was shown to perform better than possible alternatives (simple RNNs or gated recurrent units) in related prediction tasks~\cite{lstmcomparison2019}. 
To achieve good performance with acceptable model training time, we utilize AttentionLayerBahdanau \cite{7472618} as our attention class, configured with 2 layered AttentionDecoder and 1 layered BidirectionalRNNEncoder, both with 256 units.

To determine an appropriate number of training epochs, we conducted a preliminary study involving a validation set, independent of both, training and test sets that we use in our evaluation. Here we incrementally train the model, with checks after every epoch to monitor model training accuracy.
We pursue training the model till the training performance on the validation set does not improve anymore.
We found 15 epochs to be a good default for our validation sets.
Once model training is complete, we follow the procedure explained in Section \ref{subsec:PredictingFromAppendedLabels} to predict whether an unseen mutant annotation sequence is subsuming or not.

The codebase of C and Java programs with mutant information, abstracted code, and mutant annotation sequences that the encoder-decoder model trains on and predict, with mapping to the original code, are publicly available \changing{at \textbf{\url{https://github.com/garghub/Cerebro}}. In addition to our dataset, we have made available our source code and trained models as well.}

\subsection{Experimental Procedure}
\label{subsec:procedure}

In the first experimental part, we evaluate the prediction ability of our approach, answering RQ1, while in the second part, we evaluate cost-effectiveness of \our, answering RQs2-4. 

\changing{\subsubsection{First Experimental Part}}
We start by evaluating the prediction performance of \our, and the baselines, using four typical metrics, namely, \emph{Precision}, \emph{Recall}, \emph{F-measure}, and \emph{Matthews Correlation Coefficient}~(MCC)\cite{MATTHEWS1975442}. 
\changing{
A confusion matrix is computed for each one of the studied methods, \minrev{which} stores the correct and incorrect predictions. 
Given a subsuming mutant, if it is predicted as subsuming, then it is a true positive (TP); otherwise, it is a false negative (FN). 
Given a non-subsuming mutant, if it is predicted as non-subsuming, then it is a true negative (TN); otherwise, it is a false positive (FP). 
Then we can use the confusion matrix to quantitatively evaluate the prediction performance of \our and \dt prediction models. 
\begin{align*}
&\emph{Precision} = \frac{TP}{TP + FP}  \hspace{3em} \emph{Recall} = \frac{TP}{TP + FN}\\
& \\
&\emph{F-measure} = \frac{2 \times \emph{Precision} \times \emph{Recall}}{\emph{Precision} + \emph{Recall}}\\
& \\
& \emph{MCC} = \frac{TP \times TN - FP \times FN}{\sqrt{(TP + FP)(TP + FN)(TN + FP)(TN + FN)}}
\end{align*}
}

Intuitively, \emph{Precision} is the ratio of mutants truly subsuming among all the mutants predicted as subsuming. 
\emph{Recall} is the ratio of mutants correctly predicted as subsuming among all the subsuming mutants. 
\emph{F-measure} indicates the weighted harmonic mean of \emph{Precision} and \emph{Recall}. 
\emph{Matthews Correlation Coefficient}~(MCC)~\cite{MATTHEWS1975442} is a reliable metric of the quality of prediction models \cite{6824804}, that in contrast to the previous metrics, also takes into account the True Negatives (correctly predicted non-subsuming mutants). It is generally regarded as a balanced measure that can be used even when the dataset is unbalanced, \ie the classes are of very different sizes, e.g. in case of C programs, 10.2\% subsuming mutants (Positives) over 89.8\% non-subsuming mutants (Negatives). 
MCC returns a coefficient between 1 and -1. An MCC value of 1 indicates a perfect prediction, whereas a value of -1 indicates a perfect inverse prediction, i.e., a total disagreement between prediction and reality.
An MCC value equals 0 indicates that the prediction performance is equivalent to random guessing.

\changing{The mutants selected by \our are the ones predicted as subsuming. For \dt baseline, as it computes a probability of a mutant being subsuming, we followed the probability margin convention and considered those mutants whose predicted probability was higher than 0.5~\cite{ChekamPBTS20}.}

To assess the performance we perform a inter-project evaluations. We use 5-folds cross validation, where we evenly split each benchmark in 5 parts (10 programs and 2 projects per fold for C and Java benchmark, respectively). Then, for each benchmark, we repetitively use 1 fold for testing and 4 folds for training (1 part out of 4, is used for validation). 

\changing{\subsubsection{Second Experimental Part}}
To study the cost and test effectiveness of our approach and the baselines, we simulate a testing scenario where a tester selects a subset of mutants, to use for mutation analysis, and designs tests to kill them. 
\changing{
Algorithm~\ref{algo:simulation} provides the pseudo-code of the simulation process we follow in our experiments. 
It takes as input a set \texttt{M} of mutants to \minrev{analyze}, the test pool \texttt{P} and a target subsuming mutation score \texttt{tMS*}, and returns a test suite \texttt{T} that kills every mutant from \texttt{M} (or reaches the pre-specified subsuming mutation score). Additionally, it returns the subsuming mutation score obtained by the test suite \texttt{T} (\texttt{currMS*}), number of analyzed mutants (\texttt{analyzedMut}), number of equivalent mutants analyzed (\texttt{equivMut}), and number of test executions (\minrev{\texttt{tExec}}) required to generate test suite \texttt{T} during the simulated mutation testing scenario.  
}

\begin{algorithm}[H]
\caption{\changing{Pseudo-code of the simulation procedure to answer RQ2-4.}}
\label{algo:simulation}
\minrev{
\begin{algorithmic}[1]
\renewcommand{\algorithmicrequire}{\textbf{Input:}}
\renewcommand{\algorithmicensure}{\textbf{Output:}}
\Require set of mutants $\texttt{M}$
\Require test pool $\texttt{P}$
\Require target subsuming mutation score $\texttt{tMS*}$ 
\Ensure test suite $\texttt{T}$ covering mutants in $\texttt{M}$
\Ensure subsuming mutation score \texttt{currMS*} obtained by \texttt{T}
\Ensure $\texttt{analyzedMut}$ number of analyzed mutants
\Ensure $\texttt{equivMut}$ number of equivalent mutants analyzed
\Ensure $\texttt{tExec}$ number of test executions
\State $\texttt{T} \gets \emptyset$
\State $\texttt{C} \gets \texttt{M}$ \Comment{set of survived mutants}
\State $\texttt{currMS*} \gets \texttt{0}$
\While {$\texttt{currMS* < tMS*}$ and $\neg\texttt{isEmpty(C)}$}
\State $\texttt{m} \gets \texttt{pickNextMutant(C)}$ 
\State $\texttt{analyzedMut++}$
\If {the test pool P can kill mutant $\texttt{m}$}
\State $\texttt{t} \gets \texttt{randomlyPickTestKilling(m, P)}$
\State $\texttt{T} \gets \texttt{T} \cup \{\texttt{t}\}$ \Comment{add test $\texttt{t}$ to the suite}
\State $\texttt{tExec += size(C)}$ \Comment{run $\texttt{t}$ on mutants from $\texttt{C}$}
\State remove from set $\texttt{C}$ all mutants killed by $\texttt{t}$
\Else 
\State $\texttt{equivMut++}$ \Comment{$\texttt{m}$ is judged as equivalent}  
\EndIf
\State $\texttt{currMS*} \gets \texttt{calculateMS*(M,T)}$
\EndWhile
\State \Return $\texttt{T}, \texttt{currMS*}, \texttt{analyzedMut}, \texttt{equivMut}, \texttt{tExec}$
\end{algorithmic}
}
\end{algorithm}

\changing{
The simulation starts by picking (\texttt{pickNextMutant}) the top mutant \texttt{m}, according to the technique used (\our, \dt\minrev{,} and \rnd), among survived mutants from set \texttt{C} (initialized with all mutants from \texttt{M}). It then checks if there exists some test in the test pool \texttt{P} that kill \texttt{m} (this process simulates a tester picking, analyzing\minrev{,} and designing a test to kill a mutant). 
If no test kills mutant \texttt{m}, we judge it as equivalent and remove it from \texttt{C}.
Otherwise, we randomly pick one test \texttt{t} from the pool that kills \texttt{m}. 
Then, we run the test \texttt{t} on every mutant from \texttt{C} to check if the same test consequently kills other mutants (killed mutants are then removed from \texttt{C}). 
This process continues by taking the next survived mutant and finding a test to kill it until every mutant in \texttt{C} has been killed or until the desired subsuming mutation score is reached.
}
\changing{
We run this simulation with the set of mutants selected by \our, \dt\minrev{,} and \rnd, respectively, and use the reported values to compare their cost-benefit performance for answering RQ2-4. 
Since Algorithm~\ref{algo:simulation} includes some random decisions, we repeat this process 1,000 times for all the approaches.
}

\changing{To answer RQ2, we} measure the effectiveness (benefit) of the approaches in terms of the \emph{subsuming mutation score} (MS*), \ie the ratio between killed and total number of subsuming mutants\changing{, achieved by the generated test suites when analyzing the selected mutants}. The subsuming mutation score reduces the influence of redundant mutants~\cite{PapadakisHHJT16,Kurtz_2014}. 

\changing{For assessing the effectiveness of the approaches, we aim at controlling the number of mutants selected by each tool.}
In the case of \our, the mutants selected are the ones predicted as subsuming by our model. 
For \dt baseline, we rank (in descending order) the mutants according to the predicted probability of being subsuming, and follow the ranking to pick mutants (from highest probability to lowest) for analysis. 
 \rnd baseline randomly ranks the mutants to be selected. 
\changing{Initially, we consider the same number of selected mutants}  for the 3 approaches, defined as the number of mutants predicted as subsuming by \our.  \changing{For instance, if \our predicts 20 mutants as subsuming, then \dt and \rnd baselines will also select the top 20 ranked mutants.} 
Our intention is to compare the effectiveness reached by each approach, when the number of selected mutants is equal. 

Additionally, we study the number of \emph{equivalent mutants} selected by each approach (as these are an important source of redundancy during mutation testing), as well as, the \changing{required} number of mutants selected by \changing{the baselines in order} to reach the same subsuming mutation score \changing{as \our}.

\changing{To answer RQ3 and RQ4, we} study the effort (cost) required by each approach in two ways. 
We measure the human effort in terms of the number of \emph{analyzed mutants}, killable or not, that are presented to testers for analysis (\ie either designing a test to kill these or judging these as equivalent), when applying mutation testing. 
Intuitively, for a given set of mutants, the number of analyzed mutants can be considerably smaller than the entire set's size because a test designed by \minrev{analyzing} one mutant can kill other mutants as well. 
Hence, we also measure the computational effort in terms of the number of \emph{test executions} performed, during the mutation analysis procedure, \ie we count the test executions required at every step where a new test is created. 
\changing{As for RQ2, here we also study} the number of test executions and the number of mutants that require analysis by the baselines, to reach the same subsuming mutation score as \our. 



\section{Experimental Results}
\label{sec:exp-results}

\begin{figure*}[t!]
\begin{center}
\includegraphics[width=\textwidth]{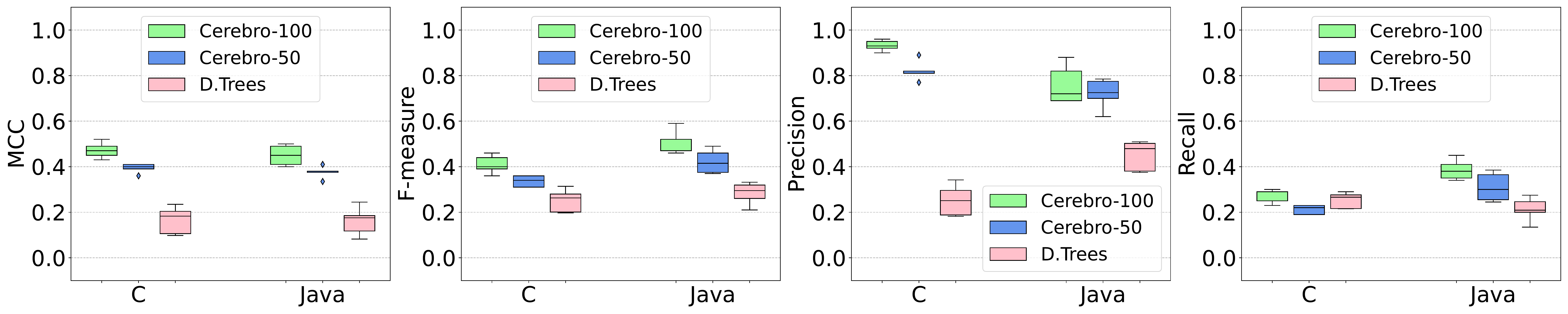}
\end{center}
\caption{(RQ1) Prediction Performance Comparison: On average, \changing{\our-100 outperforms \dt by 2.76 times, and 2.81 times higher MCC in C, and Java Benchmark.} Moreover, \changing{\our-50} outperforms \dt by 2.29 times, and 2.38 times higher MCC in C, and Java Benchmark. Overall, \our outperforms by 
\changing{2.78} times higher MCC than \dt.
}
\label{fig:rq1}
\end{figure*}

\subsection{Prediction Performance (RQ1)}
\label{subsec:RQ1-results}
Table~\ref{tab:rq1} records the average (and median) performance metrics. 
Figure~\ref{fig:rq1} shows the performance comparison in box plot format showing the distribution of performance indicators (MCC, F-measure, Precision, and Recall) for both approaches in C, and Java Benchmarks. 

\begin{table}[t!]
  \begin{center}
\caption{(RQ1) Prediction Performance of \our and \dt. On average, \our outperforms by \changing{2.78} 
times higher MCC than \dt.}
\label{tab:rq1}
\resizebox{.45\textwidth}{!}{
\begin{tabular}{| l | r | r | r | r |}
\toprule
\multicolumn{5}{c}{Average (and Median) Performance in C-Benchmark}\\
\toprule
\hline
\rule{0pt}{3ex}\textbf{Approach} & \textbf{MCC}  & \textbf{F-measure}  & \textbf{Precision}  & \textbf{Recall} \\ \hline
\rule{0pt}{3ex}\dt                & 0.17 (0.18) & 0.25 (0.26) & 0.25 (0.25) & 0.25 (0.27) \\ \hline
\rule{0pt}{3ex}\our-50 & 0.39 (0.40)  & 0.34 (0.34) & 0.82 (0.82) & 0.21 (0.22) \\ \hline 
\rule{0pt}{3ex}\changing{\our-100} & \changing{0.47 (0.47)}  & \changing{0.41 (\minrev{0.40})} & \changing{0.93 (0.93)} & \changing{0.26 (0.25)} \\ \hline
\multicolumn{5}{c}{}\\ 
\toprule
\multicolumn{5}{c}{Average (and Median) Performance in Java-Benchmark}\\\toprule
\hline
\rule{0pt}{3ex}\textbf{Approach} & \textbf{MCC}  & \textbf{F-measure}  & \textbf{Precision}  & \textbf{Recall} \\ \hline
\rule{0pt}{3ex}\dt                & 0.16 (0.18) & 0.28 (0.30) & 0.45 (0.48) & 0.21 (0.21)\\ \hline
\rule{0pt}{3ex}\our-50 & 0.38 (0.38)  & 0.42 (0.42) & 0.72 (0.73) & 0.31 (0.30) \\ \hline 
\rule{0pt}{3ex}\changing{\our-100} & \changing{0.45 (0.45)}  & \changing{0.51 (0.52)} & \changing{0.76 (0.73)} & \changing{0.39 (0.38)} \\ \hline
\end{tabular}
}
\end{center}
\end{table}

\changing{
On average, \our obtains a high Precision, i.e., 0.93 and 0.76 (\our-100), and 0.82 and 0.72 (\our-50) in C and Java benchmarks, respectively. 
Testers focusing on mutants selected by \our can be confident that these are very likely to be subsuming, providing high utility to the testing process. 
On the other hand, Recall achieved is low, i.e., 0.26 and 0.39 (\our-100), and 0.21 and 0.31 (\our-50) in C and Java benchmarks, respectively. 
This indicates that many subsuming mutants are mistakenly predicted as non-subsuming by \our. In practice these mutants can still be collaterally killed by other (mutually subsumed) subsuming \minrev{mutants} correctly predicted as subsuming by \our (which is often the case, as we will show when answering RQ2 in the following section). 
Needless to say, any complementary mutation testing and mutant selection technique can be employed to analyze the remaining mutants that are not killed by test suites designed to kill mutants selected by \our. 
}

\changing{On comparison with baselines,} we observe that \our clearly achieves much higher prediction performance in comparison to \dt in both benchmarks. 
The differences are statistically significant.\footnote{We compared the MCC values using \emph{Wilcoxon signed-rank test} and obtained a $\mathit{p-value} < 5.07\mathrm{e}{-3}$ in comparison to \dt. 
We also compared the MCC values with the \emph{Vargha-Delaney A measure}~\cite{VarghaDelaney2000} and observed that in all (100\%) cases, \our significantly outperforms baseline techniques.}

In C-Benchmark, on average, \our with its MCC of \changing{0.47 (\our-100), and} 0.39 \changing{(\our-50)} outperforms \rnd (0.0 MCC). 
\our also outperforms \dt, on average, with \changing{2.76} times higher MCC and \changing{64\%} improvement in F-measure.  
It is worth mentioning that while \our achieves \changing{3.72} times higher precision than \dt, 
\changing{\our also offers an improvement of 4\% in Recall over \dt.}

In Java-Benchmark, on average, \our with its MCC of \changing{0.45 (\our-100), and} 0.38 \changing{(\our-50)} outperforms \rnd (0.0 MCC). 
\our also outperforms \dt, on average, with \changing{2.81} 
times higher MCC, and an improvement of \changing{82\%} 
in F-measure, \changing{68.88\%} 
in Precision, and \changing{85.71\%} 
in Recall.

In summary, \our offers an improvement in prediction capability (MCC) of \changing{2.78} 
times higher than \dt. 

\subsection{Effectiveness Evaluation (RQ2)}
\label{subsec:RQ2-results}

\begin{figure*}[htp]
\begin{subfigure}[t]{0.31\textwidth}
\includegraphics[width=\textwidth]{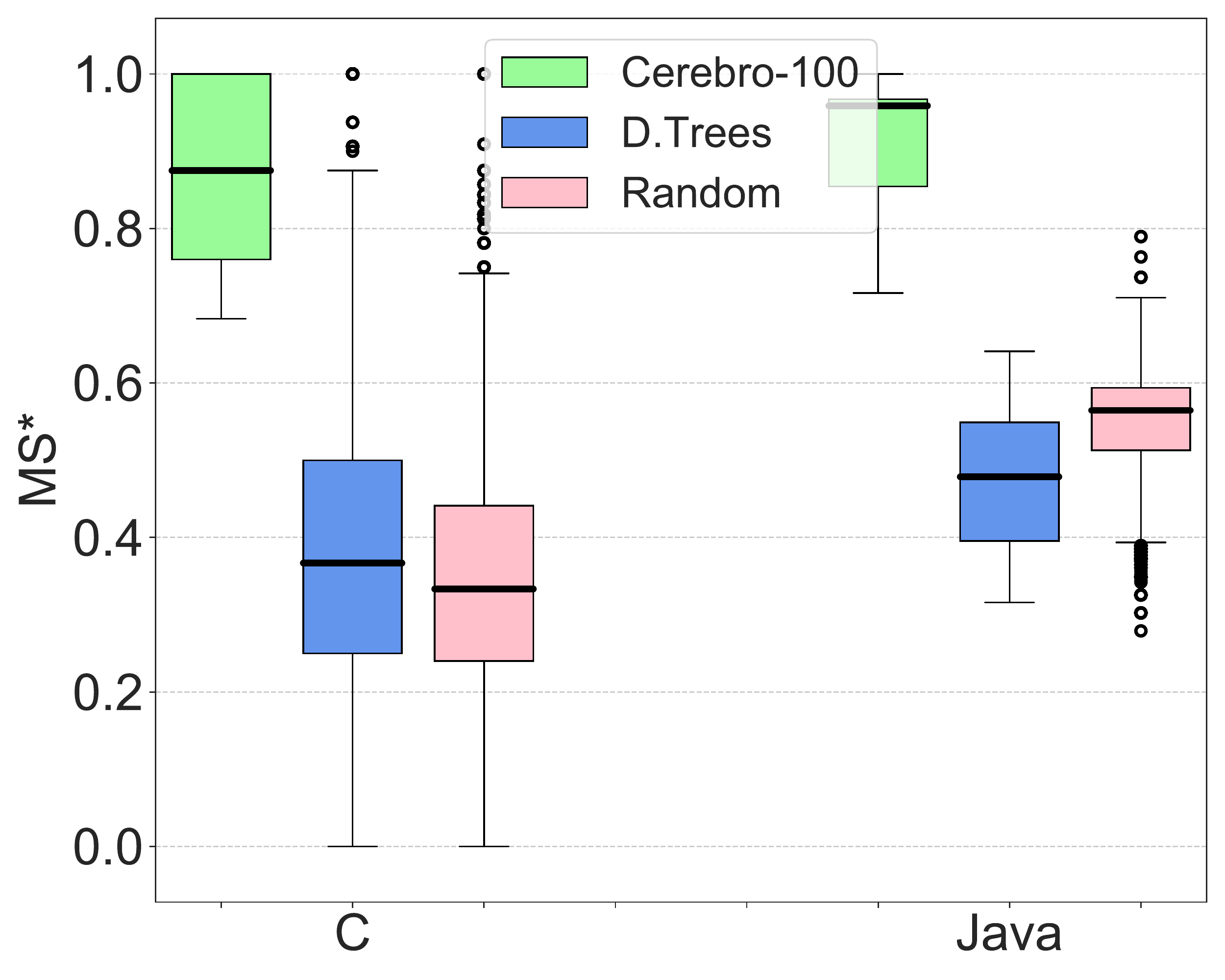}
\vspace{-1.5em}
\caption{\changing{C-Benchmark: For the same mutant selection size, \our-100 obtains an MS* of 87.50\%, while \dt, and \rnd obtains 36.67\%, and 33.33\%.\\Java-Benchmark: \our obtains, on average, an MS* of 95.90\%, while \dt, and \rnd obtains 47.85\%, and 56.45\%.}}
\label{fig:rq2_100_1}
  \end{subfigure}
\hfill
\begin{subfigure}[t]{0.32\textwidth}
\includegraphics[width=\textwidth]{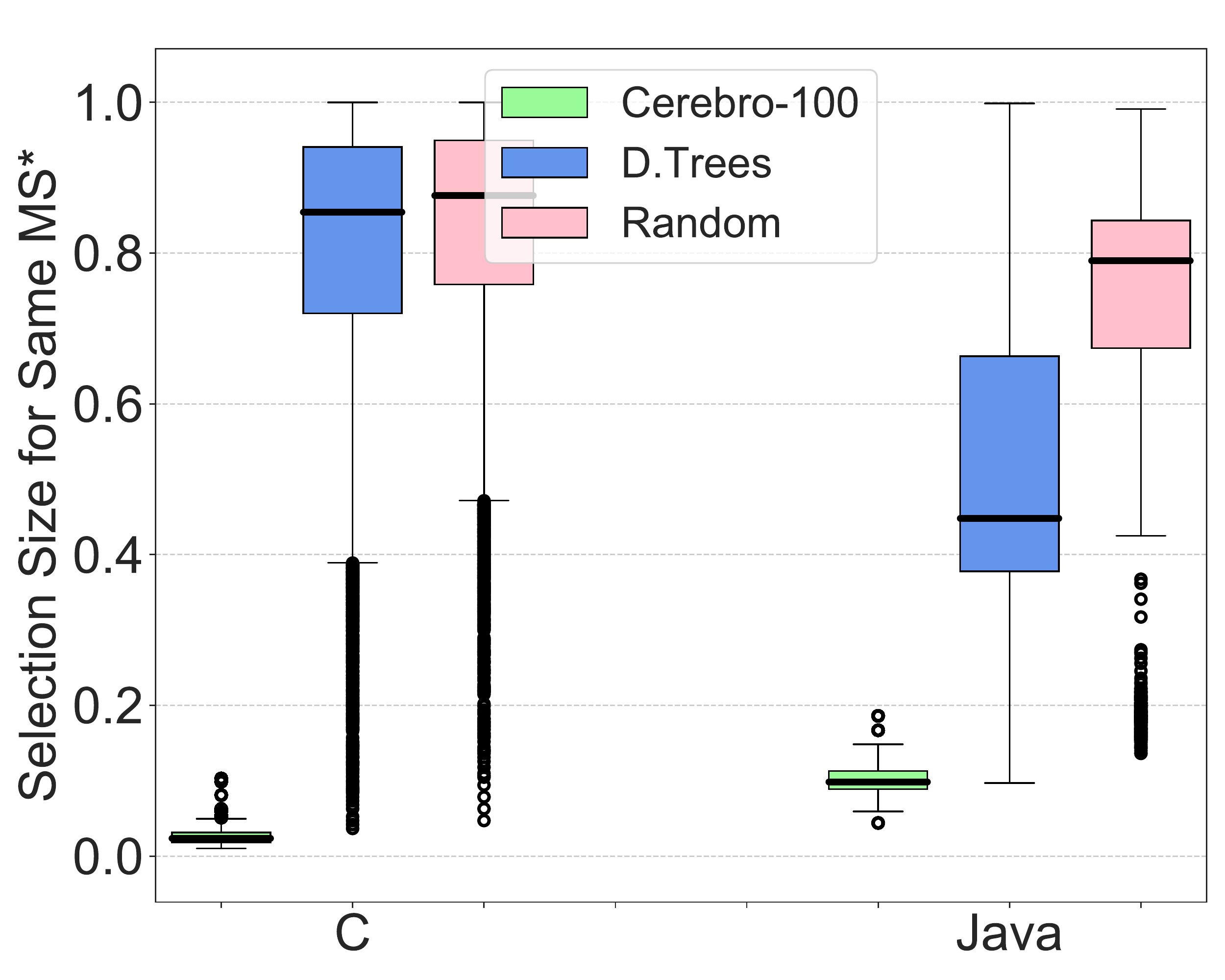}
\vspace{-1.5em}
    \caption{\changing{C-Benchmark: to reach the same MS*, \our-100 uses 2.35\% of the mutants, while \dt, and \rnd 85.42\%, and 87.61\%. \\Java-Benchmark: \our-100 uses 9.85\% of the mutants, while \dt, and \rnd use 44.80\%, and 78.97\%.}}
    \label{fig:rq2_100_2}
  \end{subfigure}
\hfill
\begin{subfigure}[t]{0.33\textwidth}
\includegraphics[width=\textwidth]{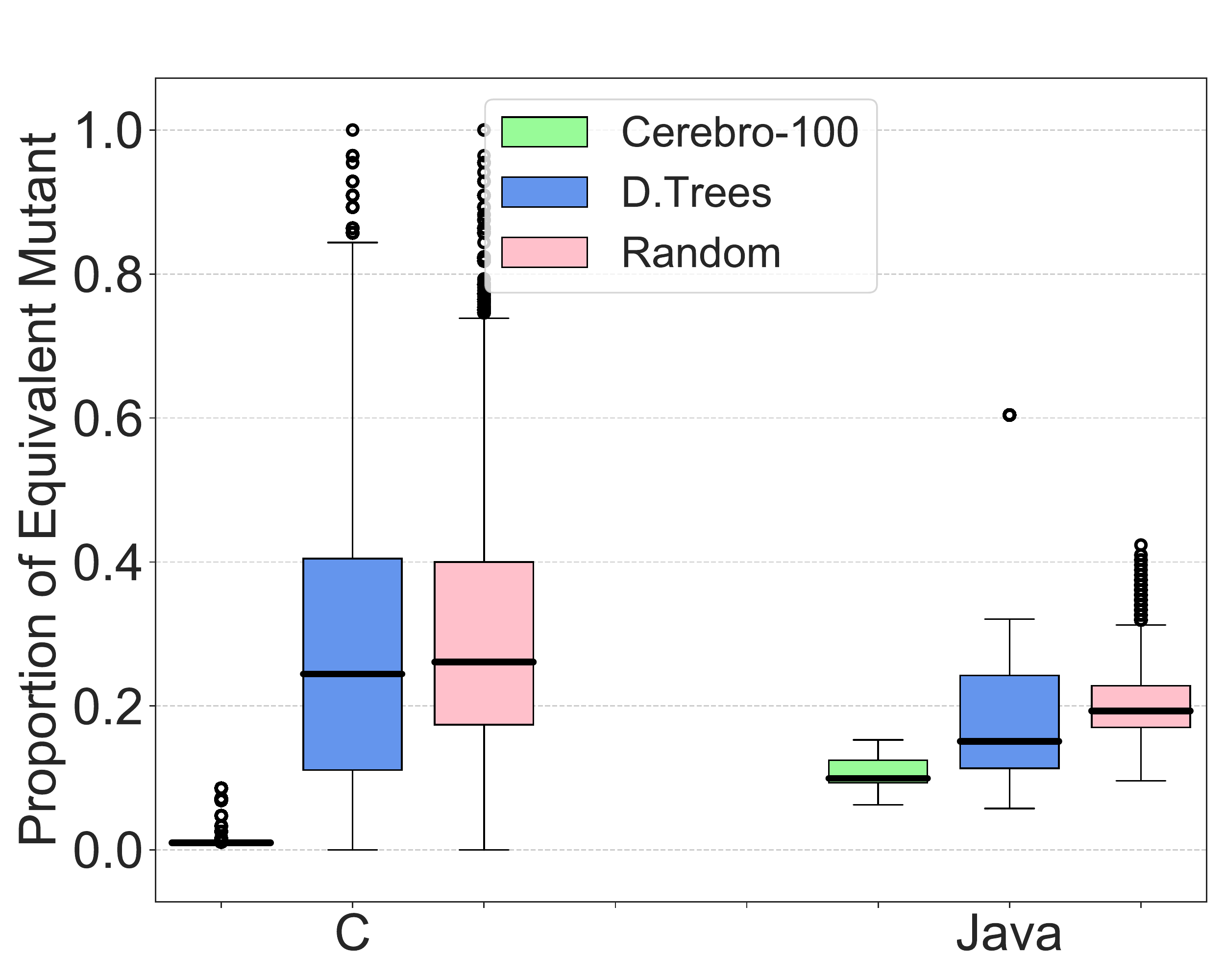}
\vspace{-1.5em}
    \caption{\changing{C-Benchmark: \our-100 selects 1.10\% equivalent mutants, while \dt, and \rnd select 24.44\%, and 26.09\%. \\Java-Benchmark: 9.95\% of mutants selected by \our-100 are equivalent, whereas 15.11\%, and 19.33\% of mutants selected by \dt, and \rnd are equivalent.}}
    \label{fig:rq2_100_3}
  \end{subfigure}
\begin{subfigure}[t]{0.31\textwidth}
\includegraphics[width=\textwidth]{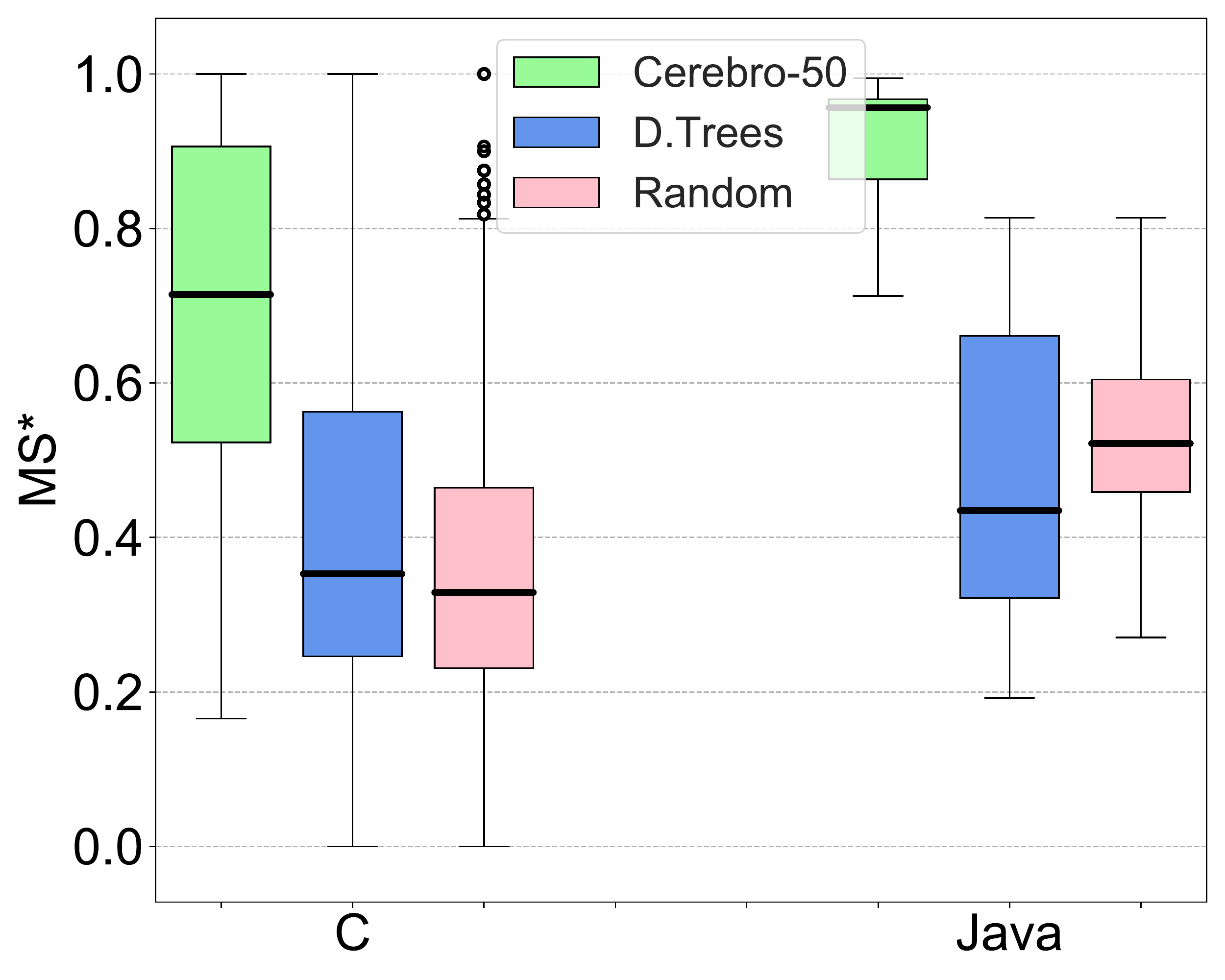}
\vspace{-1.5em}
\caption{C-Benchmark: For the same mutant selection size, \changing{\our-50} obtains an MS* of 71.43\%, while \dt, and \rnd obtains 34.23\%, and 32.88\%.\\Java-Benchmark: \changing{\our-50} obtains, on average, an MS* of 95.65\%, while \dt, and \rnd obtains 43.48\%, and 52.17\%.}
\label{fig:rq2_50_1}
  \end{subfigure}
\vspace{1em}
\hfill
\begin{subfigure}[t]{0.32\textwidth}
\includegraphics[width=\textwidth]{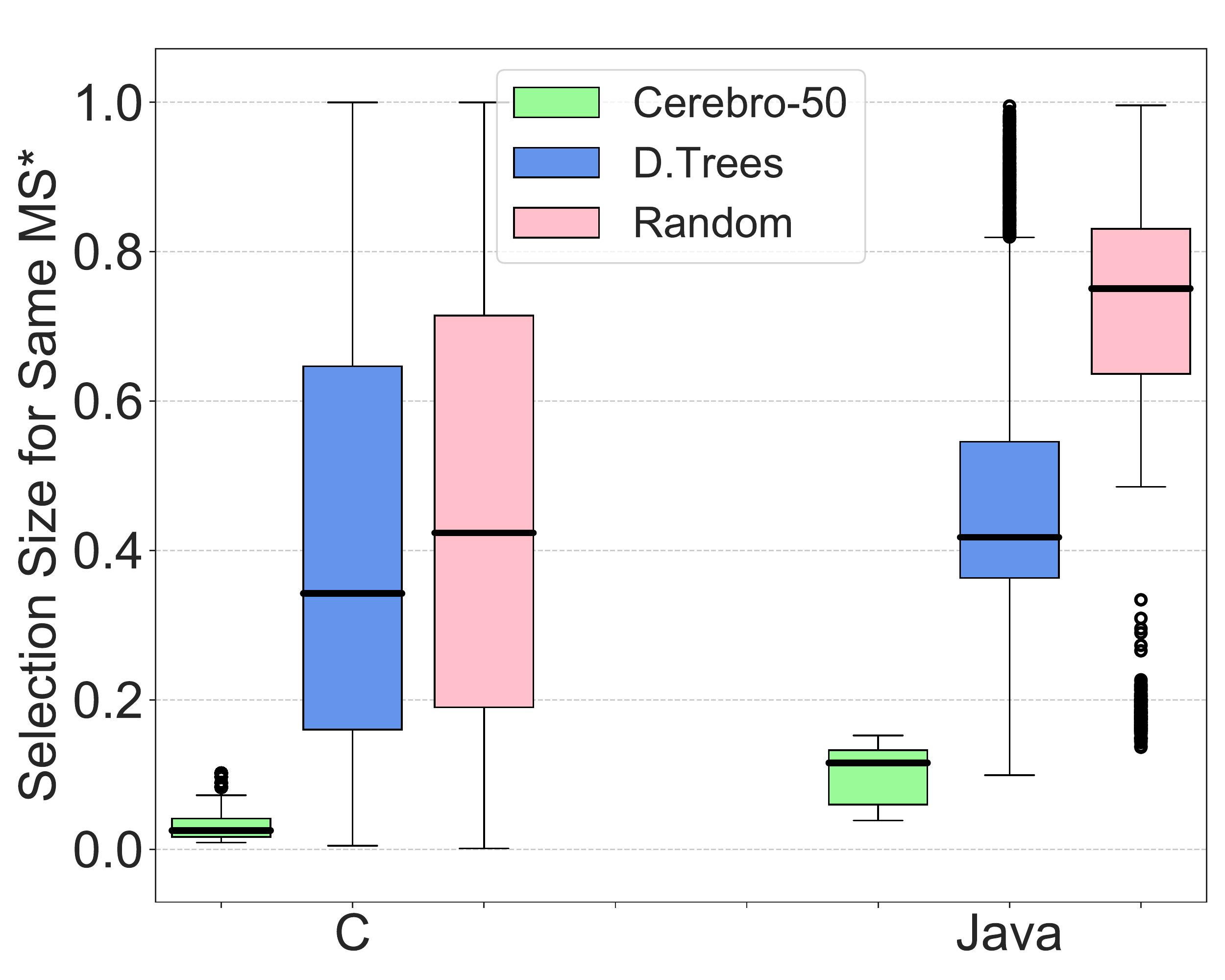}
\vspace{-1.5em}
    \caption{C-Benchmark: to reach the same MS*, \changing{\our-50} uses 2.52\% of the mutants, while \dt, and \rnd 34.24\%, and 42.37\%. \\Java-Benchmark: \changing{\our-50} uses 11.60\% of the mutants, while \dt, and \rnd use 41.77\%, and 75.09\%.}
    \label{fig:rq2_50_2}
  \end{subfigure}
\hfill
\begin{subfigure}[t]{0.33\textwidth}
\includegraphics[width=\textwidth]{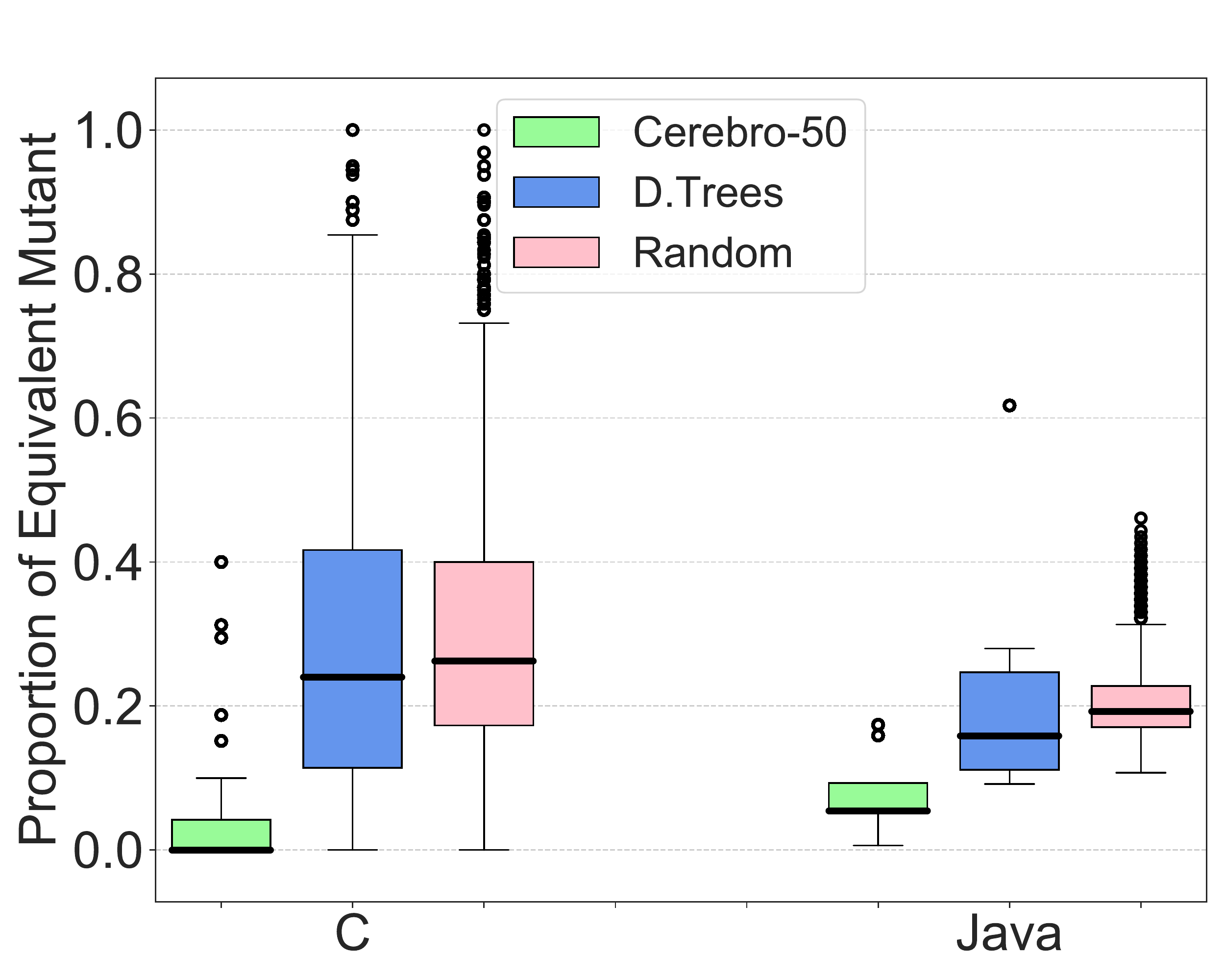}
\vspace{-1.5em}
    \caption{C-Benchmark: \changing{\our-50} selects 4.37\% equivalent mutants, while \dt, and \rnd select 24\%, and 26.23\%. \\Java-Benchmark: 5.45\% of mutants selected by \changing{\our-50} are equivalent, whereas 15.86\%, and 19.26\% of mutants selected by \dt, and \rnd are equivalent.}
    \label{fig:rq2_50_3}
  \end{subfigure}
\caption{(RQ2) Results of the Simulation - Trade off between mutant selection size and MS*.}
\label{fig:rq2}
\end{figure*}

Figure~\changing{\ref{fig:rq2_100_1} and} \ref{fig:rq2_50_1} show the average subsuming mutation score (MS*) obtained when selecting the same number of mutants  (by all techniques). 
In C-Benchmark, on average, \changing{\our-100 obtains an MS* of 87.50\%, which is 2.39 and 2.63 times higher MS* than \dt and \rnd, respectively. Moreover, \our-50} obtains an MS* of 71.43\%, which is 2.02 and 2.17 times higher MS* than \dt and \rnd, respectively.

In Java-Benchmark, on average, \changing{\our-100 obtains an MS* of 95.90\%, which is twice higher than \dt, and 69.53\% improvement over \rnd. Moreover, \our-50} obtains an MS* of 95.66\%, which is 2.20 times higher than \dt, and 83.33\% improvement over \rnd. 
The differences are statistically significant, according \changing{to} the computed $\mathit{p-value}$. 
We also compared them with the \emph{Vargha-Delaney A measure} ($\hat{A}_{12}$)~\cite{VarghaDelaney2000}, showing that \our achieves better MS* than \dt, and \rnd, in 92.4\%, and 95.7\% of the cases.

We also study the selection size needed by \dt and \rnd to achieve the same MS* obtained by \our. 
For C-Benchmark, \changing{Figure~\ref{fig:rq2_100_2} shows that while \our-100 selects only 2.35\%\ of the mutants, \dt, and \rnd need to select 85.42\% (36.35 times higher), and 87.61\% (37.28 times) of the mutants to achieve same MS* as \our. Also,} Figure~\ref{fig:rq2_50_2} shows that while \changing{\our-50} selects only 2.52\%\ of the mutants, \dt, and \rnd need to select 34.23\% (13.57 times higher), and 42.37\% (16.79 times) of the mutants, to achieve same MS* as \our.
For Java-Benchmark, while \changing{\our-100 selects 9.85\%\ of the mutants, \dt, and \rnd need to select 44.80\% (4.55 times higher), and 78.97\% (8.02 times) of the mutants, to achieve same MS* as \our-100.} Also, while \changing{\our-50} selects 11.60\%\ of the mutants, \dt, and \rnd need to select 41.77\% (3.60 times higher), and 75.09\% (6.48 times) of the mutants, to achieve same MS* as \changing{\our-50}.
We obtained a \changing{statistically} significant $\mathit{p-value}$ and $\hat{A}_{12}$ when compared these values, evidencing that \our in more than 98.5\%, and 99.1\% of the cases, selects \changing{fewer} mutants than \dt, and \rnd.

We also measure the percentage of equivalent mutants selected. 
For C-Benchmark, \changing{Figure~\ref{fig:rq2_100_3} shows that 1.10\% of mutants selected by \our-100 are equivalent, whereas 24.44\%, and 26.09\%, of the mutants selected by \dt, and \rnd, are equivalent.
Also, Figure~\ref{fig:rq2_50_3}} shows that 4.37\% of mutants selected by \changing{\our-50}  are equivalent, whereas 24\%, and 26.23\%, of the mutants selected by \dt, and \rnd, are equivalent.
In Java-Benchmark, \changing{9.95\% of the mutants selected by \our-100 are equivalent whereas for \dt, and \rnd, 15.11\% (51.86\% more), and 19.33\% (94.27\% more) selected mutants are equivalent. Also,} 5.45\% of the mutants selected by \changing{\our-50} are equivalent whereas for \dt, and \rnd, 15.86\% (2.91 times higher), and 19.26\% (3.53 times higher) selected mutants are equivalent.
The differences are statistically significant. $\hat{A}_{12}$ shows that \our in more than 90\%, and 98.4\% of the cases selects \changing{fewer} equivalent mutants than \dt, and \rnd.  These results provide evidence that our approach can reduce significantly this long-standing problem of mutation analysis.  

\begin{figure*}[htp]
\begin{subfigure}[t]{0.30\textwidth}
\includegraphics[width=\textwidth]{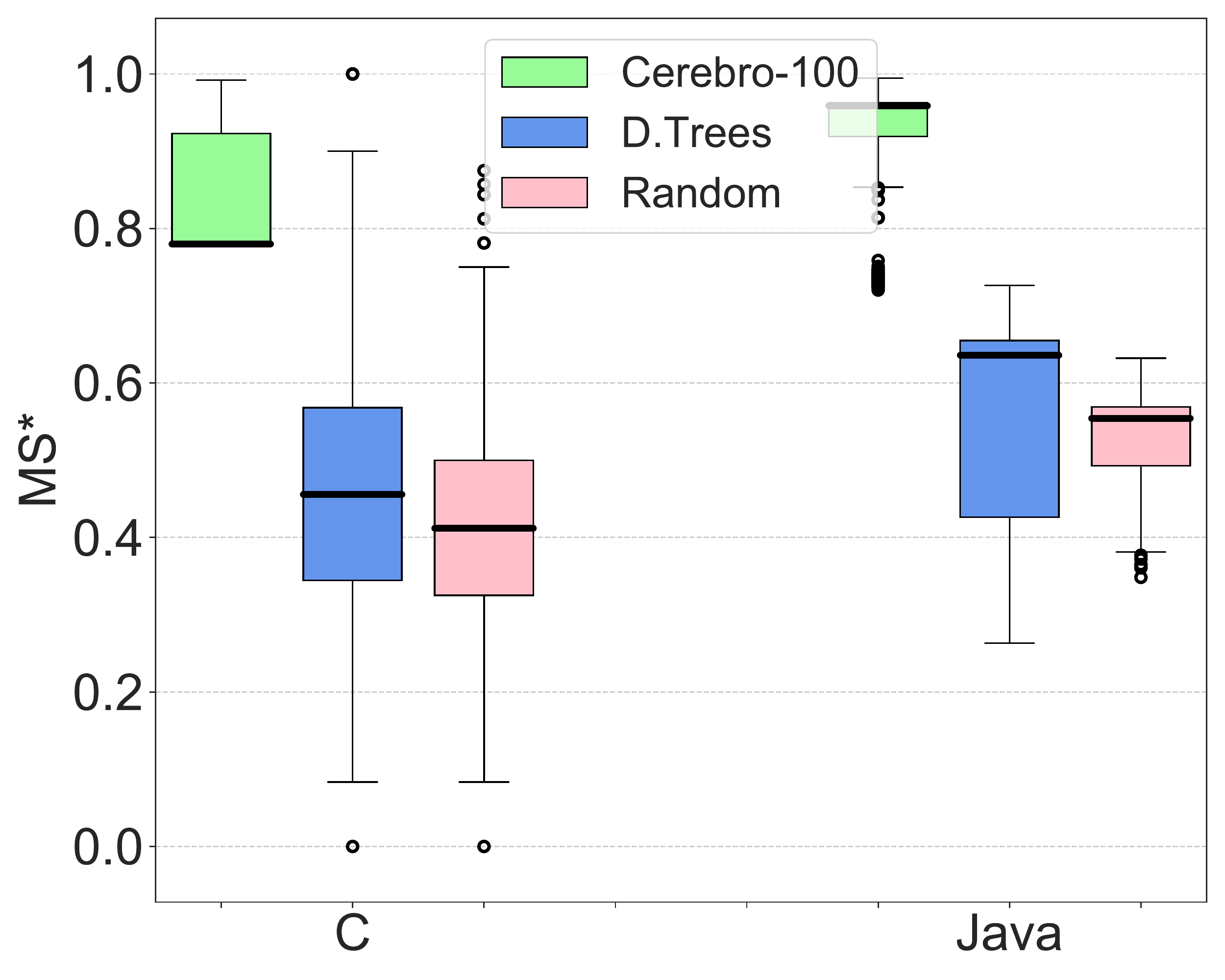}
\vspace{-1.0em}
    \caption{\changing{C-Benchmark: Same number of mutants lead to MS* of 78\%, 45.56\%, and 41.18\% for \our-100, \dt, and \rnd.\\Java-Benchmark: \our-100 reaches MS* of 94.90\%, whereas \dt, and \rnd reach 63.59\%, and 55.43\%.}}
    \label{fig:rq3_100_1}
  \end{subfigure}
\hfill
\begin{subfigure}[t]{0.32\textwidth}
\includegraphics[width=\textwidth]{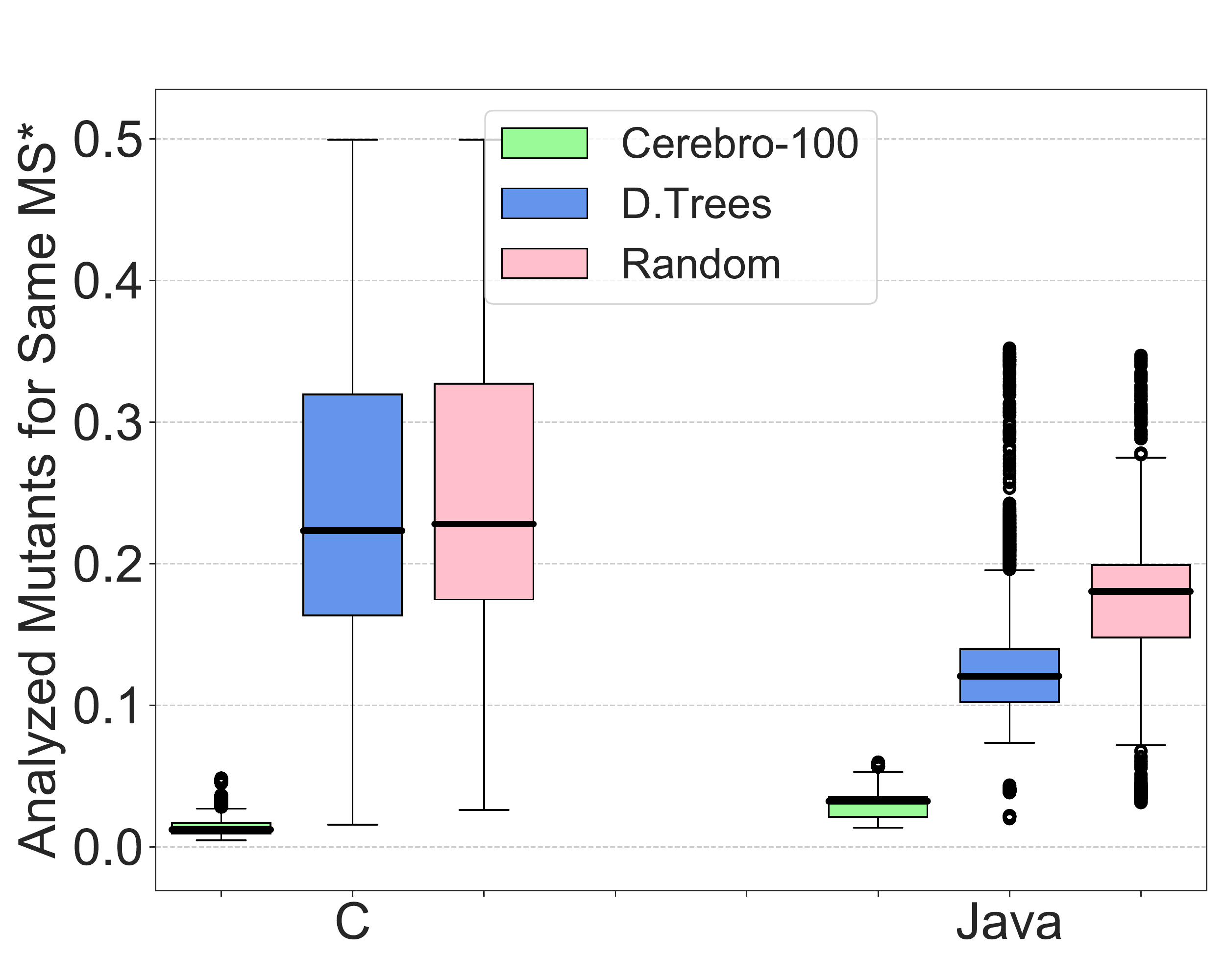}
\vspace{-1.0em}
    \caption{\changing{C-Benchmark: \our-100 analyzes 1.21\% mutants, whereas \dt, and \rnd analyze 22.33\%, and 22.80\% to reach the same MS* as \our-100.\\Java-Benchmark: \our-100 analyze 3.22\% mutants, whereas \dt, and \rnd analyze 12.07\% and 18.05\% to reach same MS*.}}
    \label{fig:rq3_100_2}
  \end{subfigure}
  \hfill
\begin{subfigure}[t]{0.32\textwidth}
\includegraphics[width=\textwidth]{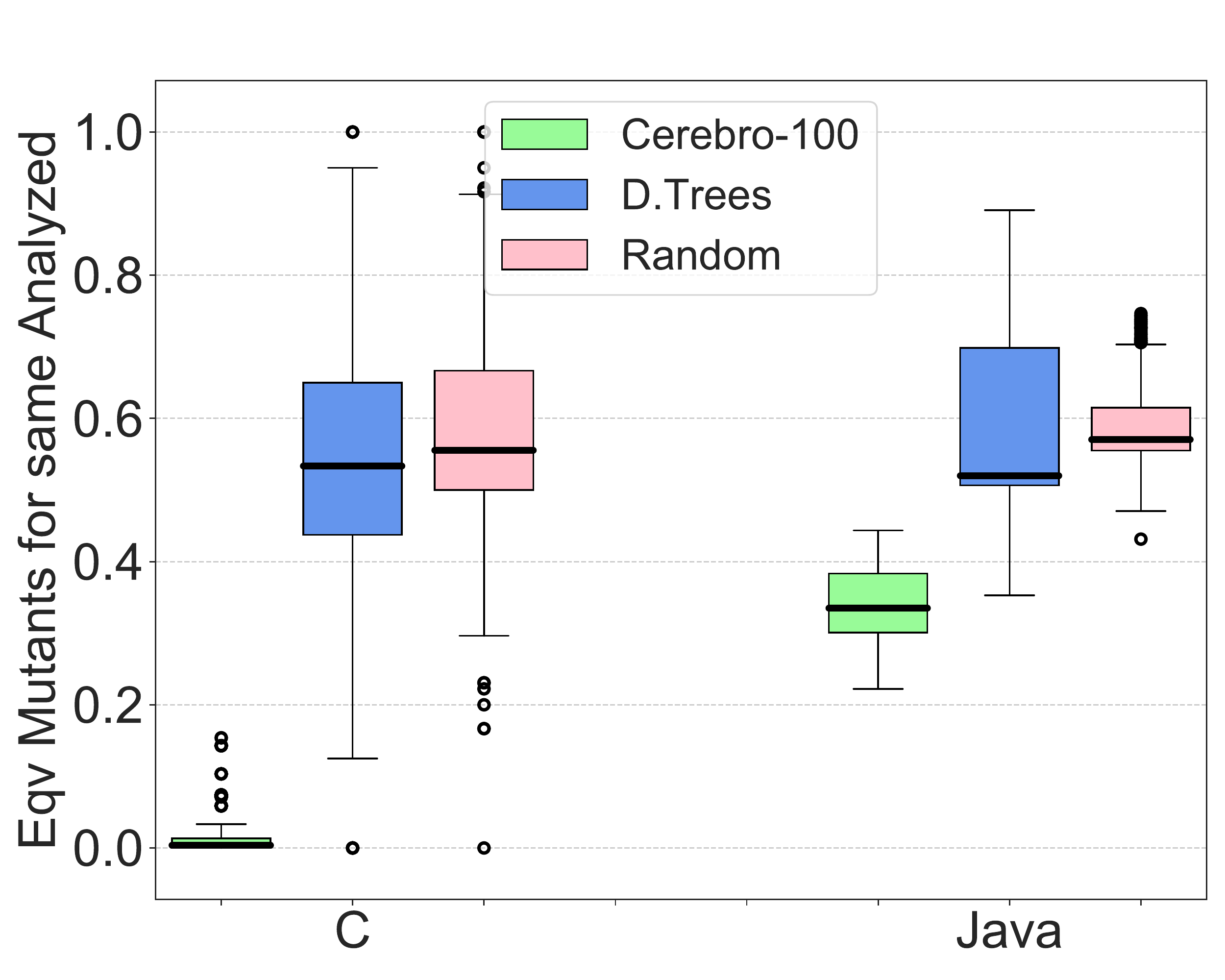}
\vspace{-1.0em}
    \caption{\changing{C-Benchmark: 3.70\%, 53.33\%, and 55.56\% of the mutants selected by \our-100, \dt, and \rnd are equivalent. \\Java-Benchmark:  33.48\%, 52\%, and 57.04\% of the mutants selected by \our-100, \dt, and \rnd are equivalent.}}
    \label{fig:rq3_100_3}
  \end{subfigure}
\begin{subfigure}[t]{0.30\textwidth}
\includegraphics[width=\textwidth]{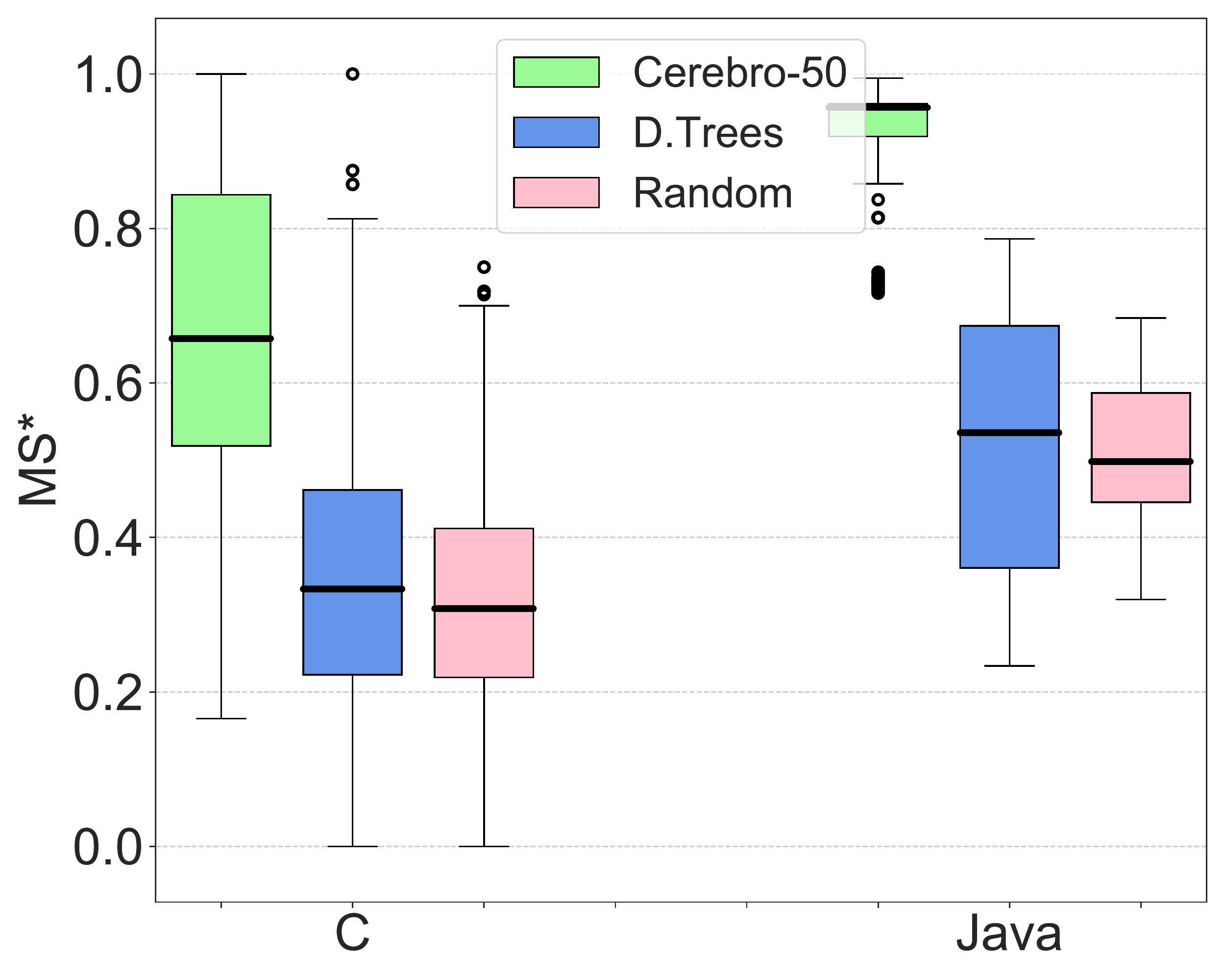}
\vspace{-1.0em}
    \caption{C-Benchmark: Same number of mutants lead to MS* of 65.75\%, 33.33\%, and 30.77\% for \changing{\our-50}, \dt, and \rnd.\\Java-Benchmark: \changing{\our-50} reaches MS* of 95.65\%, whereas \dt, and \rnd reach 53.54\%, and 49.83\%.}
    \label{fig:rq3_50_1}
  \end{subfigure}
\vspace{1.0em}
\hfill
\begin{subfigure}[t]{0.32\textwidth}
\includegraphics[width=\textwidth]{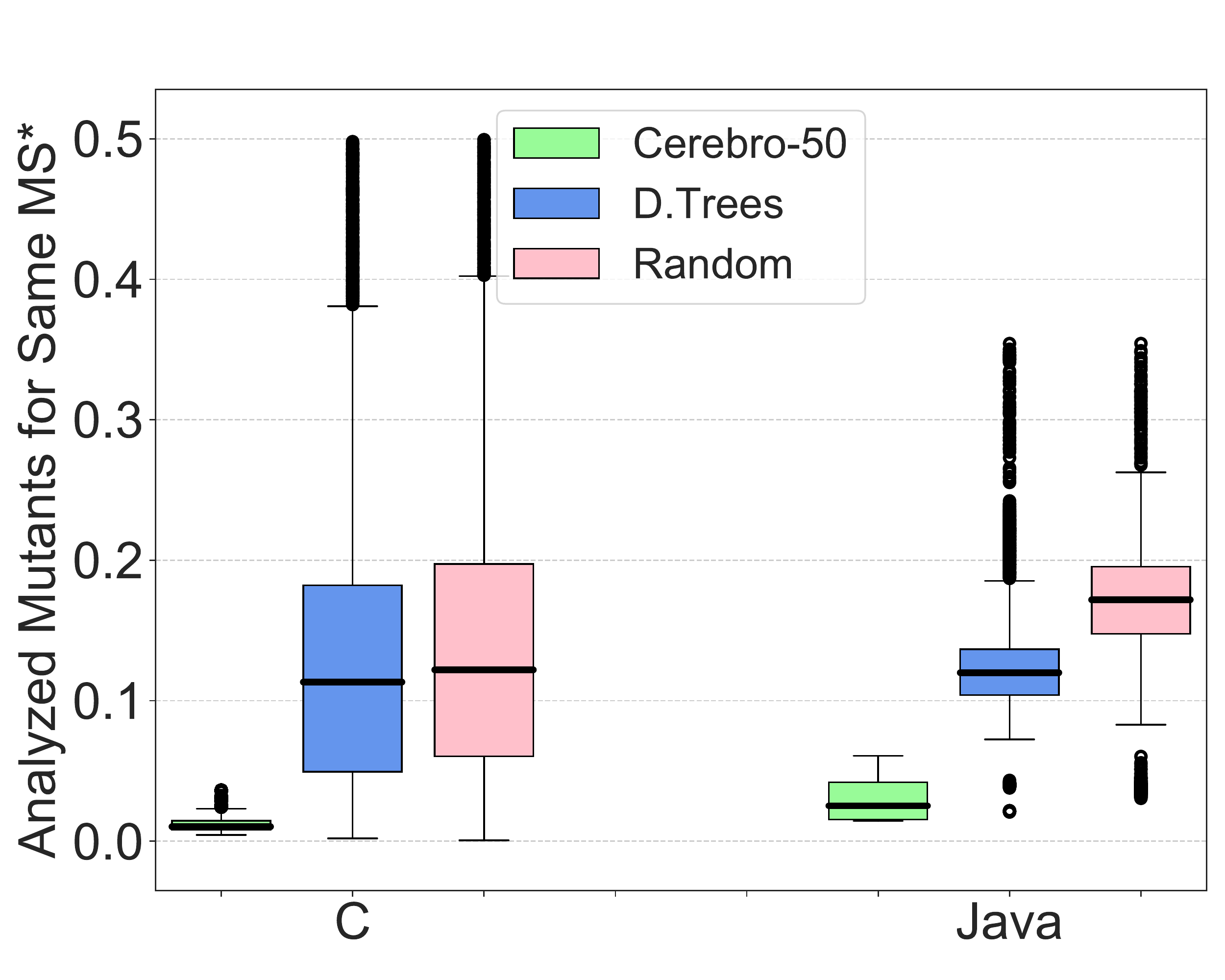}
\vspace{-1.0em}
    \caption{C-Benchmark: \changing{\our-50} analyzes 1.02\% mutants, whereas \dt, and \rnd analyze 11.92\%, and 13.17\% to reach the same MS* as \changing{\our-50}.\\Java-Benchmark: \changing{\our-50} analyze 2.52\% mutants, whereas \dt, and \rnd analyze 12\% and 17.19\% to reach same MS*.}
    \label{fig:rq3_50_2}
  \end{subfigure}
  \hfill
\begin{subfigure}[t]{0.32\textwidth}
\includegraphics[width=\textwidth]{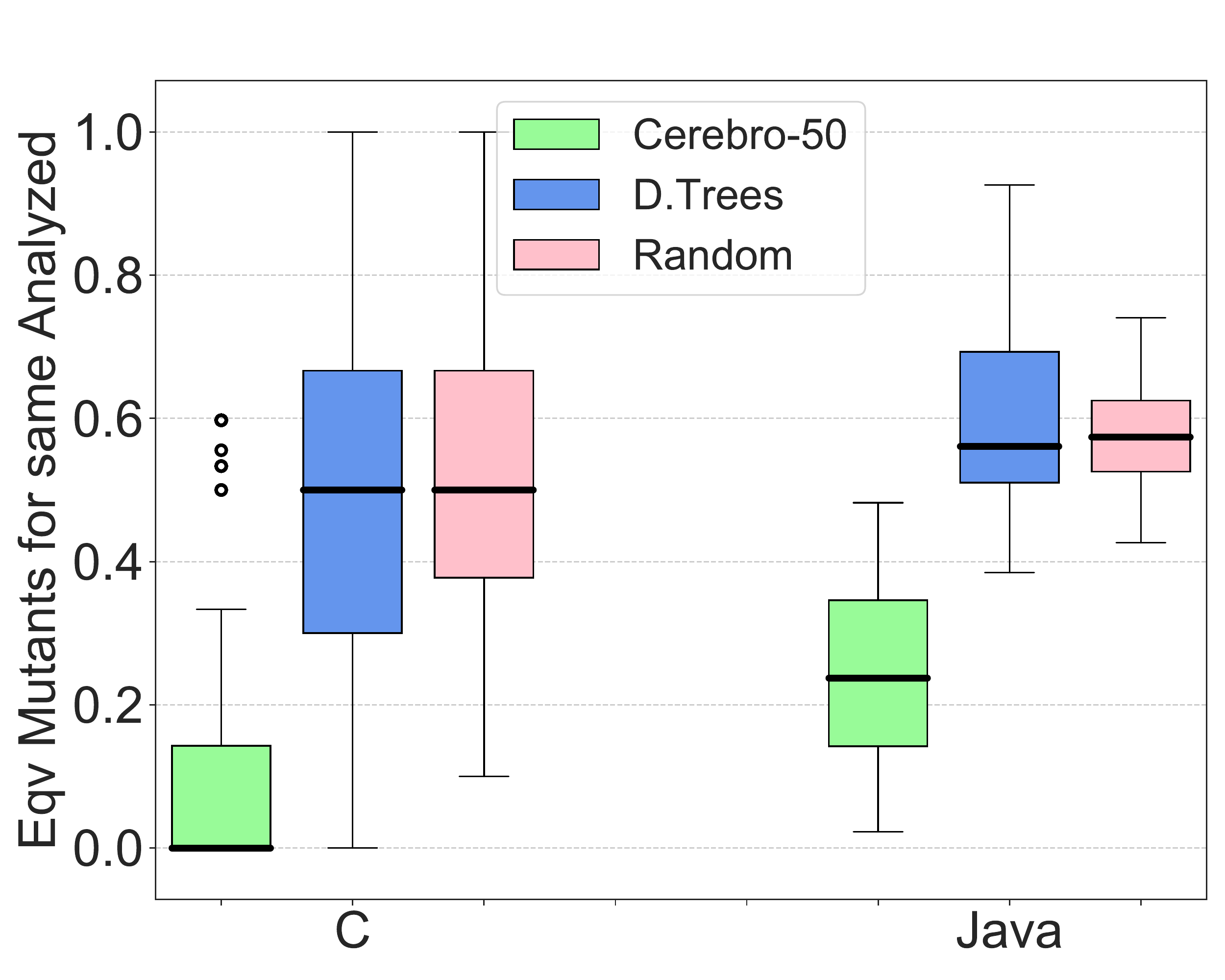}
\vspace{-1.0em}
    \caption{C-Benchmark: 11.31\%, 50\%, and 50\% of the mutants selected by \changing{\our-50}, \dt, and \rnd are equivalent. \\Java-Benchmark:  23.72\%, 56.08\%, and 57.38\% of the mutants selected by \changing{\our-50}, \dt, and \rnd are equivalent.}
    \label{fig:rq3_50_3}
  \end{subfigure}

\caption{(RQ3) Results of the Simulation - Trade off between percentage of mutants analyzed and MS*.}
\label{fig:rq3}
\vspace{-0.75em}
\end{figure*}

\subsection{Number of Analyzed Mutants (RQ3)}
\label{subsec:RQ3-results}

Figures~\ref{fig:rq3_100_1} and \ref{fig:rq3_50_1} show the average subsuming mutation score (MS*) obtained by each technique for the same number of analyzed mutants. 
In C-Benchmark, on average, \changing{\our-100 achieved an MS* of 78\%, which is an improvement of 89.41\%, and 71.20\% over the MS* of \rnd, and \dt, respectively. Moreover, \our-50} achieved an MS* of 65.75\%, which is 2.14 times higher than \rnd and an improvement of 97\% over \dt
In Java-Benchmark, on average, \changing{\our-100 achieved an MS* of 94.90\%, an improvement of 49.24\% and 71.21\% over \dt and \rnd, respectively. Moreover, \our-50} achieved an MS* of 95.65\%, an improvement of 78.65\% and 91.94\% over \dt and \rnd, respectively. 
The differences are \changing{statistically} significant, according to the computed $\mathit{p-value}$ and $\hat{A}_{12}$. 
We observed that \our in more than 96.2\%, and 98.4\%, of the cases is better than \dt, and \rnd. 

We also study what should be the percentage of mutants to be analyzed by \dt and \rnd to achieve the same MS* as \our.
For C-Benchmark, \changing{Figure~\ref{fig:rq3_100_2} shows that while \our-100 analyzes 1.21\% mutants, \dt, and \rnd need to analyze 22.33\% (18.45 times higher), and 22.80\% (18.84 times higher) of mutants to reach same MS* as \our-100. Also,} Figure~\ref{fig:rq3_50_2} shows that while \changing{\our-50} analyzes 1.02\% mutants, \dt, and \rnd need to analyze 11.92\% (11.58 times higher), and 13.17\% (12.78 times higher) of mutants to reach same MS* as \changing{\our-50}.
In Java-Benchmark, while \changing{\our-100 analyzes 3.22\% mutants, \dt, and \rnd need to analyze 12.07\% (3.75 times higher), and 18.05\% (5.61 times higher) of mutants to reach same MS* as \our-100. Moreover,} while \changing{\our-50} analyzes 2.52\% mutants, \dt, and \rnd need to analyze 12.00\% (4.76 times higher), and 17.19\% (6.82 times) of mutants to reach same MS* as \changing{\our-50}.
We obtained a \changing{statistically} significant $\mathit{p-value}$ and $\hat{A}_{12}$, showing that \our in more than 99\% of the cases analyzes less mutants than \dt and \rnd.

\begin{figure*}[htp]
\begin{subfigure}[t]{0.23\textwidth}
\includegraphics[width=\textwidth]{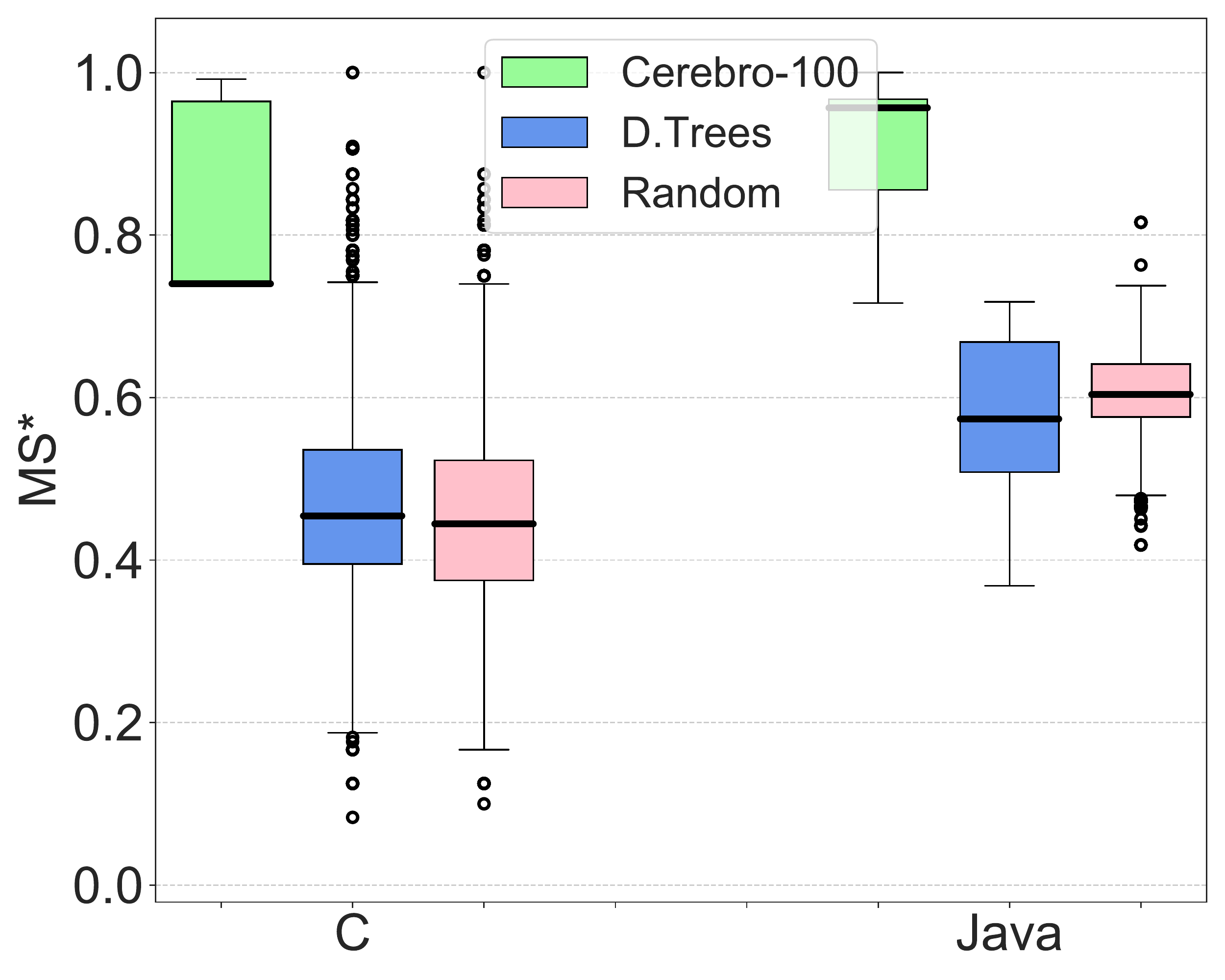}
\vspace{-1.5em}
    \caption{\changing{C-Benchmark: For same number of test executions, \our-100 obtains an MS* of 74\%, while \dt, and \rnd obtain 45.45\%, and 44.44\%.\\Java-Benchmark: \our-100 obtains MS* of 95.65\%, while \dt, and \rnd obtain 57.38\%, and 60.40\%.}}
    \label{fig:rq4_100_1}
  \end{subfigure}
\hfill
\begin{subfigure}[t]{0.235\textwidth}
\includegraphics[width=\textwidth]{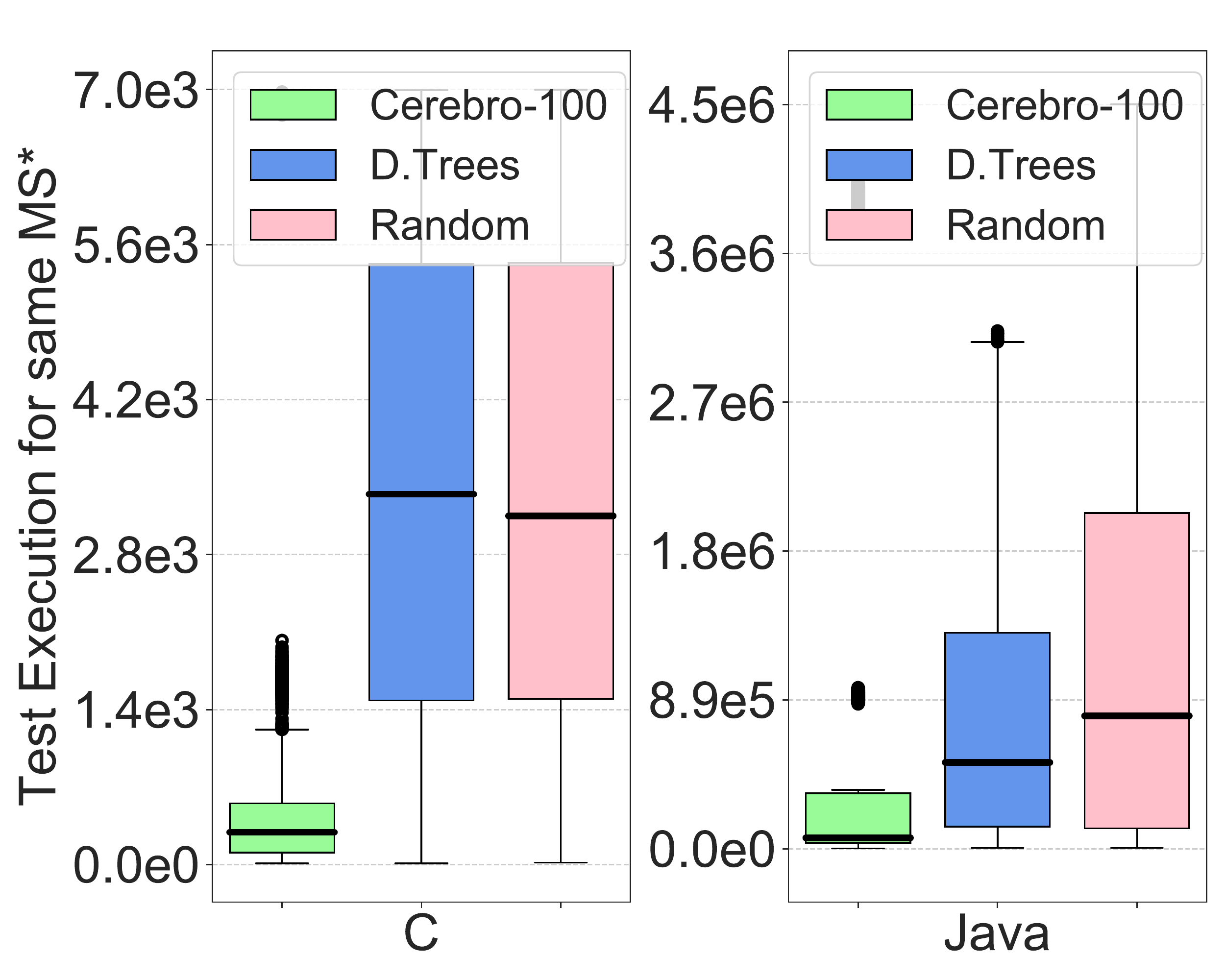}
\vspace{-1.5em}
    \caption{\changing{C-Benchmark: \our-100 requires 291 test executions, while \dt, and \rnd require 3,345, and 3,149 to reach the same MS* as \our-100.\\Java-Benchmark: 65,741 test executions are required by \our-100, while \dt, and \rnd require 517,040, and 795,304.}}
    \label{fig:rq4_100_2}
  \end{subfigure}
\hfill
\begin{subfigure}[t]{0.23\textwidth}
\includegraphics[width=\textwidth]{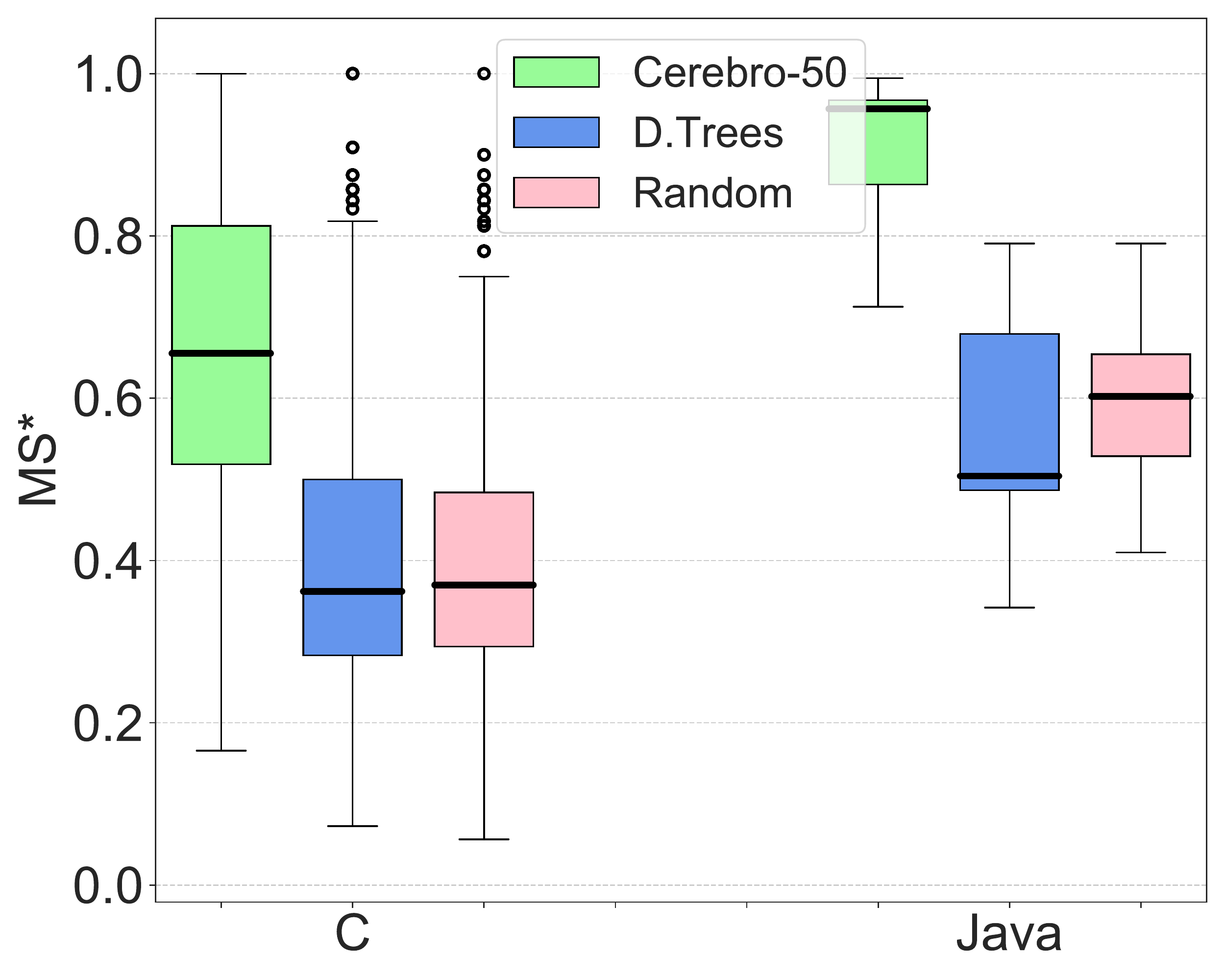}
\vspace{-1.5em}
    \caption{C-Benchmark: For same number of test executions, \changing{\our-50} obtains an MS* of 65.52\%, while \dt, and \rnd obtain 36.20\%, and 36.99\%.\\Java-Benchmark: \changing{\our-50} obtains MS* of 95.65\%, while \dt, and \rnd obtain 50.41\%, and 60.21\%.}
    \label{fig:rq4_50_1}
  \end{subfigure}
\hfill
\begin{subfigure}[t]{0.235\textwidth}
\includegraphics[width=\textwidth]{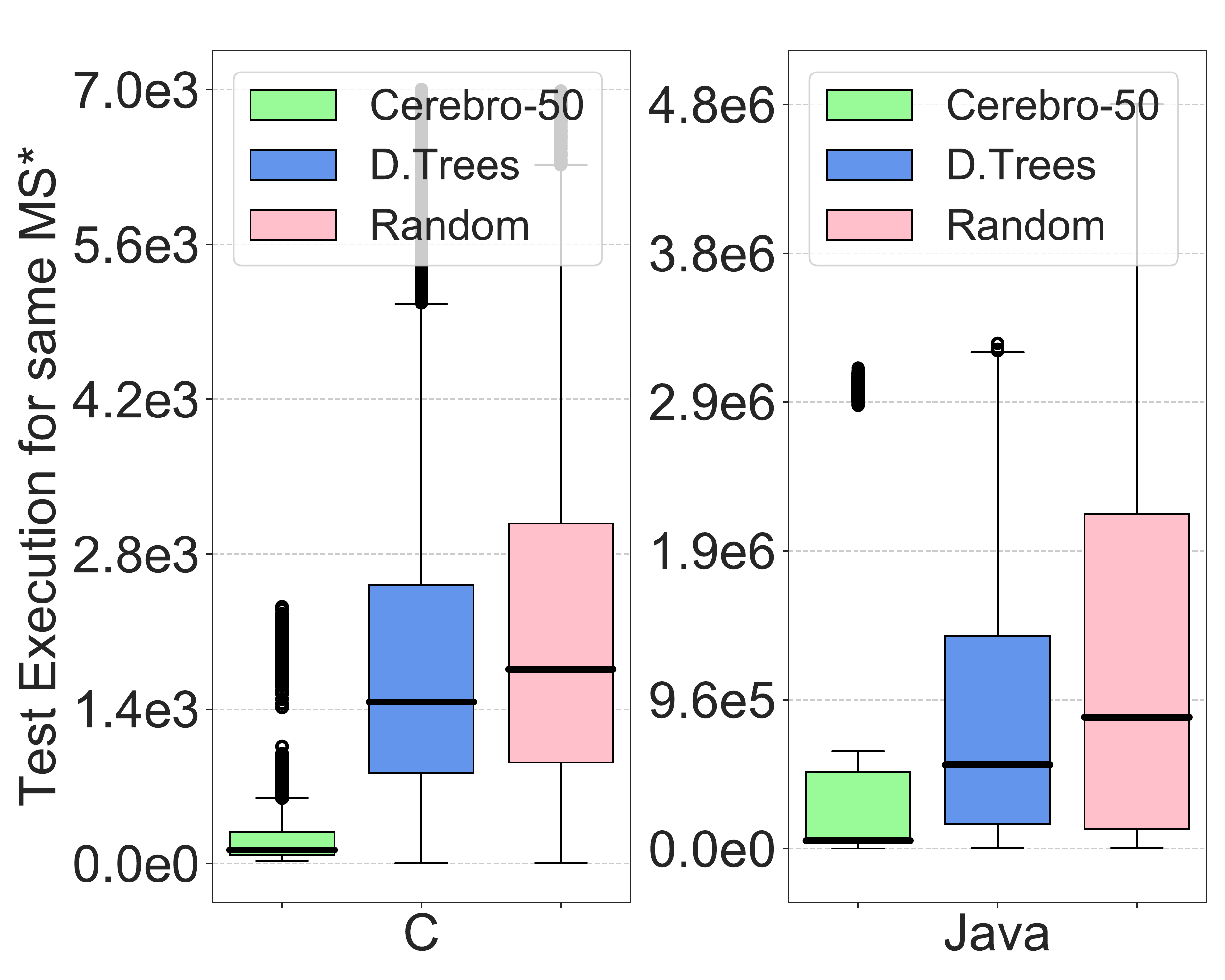}
\vspace{-1.5em}
    \caption{C-Benchmark: \changing{\our-50} requires 125 test executions, while \dt, and \rnd require 1,785, and 2,182 to reach the same MS* as \changing{\our-50}.\\Java-Benchmark: 50,622 test executions are required by \changing{\our-50}, while \dt, and \rnd require 560,866, and 894,494.}
    \label{fig:rq4_50_2}
  \end{subfigure}
\caption{(RQ4) Results of the Simulation - Trade off between number of test executions and MS*.}
\label{fig:rq4}
\end{figure*}

We also measure the percentage of equivalent mutants analyzed by each technique. 
For C-Benchmark, \changing{Figure~\ref{fig:rq3_100_3}} shows that, on average, \changing{\our-100 analyzes 3.70\% equivalent mutants, while 53.33\% (14.41 times higher), and 55.56\% (15.02 times higher) of the mutants analyzed by \dt, and \rnd are equivalent. Also,} \ref{fig:rq3_50_3} shows that \changing{\our-50} analyzes 11.31\% equivalent mutants, while 50\% (4.42 times higher) of the mutants analyzed by \dt and \rnd are equivalent. 
For Java-Benchmark, on average, \changing{33.48\% of the mutants analyzed by \our-100 are equivalent, while \dt, and \rnd analyze 52\% (55.31\% more), and 57.04\% (70.37\% more) equivalent mutants. Also,} 23.72\% of the mutants analyzed by \changing{\our-50} are equivalent, while \dt, and \rnd analyze 56.08\% (2.36 times higher), and 57.38\% (2.42 times higher) equivalent mutants.
This indicates that the baselines suggest the consumption of a large effort to analyze redundant mutants, in comparison to \our.
The differences are \changing{statistically} significant. 
$\hat{A}_{12}$ suggests that \our in more than 98\% of the cases analyzes \changing{fewer} equivalent mutants than \dt, and \rnd.

\subsection{Number of Test Executions (RQ4)}
\label{subsec:RQ4-results}

Figure~\changing{\ref{fig:rq4_100_1}} and \ref{fig:rq4_50_1} show the average subsuming mutation score (MS*) when the number of test executions are fixed. 
In C-Benchmark, on average, \changing{\our-100 achieves an MS* of 74\%, outperforming \dt, and \rnd by 62.82\%, and 66.52\% (\dt, and \rnd achieve 45.45\%, and 44.44\% of MS*). Also, \our-50} achieves an MS* of 65.52\%, outperforming \dt, and \rnd by 80.95\%, and 77.14\% (\dt, and \rnd achieve 36.21\%, and 36.99\% of MS*).
In Java-Benchmark, on average, \changing{\our-100 and \our-50 achieve an MS* of 95.65\% in both simulations, an improvement of approx. 67\%, and 58\% over \dt, and \rnd (\dt, and \rnd achieve 57.38\%, and 60.40\% of MS* in first simulation when compared against \our-100, and 50.41\%, and 60.21\% of MS* in the second comparison simulation against \our-50).} 
We obtained a \changing{statistically} significant $\mathit{p-value}$. 
Also $\hat{A}_{12}$ suggests that \our in 94.15\%, and 95.7\%, of the cases is better than \dt, and \rnd.

We also measure the test executions required by the baselines to achieve the same MS* as \our. 
\changing{Figure~\ref{fig:rq4_100_2} shows that, in C-Benchmark, \our-100 requires 291 test executions (median), while \dt, and \rnd require 3,345, and 3,149. Also,} Figure~\ref{fig:rq4_50_2} shows that \our-50 requires 125 test executions (median), while \dt, and \rnd require 1,785, and 2,182.
This shows that \changing{\our-100 is 10-12 times less and \our-50} is 14-17 times less expensive (computationally) than the baselines. 

\changing{In Java-Benchmark, \dt, and \rnd require 517,040, and 795,304 test executions (median) to achieve the same MS* as \our-100, for which 65,741 test executions are required. Moreover,} \dt, and \rnd require 560,866, and 894,494 test executions to achieve the same MS* as \our-50, for which 50,622 test executions are required.
This shows that the baselines require \changing{7 to 12 times, and} 11 to 17 times higher computational effort than \changing{\our-100, and} \our-50. 

These differences are \changing{statistically} significant. 
$\hat{A}_{12}$ value indicates that in more than 98.7\% of the cases, \our executes \changing{fewer} tests than \dt and \rnd.

\section{Discussion}
\label{sec:discussion}
\changing{
\our is a learning-based method, and thus its performance depends on a number of parameters and design decisions we made. To this end, we discuss the key (intuitive) parameters that make the Machine Translation approach we use effective \minrev{(Section~\ref{sec:WhyCerebro})}, together with empirical results demonstrating the potential impact on \minrev{the model's} performance \minrev{given} the design decisions of using unabstracted code sequences \minrev{(Section~\ref{sec:unabstracted-training})}, sequences with decreased length during training \minrev{(Section~\ref{sec:reducing-length-training}),} and the impact of assuming unkilled mutants as \minrev{equivalent mutants} during testing \minrev{(Section~~\ref{sec:noise-equivalent-mutants})}. 


}
\changing{\subsection{Why \our is a good candidate for subsuming mutant prediction?}
\label{sec:WhyCerebro}
There are three main factors that make Machine Translation a good candidate for subsuming mutant prediction. The first one is that it learns to select mutants using the exact local context (entire code snippet composed of 50-100 tokens, represented as a sequence), while previous work considers AST and data-flow abstractions\minrev{~\cite{ChekamPBTS20}}, ignoring the exact formulation of the code snippet. In a sense, the key determining factor is the sequence that code tokens appear in the local context (considered code snippet). The second reason is that the machine translator includes a powerful self-attention mechanism, which together with the encoder-decoder architecture makes the learning resistant to noise \cite{TangMRS18}, and able to learn out of imbalanced data. Overall, previous research has shown that this architecture often makes \minrev{the best} predictions \minrev{for} many NLP tasks~\cite{devlin2018bert}. This is actually the reason why Machine Translation has been successfully used in code analysis tasks such as mutant generation, code clone detection, test assertions generation, etc. The third reason is the diversity of the selected mutants, i.e., Cerebro selects a few mutants per code block, which allows eliminating local redundancies, while spreading testing across the entire code-base.}

\changing{\subsection{Impact of removing code abstraction}
\label{sec:unabstracted-training}

We analyzed the impact of using unabstracted code sequences to train our models instead of proposed abstracted code sequences and how it affects the model prediction performance (RQ1). In this experiment, we just removed the code comments and kept everything else as it is. We found a prediction performance reduction for projects in both C and Java benchmarks. For C-Benchmark, the model performance deteriorated by 18.9\% in MCC, 14.4\% in Precision and 18.1\% in Recall. For Java-Benchmark, although we found an improvement of 15.4\% in Recall, the overall performance deteriorated by 17.9\% in MCC and 22.5\% in Precision. 
}
\begin{table}[t!]
  \begin{center}
\caption{
\changing{Impact of the abstraction process and sequence length in \our's prediction performance: 
On average, MCC is decreased by 18\% with unabstracted code and decreased by 24\% with sequence length 25.
}
}
\label{tab:rq1_all}
\resizebox{.45\textwidth}{!}{
\begin{tabular}{| l | r | r | r | r |}
\toprule
\multicolumn{5}{c}{\changing{Average (and Median) Performance in C-Benchmark}}\\
\toprule
\hline
\rule{0pt}{3ex}\textbf{\changing{Approach}} & \textbf{\changing{MCC}}  & \textbf{\changing{F-measure}}  & \textbf{\changing{Precision}}  & \textbf{\changing{Recall}} \\ \hline
\rule{0pt}{3ex}\changing{\our-100}               & \changing{0.47 (0.47)}  & \changing{0.41 (0.40)} & \changing{0.93 (0.93)} & \changing{0.26 (0.25)} \\ \hline 
\rule{0pt}{3ex}\changing{\our-50}               & \changing{0.39 (0.40)}  & \changing{0.34 (0.34)} & \changing{0.82 (0.82)} & \changing{0.21 (0.22)} \\ \hline 
\rule{0pt}{3ex}\changing{\our-unabs}               & \changing{0.32 (0.31)}  & \changing{0.28 (0.27)} & \changing{0.70 (0.73)} & \changing{0.17 (0.16)} \\ \hline 
\rule{0pt}{3ex}\changing{\our-25}               & \changing{0.30 (0.29)}  & \changing{0.27 (0.28)} & \changing{0.64 (0.61)} & \changing{0.17 (0.18)} \\ \hline 
\multicolumn{5}{c}{}\\ 
\toprule
\multicolumn{5}{c}{\changing{Average (and Median) Performance in Java-Benchmark}}\\\toprule
\hline
\rule{0pt}{3ex}\textbf{\changing{Approach}} & \textbf{\changing{MCC}}  & \textbf{\changing{F-measure}}  & \textbf{\changing{Precision}}  & \textbf{\changing{Recall}} \\ \hline
\rule{0pt}{3ex}\changing{\our-100}               & \changing{0.45 (0.45)}  & \changing{0.51 (0.52)} & \changing{0.76 (0.73)} & \changing{0.39 (0.38)} \\ \hline 
\rule{0pt}{3ex}\changing{\our-50}               & \changing{0.38 (0.38)}  & \changing{0.42 (0.42)} & \changing{0.72 (0.73)} & \changing{0.31 (0.30)} \\ \hline 
\rule{0pt}{3ex}\changing{\our-unabs}               & \changing{0.31 (0.34)}  & \changing{0.43 (0.41)} & \changing{0.56 (0.53)} & \changing{0.36 (0.38)} \\ \hline 
\rule{0pt}{3ex}\changing{\our-25}               & \changing{0.29 (0.32)}  & \changing{0.42 (0.41)} & \changing{0.51 (0.45)} & \changing{0.36 (0.37)} \\ \hline 
\end{tabular}
}
\end{center}
\end{table}

\begin{figure*}[t!]
\begin{center}
\includegraphics[width=\textwidth]{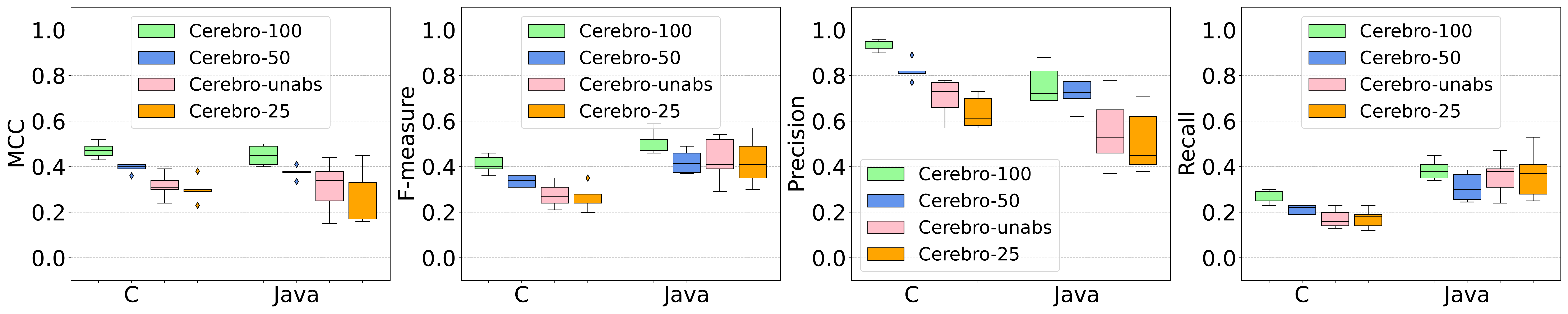}
\end{center}
\caption{\changing{Impact of the abstraction process and sequence length in \our's prediction performance: 
On average, MCC is decreased by 18\% with unabstracted code and decreased by 24\% with sequence length 25.
}}
\label{fig:rq1_all}
\end{figure*}

\changing{\subsection{Impact of reducing the sequence length}
\label{sec:reducing-length-training}


We also analyzed the impact of reducing the length of sequences that we use to train our models and how it affects the model prediction performance (RQ1). In this experiment, we reduced the sequence length from 50 tokens per sequence to 25 tokens per sequence. Figure~\ref{fig:rq1_all} and Table~\ref{tab:rq1_all} shows the average and median scores achieved by the models. For simulation details on Effectiveness Evaluation (RQ2), Number of Analyzed Mutants (RQ3) and Number of Test Executions (RQ4), please refer to our online repository. From these results we found that reducing the length of sequences used by the models to train also deteriorated the model prediction performance for projects in both C and Java benchmarks. For C-Benchmark, the model performance deteriorated by 23.5\% in MCC, 22.2\% in Precision and 18.1\% in Recall. For Java-Benchmark, although we found an improvement of 18.7\% in Recall, the overall performance deteriorated by 24.7\% in MCC and 28.6\% in Precision.
}

\begin{table}[t!]
  \begin{center}
\caption{
\changing{Impact of noise in evaluation on all approaches' performance (MS*): \our's and \dt' performances are more or less inversely related to the noise in evaluation. For \rnd selection, the performance also deteriorated in most of the cases, with exceptions of 10\% noise in C benchmark, and 6\% and 8\% noise in Java benchmark where \rnd's performance improved by 6.48\%, and 0.23\% and 1.26\% improved MS*, respectively.}
}
\label{tab:noise}
\resizebox{.45\textwidth}{!}{
\begin{tabular}{| r | r | r | r |}
\toprule
\multicolumn{4}{c}{\changing{Performance Change \% (Median) in MS* w.r.t. noise}}\\
\multicolumn{4}{c}{\changing{for C-Benchmark}}\\
\toprule
\hline
\rule{0pt}{3ex}\textbf{\changing{Noise (\%)}} & \textbf{\changing{\our}}  & \textbf{\changing{\dt}}  & \textbf{\changing{\rnd}} \\ \hline
\rule{0pt}{2ex}\changing{2\%} & \changing{$\downarrow$ -2.24\%} & \changing{$\downarrow$ -3.61\%} & \changing{$\downarrow$ -13.91\%} \\ \hline
\rule{0pt}{2ex}\changing{4\%} & \changing{$\downarrow$ -3.10\%} & \changing{$\downarrow$ -3.59\%} & \changing{$\downarrow$ -8.89\%} \\ \hline
\rule{0pt}{2ex}\changing{6\%} & \changing{$\downarrow$ -4.67\%} & \changing{$\downarrow$ -3.83\%} & \changing{$\downarrow$ -0.95\%} \\ \hline
\rule{0pt}{2ex}\changing{8\%} & \changing{$\downarrow$ -5.78\%} & \changing{$\downarrow$ -7.40\%} & \changing{$\downarrow$ -8.17\%} \\ \hline
\rule{0pt}{2ex}\changing{10\%} & \changing{$\downarrow$ -7.06\%} & \changing{$\downarrow$ -6.53\%} & \changing{$\uparrow$ +6.48\%} \\ \hline
\multicolumn{4}{c}{}\\ 
\toprule
\multicolumn{4}{c}{\changing{Performance Change \% (Median) in MS* w.r.t. noise}}\\
\multicolumn{4}{c}{\changing{for Java-Benchmark}}\\
\toprule
\hline
\rule{0pt}{3ex}\textbf{\changing{Noise (\%)}} & \textbf{\changing{\our}}  & \textbf{\changing{\dt}}  & \textbf{\changing{\rnd}} \\ \hline
\rule{0pt}{2ex}\changing{2\%} & \changing{$\downarrow$ -2.19\%} & \changing{$\downarrow$ -1.17\%} & \changing{$\downarrow$ -0.16\%} \\ \hline
\rule{0pt}{2ex}\changing{4\%} & \changing{$\downarrow$ -4.55\%} & \changing{$\downarrow$ -1.89\%} & \changing{$\downarrow$ -0.39\%} \\ \hline
\rule{0pt}{2ex}\changing{6\%} & \changing{$\downarrow$ -6.15\%} & \changing{$\downarrow$ -2.76\%} & \changing{$\uparrow$ +0.23\%} \\ \hline
\rule{0pt}{2ex}\changing{8\%} & \changing{$\downarrow$ -7.50\%} & \changing{$\downarrow$ -3.76\%} & \changing{$\uparrow$ +1.26\%} \\ \hline
\rule{0pt}{2ex}\changing{10\%} & \changing{$\downarrow$ -8.61\%} & \changing{$\downarrow$ -4.63\%} & \changing{$\downarrow$ -2.80\%} \\ \hline
\end{tabular}
}
\end{center}
\end{table}

\begin{figure*}[t!]
\begin{center}
\includegraphics[width=\textwidth]{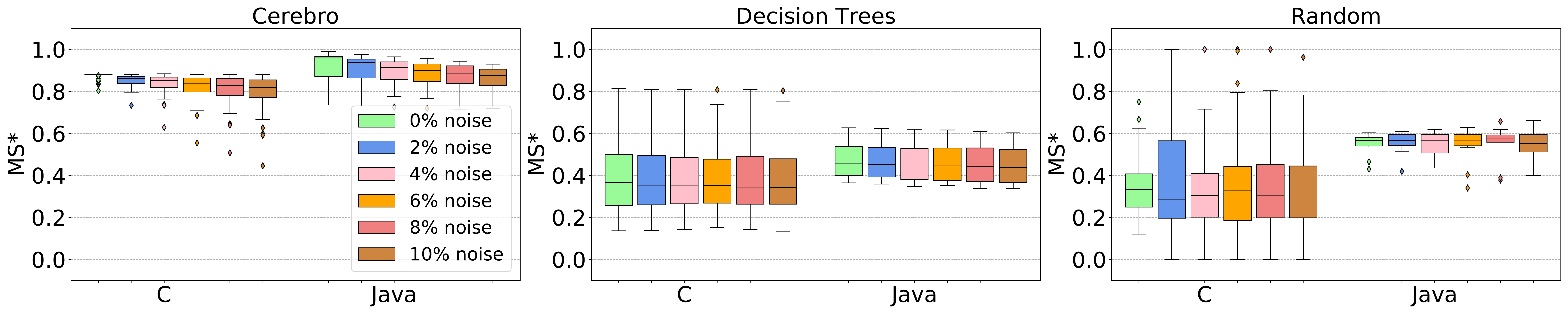}
\end{center}
\caption{\changing{Impact of noise in evaluation on all approaches' performance (MS*): \our's and \dt' performances are more or less inversely related to the noise in evaluation. For \rnd selection, the performance also deteriorated in most of the cases, with exceptions of 10\% noise in C benchmark, and 6\% and 8\% noise in Java benchmark where \rnd's performance improved by 6.48\%, and 0.23\% and 1.26\% improved MS*, respectively.
}}
\label{fig:noise}
\end{figure*}

\changing{\subsection{Impact of considering equivalent, mutants that are subsuming, \ie~impact of potential mistakes in our evaluation}
\label{sec:noise-equivalent-mutants}

In our experiments, we considered the mutants that were not killed by our test suite as unkillable a.k.a. equivalent. Although this being an undecidable problem (as we elaborated in Section~\ref{subsec:problem-of-mutant-equivalence}), we analyzed the impact of what would have happened if the mutants that we considered as equivalent were subsuming instead. Hence, we addressed this by introducing noise in our evaluation, \ie~we assumed 2\% equivalent mutants in our evaluation set as subsuming and analyzed the change in performance (MS* achieved) for all the approaches (\our, \dt and \rnd). We gradually increased the noise percentage from 2\% till 10\% (\ie~2\%, 4\%, 6\%, 8\%, 10\%) and analyzed the change in behaviour for all the approaches (\ie~change in MS*), if it increases or decreases with increase in noise.

We found that \our's and \dt' performances are more or less inversely related to the noise in evaluation (Figure~\ref{fig:noise}). Higher the noise, lower the MS* achieved by both the approaches (with an exception of 10\% noise in C benchmark for \dt where \dt performed better than in case of 8\% noise, as detailed in Table~\ref{tab:noise}). For \rnd selection, the performance also deteriorated in most of the cases, with an exception of 10\% noise in C benchmark, and 6\% and 8\% noise in Java benchmark where \rnd's performance improved by 6.48\%, and 0.23\% and 1.26\% improved MS*, respectively. Despite the reduction in performance due to introduced noise, \our still achieves higher MS* than the baselines (Figure~\ref{fig:noise}).
}

\section{Threats to Validity}
\label{sec:threats-validity}

\textit{External Validity}: Threats may relate to the subjects we used. 
Although our evaluation expands to both C and Java projects of different sizes, the results may not generalize to other projects or programming languages. We consider this threat as low since we have a large sample of programs, i.e., we perform one of the largest mutation testing studies \changing{to} date. 

Other external threat lies in the operators we used, since our prediction approach might not work for different types of mutants. 
To reduce this threat, we employ modern mutation tools, for both C and Java that implement a large variety of mutation operators. For the C-Benchmark, taken from~\cite{Chekam+2021}, 816 simple operators across 18 categories were considered; while for creating our Java-Benchmark, we consider the group ``ALL'' of mutation operators provided by Pitest~\cite{Coles+2016}, resulting in 112 simple operators across 29 categories.

\textit{Internal Validity}: Threats may relate to the restriction that we impose on sequence length, \ie a maximum of \changing{\emph{100}} tokens. This was done to enable reasonable model training time, approximately \changing{\emph{740}} hours. \changing{Moreover, restricting the sequence length to \emph{50} assisted to reach an appropriate training time of 360 hours. 
\minrev{However, it resulted in a prediction performance deterioration of approximately 15\%, as discussed in Section~\ref{sec:discussion}.} 
} 
Other threats maybe due to the use of machine translation for classification. This choice was made for simplicity, to use the related framework out of the box, similar to the related studies \cite{tufano_icsme_2019,tufano_tosem_2019}. Still a potential ``sequence to class classifier'' may \changing{yield} better results, though such improvements should be marginal given the low number of unexpected labels we get, i.e., on average, 2.15\% of the mutants do not get a valid label (4.2\% in C and 0.1\% in Java). 

Threats may also relate to the features we implemented for training the \dt baseline.  
We follow the guidelines provided in~\cite{ChekamPBTS20}, to extract the 16 features for our Java dataset. 
Unfortunately, many of the 28 features for C programs presented in~\cite{ChekamPBTS20} depend on the semantic of the C language, that we found unfeasible to be replicated in Java. However, the prediction performance of \dt in Java are in line with the results obtained for C, indicating that the impact of this threat is low. 

Other internal validity threats could be related to the test suites we used \minrev{and the mutants considered as subsuming and equivalent}. 
To deal with this issue, we used well-tested programs and state-of-the-art tool \minrev{to generate extensive pools of tests} (KLEE\cite{Cadar+2008}, SEMu\cite{Chekam+2021}, and EvoSuite\cite{FraserZeller2010}). 
\minrev{Since identifying subsuming and equivalent mutants is an undecidable problem, in our experimental setup, we approximate them through an extensive pool of tests. 
This has been a typical process followed in related mutation testing studies~\cite{JiaH09, Ammann_2014, PapadakisHHJT16, Kurtz+2016, PapadakisCT18}. 
To be more accurate, our underlying assumption is that the extensive pool of tests used in our experiments are a valid representation of all possible tests that a tester can manually or automatically generate. This assumption allowed us to identify the minimal set of mutants (i.e., subsuming mutants) that a tester needs to kill in order to kill every other killable mutant (i.e., subsumed mutants).  
Also, we assumed that unkilled mutants are equivalent. 
}
Even if this may not be the case, it is likely that the testers guided by mutation won't be able to kill all the killable mutants. Here it must be noted that since Cerebro is quite precise, its feeding with less noisy data, i.e., correct labels, will make it perform better, i.e., more accurate labelling in training will result in better predictions. Nevertheless, we also investigate the impact of having such noisy data and found minor discrepancies, please refer to Section \ref{sec:noise-equivalent-mutants}.

\our's use may also pose additional threats. In particular, \our required approximately 5 minutes for preprocessing of the projects and 5 minutes for classification (decoding results). While this time overhead is low, compared to the hours of test executions, it may still be important. Although our implementation is non-optimal and involves no parallelism, however our encoding and decoding can easily be parallelized, since mutant instances are independent of one another. 

\textit{Construct Validity}: 
Our assessment metrics, subsuming mutation score, number of equivalent mutants and number of test executions may not reflect the actual testing cost~/~effectiveness values. These metrics have been suggested by literature \cite{Papadakis+2019,Andrews+2006,Kurtz+2016} and are intuitive, i.e., number of selected and analyzed mutants essentially simulate the manual effort involved by testers, subsuming mutation score the level of covering the test requirements \cite{Ammann_2014, PapadakisHHJT16}, and number of test executions capture the computational effort involved. Here it should be noted that automated test generation tools may reduce this cost but they require testers to check the related test oracles. Similarly, equivalent detection techniques and related heuristics may also reduce the manual effort involved \cite{KintisPM15}. Though, in C we applied TCE \changing{(Trivial Compiler Equivalence)} \cite{Kintis+2018, HaririSFMM19} to filter out equivalent and duplicated mutants and our approach still provided significant benefits.  
Similarly, the use of test executions capture the computational effort involved independently of the test execution framework and optimizations used \cite{WangXSZH17, ZhangMK13, ChenZ18, Papadakis+2019}, the machines and the level of parallelization used during test execution. Nevertheless, the differences are substantial making such threats unlikely to happen.    
Overall, we mitigate these threats by following suggestions from mutation testing literature \cite{Papadakis+2019,Andrews+2006,Kurtz+2016}, using state-of-the-art tools, performing several simulations, forming very large and diverse test pools, and got consistent and stable results across our subjects. 

\section{Related Work}
\label{sec:related-work}
\changing{Mutation testing has been established as one of the strongest test criteria \cite{ChekamPTH17, Ammann_2014}. Despite its potential, mutation is considered to be expensive since it introduces too many mutants. To this end, 
random mutant sampling~\cite{HintTestDataSelection1978, PapadakisM10} and selective mutation \cite{OffuttLRUZ96} (restricting mutant instances according to their types) have been proposed as potential solutions. Unfortunately, these approaches fail to capture relevant program semantics and performing similarly to random mutant sampling \cite{ZhangGMK13,Kurtz+2016,ChekamPBTS20}.}

Other attempts regard the selection of relevant program locations, which should be mutated. Sun et al. \cite{SunXLZ17} proposed selecting mutants that reside in diverse static control flow graph paths. Gong et al. \cite{GongZYM17} identified dominator nodes (using static control flow graph) to select mutants. 

More recent attempts regard the identification of interesting mutants (pairs of mutant types and related locations). Petrovic and Ivankovic  \cite{Goran} and Just et al. \cite{JustKA17} proposed using the code AST in order to identify ``useful'' mutants.  Petrovic and Ivankovic used what they called arid nodes (special AST nodes), while Just et al. used the AST parent and child nodes, in order to identify high utility mutants. Mirshokraie et al. \cite{Mirshokraie0P15} employed complexity metrics together with test executions to select killable mutants. Similarly, Titcheu et al. \cite{ChekamPBTS20} employed static features, including data flow analysis, complexity and AST information, in order to perform mutant selection, wrt mutants linked with real faults. 

In our analysis we approximate the performance of the above approaches through the two baselines we adopt and show that our approach significantly outperforms these. Random mutant sampling is performing comparably to operator mutant selection \cite{ZhangGMK13}, while the supervised baseline we consider simulates the AST-based and complexity-based approaches. 

Perhaps the closest work to ours, is from Marcozzi et al. \cite{MarcozziBKPPC18}, which attempts to identify subsumed mutants using verification techniques (such as weakest precondition). While  Marcozzi et al.\changing{'s} approach is particularly powerful, it targets weak mutation and not  strong as we do.  This results in several false positives in the strong mutation case due  to failed error propagation \cite{ChekamPTH17}.  Moreover, Marcozzi et al.\changing{'s} approach is time consuming, requires complex computations and infrastructure while \our is fast and simple. Nevertheless, future research  should attempt to combine these methods. 

Tufano et al. \cite{tufano_icsme_2019} proposed using Neural Machine Translation to learn mutations from bug fixes 
with the aim of introducing mutations that are syntactically similar to real bugs. \our relies on the same technology, though it targets a different problem; the identification of high utility mutants, among those given by regular mutation testing tools, while Tufano et al. aim at generating mutants regardless of their potential. This indicates that \our can complement Tufano et al by selecting relevant mutants. Nevertheless, we focus on subsuming mutants, that could help measuring test adequacy and designing test suites, which are unlikely to be supported by Tufano et al. as there is no notion of subsumption in the bug-fixing sets they use. Moreover, we make no assumption on the availability and repetitiveness of historical bugs and their fixes. 

Predictive mutation testing (PMT) \cite{0050ZHH0019} attempts to predict whether a given test can kill a given mutant without performing any mutant execution. The approach relies on a set of both static and dynamic features (relying on coverage and code attributes) and achieves relatively good results (on average with 10\% error). Though, PMT mainly targets intra-project predictions, while \our targets inter-project. Nevertheless, PMT is incomparable to \our since it aims at evaluating test execution results, while we do mutant selection prior to any test execution. In other words, we aim at identifying the mutants to be used for test design/generation, while PMT  to verify  whether mutants are killed by some tests. Therefore, the two methods target different but complementary problems. 

\changing{Evolutionary Mutation Testing (EMT) \cite{Dominguez-JimenezEGM11} utilises dynamic features (execution traces) in order to identify interesting locations and mutant types. As such\minrev{,} EMT requires tests and user feedback, which make it different but complementary to ours; \our can set a starting point for EMT or integrate its predictions within EMT's fitness function. Higher-order mutation \cite{JiaH09} aims at dynamically optimizing mutants based on given test suites. This means that Cerebro can be directly applied to support test generation prior to any test generation, while higher-order mutation is only applicable after test generation. Perhaps more importantly, Cerebro does not introduce any expensive dynamic mutant execution, while higher-order mutation introduces major mutant execution overheads.}

\section{Conclusion and Future Work}
\label{sec:conclusion}
We presented \our, a method that learns to select subsuming mutants (subset of mutants that subsumes the others, i.e., tests killing them also kill all the mutants of the given mutant set) from given mutant sets.
Experiments with 58 programs showed that \our identified subsuming mutants with \changing{0.85} 
precision and \changing{0.33} 
recall at an inter-project scenario (trained on different projects than the ones it was evaluated). These predictions enable testers designing test cases capable of killing more than two times the subsuming mutants that they would kill if they were using either randomly selected mutants or another previously proposed machine learning-based mutant selection  technique. At the same time \our entails the analysis of 
\changing{66\%~fewer} equivalent mutants and \changing{90\%} 
less mutant executions, indicating a large reduction on the practical effort/cost of the approach.

\changing{Recently, it has become increasingly common to pre-train the entire model on a data-rich task\minrev{,} which causes the model to develop general-purpose abilities and knowledge that can then be transferred to downstream tasks~\cite{raffel2019exploring}. In this practice \aka Transfer Learning and its applications to computer vision~\cite{oquab2014learning, vedaldi2014convolutional}, pre-training is typically done via supervised learning on a large labeled data set like ImageNet~\cite{russakovsky2015imagenet}. In contrast, modern techniques for transfer learning in Natural Language Processing (\textit{NLP}) often pre-train using unsupervised learning on unlabeled data~\cite{devlin2018bert, liu2021robustly}. The resulting pre-trained models are further trained on specialized datasets to accomplish the desired tasks. Unsupervised pre-training for \textit{NLP} is attractive and seems a good fit for neural networks as it have been shown to exhibit remarkable scalability, \ie it is often possible to achieve better performance simply by training a larger model on a larger data set~\cite{hestness2017deep, shazeer2017outrageously, jozefowicz2016exploring, mahajan2018exploring}. It will be worthwhile to explore such available pre-trained models~\cite{DBLP:journals/corr/abs-2102-02017, DBLP:conf/emnlp/FengGTDFGS0LJZ20} and if these can be further refined to address our specific prediction task.

On the other hand, as we have shown that \our is proficient in capturing the silent features and patterns of the code context, it is promising to explore \our in security\minrev{-}specific task such as prediction of zero-day vulnerabilities\minrev{,} which pose a very high risk~\cite{zero-day}. Vulnerabilities are fewer in comparison to defects, limiting the information one can learn from. Also, their identification requires an attacker's mindset~\cite{morrison2015challenges}, which developers or code reviewers may not possess. Lastly, the continuous growth of codebases makes it difficult to investigate them entirely and track all code changes. For instance, Linux kernel, which is one of the projects with \minrev{the} highest number of publicly reported vulnerabilities, reached 27.8 million LoC (Lines of Codes) at the beginning of 2020~\cite{linux-news}. Hence, it will also be rewarding to explore \our in this line of work.}

\section*{Acknowledgment}

This work is supported by the Luxembourg National Research Funds (FNR) through the INTER project grant, INTER/ANR/18/12632675/SATOCROSS.

\bibliographystyle{plain}
\bibliography{Bibliography}

\begin{thebibliography}{10}

\bibitem{tensorflow2015-whitepaper}
Mart\'{\i}n Abadi, Ashish Agarwal, Paul Barham, Eugene Brevdo, Zhifeng Chen,
  Craig Citro, Greg~S. Corrado, Andy Davis, Jeffrey Dean, Matthieu Devin,
  Sanjay Ghemawat, Ian Goodfellow, Andrew Harp, Geoffrey Irving, Michael Isard,
  Yangqing Jia, Rafal Jozefowicz, Lukasz Kaiser, Manjunath Kudlur, Josh
  Levenberg, Dandelion Man\'{e}, Rajat Monga, Sherry Moore, Derek Murray, Chris
  Olah, Mike Schuster, Jonathon Shlens, Benoit Steiner, Ilya Sutskever, Kunal
  Talwar, Paul Tucker, Vincent Vanhoucke, Vijay Vasudevan, Fernanda Vi\'{e}gas,
  Oriol Vinyals, Pete Warden, Martin Wattenberg, Martin Wicke, Yuan Yu, and
  Xiaoqiang Zheng.
\newblock {TensorFlow}: Large-scale machine learning on heterogeneous systems,
  2015.
\newblock Software available from tensorflow.org.

\bibitem{Acree80}
Allen~Troy Acree.
\newblock {\em {O}n {M}utation}.
\newblock Phd thesis, School of Information and Computer Science, Georgia
  Institute of Technology, Atlanta, Georgia, 1980.

\bibitem{Ammann_2014}
P.~{Ammann}, M.~E. {Delamaro}, and J.~{Offutt}.
\newblock Establishing theoretical minimal sets of mutants.
\newblock In {\em 2014 IEEE Seventh International Conference on Software
  Testing, Verification and Validation}, pages 21--30, 2014.

\bibitem{AmmannOffutt2008}
Paul Ammann and Jeff Offutt.
\newblock {\em Introduction to Software Testing}.
\newblock Cambridge University Press, USA, 1 edition, 2008.

\bibitem{Andrews+2006}
J.~H. {Andrews}, L.~C. {Briand}, Y.~{Labiche}, and A.~S. {Namin}.
\newblock Using mutation analysis for assessing and comparing testing coverage
  criteria.
\newblock {\em IEEE Transactions on Software Engineering}, 32(8):608--624,
  2006.

\bibitem{bahdanau_arxiv_2014}
Dzmitry Bahdanau, Kyunghyun Cho, and Yoshua Bengio.
\newblock Neural machine translation by jointly learning to align and
  translate.
\newblock In Yoshua Bengio and Yann LeCun, editors, {\em 3rd International
  Conference on Learning Representations, {ICLR} 2015, San Diego, CA, USA, May
  7-9, 2015, Conference Track Proceedings}, 2015.

\bibitem{7472618}
Dzmitry Bahdanau, Jan Chorowski, Dmitriy Serdyuk, Philémon Brakel, and Yoshua
  Bengio.
\newblock End-to-end attention-based large vocabulary speech recognition.
\newblock In {\em 2016 IEEE International Conference on Acoustics, Speech and
  Signal Processing (ICASSP)}, pages 4945--4949, 2016.

\bibitem{britz_arxiv_2017}
Denny Britz, Anna Goldie, Minh-Thang Luong, and Quoc Le.
\newblock Massive exploration of neural machine translation architectures.
\newblock In {\em Proceedings of the 2017 Conference on Empirical Methods in
  Natural Language Processing}, pages 1442--1451, Copenhagen, Denmark,
  September 2017. Association for Computational Linguistics.

\bibitem{BuddA82}
Timothy~Alan Budd and Dana Angluin.
\newblock {T}wo {N}otions of {C}orrectness and {T}heir {R}elation to {T}esting.
\newblock {\em {A}cta {I}nformatica}, 18(1):31--45, March 1982.

\bibitem{Cadar+2008}
Cristian Cadar, Daniel Dunbar, and Dawson Engler.
\newblock Klee: Unassisted and automatic generation of high-coverage tests for
  complex systems programs.
\newblock In {\em Proceedings of the 8th USENIX Conference on Operating Systems
  Design and Implementation}, OSDI'08, page 209–224, USA, 2008. USENIX
  Association.

\bibitem{ChekamPBTS20}
Thierry~Titcheu Chekam, Mike Papadakis, Tegawend{\'{e}}~F. Bissyand{\'{e}},
  Yves~Le Traon, and Koushik Sen.
\newblock Selecting fault revealing mutants.
\newblock {\em Empirical Software Engineering}, 25(1):434--487, 2020.

\bibitem{Chekam+2021}
Thierry~Titcheu Chekam, Mike Papadakis, Maxime Cordy, and Yves~Le Traon.
\newblock Killing stubborn mutants with symbolic execution.
\newblock {\em ACM Trans. Softw. Eng. Methodol.}, 30(2), January 2021.

\bibitem{Chekam+2019}
Thierry~Titcheu Chekam, Mike Papadakis, and Yves Le~Traon.
\newblock Mart: A mutant generation tool for llvm.
\newblock In {\em Proceedings of the 2019 27th ACM Joint Meeting on European
  Software Engineering Conference and Symposium on the Foundations of Software
  Engineering}, ESEC/FSE 2019, page 1080–1084, New York, NY, USA, 2019.
  Association for Computing Machinery.

\bibitem{ChekamPTH17}
Thierry~Titcheu Chekam, Mike Papadakis, Yves~Le Traon, and Mark Harman.
\newblock An empirical study on mutation, statement and branch coverage fault
  revelation that avoids the unreliable clean program assumption.
\newblock In Sebasti{\'{a}}n Uchitel, Alessandro Orso, and Martin~P. Robillard,
  editors, {\em Proceedings of the 39th International Conference on Software
  Engineering, {ICSE} 2017, Buenos Aires, Argentina, May 20-28, 2017}, pages
  597--608. {IEEE} / {ACM}, 2017.

\bibitem{ChenZ18}
Lingchao Chen and Lingming Zhang.
\newblock Speeding up mutation testing via regression test selection: An
  extensive study.
\newblock In {\em 11th {IEEE} International Conference on Software Testing,
  Verification and Validation, {ICST} 2018, V{\"{a}}ster{\aa}s, Sweden, April
  9-13, 2018}, pages 58--69. {IEEE} Computer Society, 2018.

\bibitem{Cho+2014}
Kyunghyun Cho, Bart van Merri{\"e}nboer, Caglar Gulcehre, Dzmitry Bahdanau,
  Fethi Bougares, Holger Schwenk, and Yoshua Bengio.
\newblock Learning phrase representations using {RNN} encoder{--}decoder for
  statistical machine translation.
\newblock In {\em Proceedings of the 2014 Conference on Empirical Methods in
  Natural Language Processing ({EMNLP})}, pages 1724--1734, Doha, Qatar,
  October 2014. Association for Computational Linguistics.

\bibitem{Coles+2016}
Henry Coles, Thomas Laurent, Christopher Henard, Mike Papadakis, and Anthony
  Ventresque.
\newblock Pit: A practical mutation testing tool for java (demo).
\newblock In {\em Proceedings of the 25th International Symposium on Software
  Testing and Analysis}, ISSTA 2016, page 449–452, New York, NY, USA, 2016.
  Association for Computing Machinery.

\bibitem{7816536}
M.~L. {Collard} and J.~I. {Maletic}.
\newblock srcml 1.0: Explore, analyze, and manipulate source code.
\newblock In {\em 2016 IEEE International Conference on Software Maintenance
  and Evolution (ICSME)}, pages 649--649, 2016.

\bibitem{HintTestDataSelection1978}
R.~A. DeMillo, R.~J. Lipton, and F.~G. Sayward.
\newblock Hints on test data selection: Help for the practicing programmer.
\newblock {\em IEEE Computer}, 11(4):34--41, April 1978.

\bibitem{devlin2018bert}
Jacob Devlin, Ming{-}Wei Chang, Kenton Lee, and Kristina Toutanova.
\newblock {BERT:} pre-training of deep bidirectional transformers for language
  understanding.
\newblock {\em Proceedings of the 2019 Conference of the North American Chapter
  of the Association for Computational Linguistics: Human Language
  Technologies, {NAACL-HLT} 2019, Minneapolis, MN, USA, June 2-7, 2019, Volume
  1 (Long and Short Papers)}, pages 4171--4186, 2019.

\bibitem{Dominguez-JimenezEGM11}
Juan~Jos{\'{e}} Dom{\'{\i}}nguez{-}Jim{\'{e}}nez, Antonia Estero{-}Botaro,
  Antonio Garc{\'{\i}}a{-}Dom{\'{\i}}nguez, and Inmaculada Medina{-}Bulo.
\newblock Evolutionary mutation testing.
\newblock {\em Inf. Softw. Technol.}, 53(10):1108--1123, 2011.

\bibitem{DBLP:conf/emnlp/FengGTDFGS0LJZ20}
Zhangyin Feng, Daya Guo, Duyu Tang, Nan Duan, Xiaocheng Feng, Ming Gong, Linjun
  Shou, Bing Qin, Ting Liu, Daxin Jiang, and Ming Zhou.
\newblock Codebert: {A} pre-trained model for programming and natural
  languages.
\newblock In Trevor Cohn, Yulan He, and Yang Liu, editors, {\em Findings of the
  Association for Computational Linguistics: {EMNLP} 2020, Online Event, 16-20
  November 2020}, volume {EMNLP} 2020 of {\em Findings of {ACL}}, pages
  1536--1547. Association for Computational Linguistics, 2020.

\bibitem{FraserZeller2010}
Gordon Fraser and Andreas Zeller.
\newblock Mutation-driven generation of unit tests and oracles.
\newblock In {\em Proceedings of the ACM International Symposium on Software
  Testing and Analysis}, ISSTA '10, pages 147--158, New York, NY, USA, 2010.
  ACM.

\bibitem{GongZYM17}
Dunwei Gong, Gongjie Zhang, Xiangjuan Yao, and Fanlin Meng.
\newblock Mutant reduction based on dominance relation for weak mutation
  testing.
\newblock {\em Information {\&} Software Technology}, 81:82--96, 2017.

\bibitem{HaririSFMM19}
Farah Hariri, August Shi, Vimuth Fernando, Suleman Mahmood, and Darko Marinov.
\newblock Comparing mutation testing at the levels of source code and compiler
  intermediate representation.
\newblock In {\em 12th {IEEE} Conference on Software Testing, Validation and
  Verification, {ICST} 2019, Xi'an, China, April 22-27, 2019}, pages 114--124.
  {IEEE}, 2019.

\bibitem{hestness2017deep}
Joel Hestness, Sharan Narang, Newsha Ardalani, Gregory~F. Diamos, Heewoo Jun,
  Hassan Kianinejad, Md. Mostofa~Ali Patwary, Yang Yang, and Yanqi Zhou.
\newblock Deep learning scaling is predictable, empirically.
\newblock {\em CoRR}, abs/1712.00409, 2017.

\bibitem{vedaldi2014convolutional}
Yangqing Jia, Evan Shelhamer, Jeff Donahue, Sergey Karayev, Jonathan Long,
  Ross~B. Girshick, Sergio Guadarrama, and Trevor Darrell.
\newblock Caffe: Convolutional architecture for fast feature embedding.
\newblock In Kien~A. Hua, Yong Rui, Ralf Steinmetz, Alan Hanjalic, Apostol
  Natsev, and Wenwu Zhu, editors, {\em Proceedings of the {ACM} International
  Conference on Multimedia, {MM} '14, Orlando, FL, USA, November 03 - 07,
  2014}, pages 675--678. {ACM}, 2014.

\bibitem{JiaH09}
Yue Jia and Mark Harman.
\newblock Higher order mutation testing.
\newblock {\em Inf. Softw. Technol.}, 51(10):1379--1393, 2009.

\bibitem{jozefowicz2016exploring}
Rafal J{\'{o}}zefowicz, Oriol Vinyals, Mike Schuster, Noam Shazeer, and Yonghui
  Wu.
\newblock Exploring the limits of language modeling.
\newblock {\em CoRR}, abs/1602.02410, 2016.

\bibitem{JustKA17}
Ren{\'{e}} Just, Bob Kurtz, and Paul Ammann.
\newblock Inferring mutant utility from program context.
\newblock In {\em Proceedings of the 26th {ACM} {SIGSOFT} International
  Symposium on Software Testing and Analysis, Santa Barbara, CA, USA, July 10 -
  14, 2017}, pages 284--294, 2017.

\bibitem{5693206}
M.~{Kintis}, M.~{Papadakis}, and N.~{Malevris}.
\newblock Evaluating mutation testing alternatives: A collateral experiment.
\newblock In {\em 2010 Asia Pacific Software Engineering Conference}, pages
  300--309, 2010.

\bibitem{Kintis+2018}
Marinos Kintis, Mike Papadakis, Yue Jia, Nicos Malevris, Yves~Le Traon, and
  Mark Harman.
\newblock Detecting trivial mutant equivalences via compiler optimisations.
\newblock {\em {IEEE} Trans. Software Eng.}, 44(4):308--333, 2018.

\bibitem{KintisPM15}
Marinos Kintis, Mike Papadakis, and Nicos Malevris.
\newblock Employing second-order mutation for isolating first-order equivalent
  mutants.
\newblock {\em Softw. Test. Verification Reliab.}, 25(5-7):508--535, 2015.

\bibitem{Kurtz_2014}
B.~{Kurtz}, P.~{Ammann}, M.~E. {Delamaro}, J.~{Offutt}, and L.~{Deng}.
\newblock Mutant subsumption graphs.
\newblock In {\em 2014 IEEE Seventh International Conference on Software
  Testing, Verification and Validation Workshops}, pages 176--185, 2014.

\bibitem{Kurtz+2016}
Bob Kurtz, Paul Ammann, Jeff Offutt, M\'{a}rcio~E. Delamaro, Mariet Kurtz, and
  Nida G\"{o}k\c{c}e.
\newblock Analyzing the validity of selective mutation with dominator mutants.
\newblock In {\em Proceedings of the 2016 24th ACM SIGSOFT International
  Symposium on Foundations of Software Engineering}, FSE 2016, page 571–582,
  New York, NY, USA, 2016. Association for Computing Machinery.

\bibitem{linux-news}
Linux in 2020: 27.8 million lines of code in the kernel, 1.3 million in
  systemd.
\newblock
  \url{https://www.theregister.com/2020/01/06/linux_2020_kernel_systemd_code/},
  (accessed October 12, 2021).

\bibitem{liu2021robustly}
Zhuang Liu, Wayne Lin, Ya~Shi, and Jun Zhao.
\newblock A robustly optimized bert pre-training approach with post-training.
\newblock In {\em China National Conference on Chinese Computational
  Linguistics}, pages 471--484. Springer, 2021.

\bibitem{mahajan2018exploring}
Dhruv Mahajan, Ross Girshick, Vignesh Ramanathan, Kaiming He, Manohar Paluri,
  Yixuan Li, Ashwin Bharambe, and Laurens Van Der~Maaten.
\newblock Exploring the limits of weakly supervised pretraining.
\newblock In {\em Proceedings of the European conference on computer vision
  (ECCV)}, pages 181--196, 2018.

\bibitem{MarcozziBKPPC18}
Micha{\"{e}}l Marcozzi, S{\'{e}}bastien Bardin, Nikolai Kosmatov, Mike
  Papadakis, Virgile Prevosto, and Lo{\"{\i}}c Correnson.
\newblock Time to clean your test objectives.
\newblock In Michel Chaudron, Ivica Crnkovic, Marsha Chechik, and Mark Harman,
  editors, {\em Proceedings of the 40th International Conference on Software
  Engineering, {ICSE} 2018, Gothenburg, Sweden, May 27 - June 03, 2018}, pages
  456--467. {ACM}, 2018.

\bibitem{DBLP:journals/corr/abs-2102-02017}
Antonio Mastropaolo, Simone Scalabrino, Nathan Cooper, David Nader{-}Palacio,
  Denys Poshyvanyk, Rocco Oliveto, and Gabriele Bavota.
\newblock Studying the usage of text-to-text transfer transformer to support
  code-related tasks.
\newblock {\em CoRR}, abs/2102.02017, 2021.

\bibitem{MATTHEWS1975442}
B.W. Matthews.
\newblock Comparison of the predicted and observed secondary structure of t4
  phage lysozyme.
\newblock {\em Biochimica et Biophysica Acta (BBA) - Protein Structure},
  405(2):442 -- 451, 1975.

\bibitem{Mirshokraie0P15}
Shabnam Mirshokraie, Ali Mesbah, and Karthik Pattabiraman.
\newblock Guided mutation testing for javascript web applications.
\newblock {\em {IEEE} Trans. Software Eng.}, 41(5):429--444, 2015.

\bibitem{morrison2015challenges}
Patrick Morrison, Kim Herzig, Brendan Murphy, and Laurie Williams.
\newblock Challenges with applying vulnerability prediction models.
\newblock In {\em Proceedings of the 2015 Symposium and Bootcamp on the Science
  of Security}, pages 1--9, 2015.

\bibitem{OffuttLRUZ96}
A.~Jefferson Offutt, Ammei Lee, Gregg Rothermel, Roland~H. Untch, and Christian
  Zapf.
\newblock An experimental determination of sufficient mutant operators.
\newblock {\em {ACM} Trans. Softw. Eng. Methodol.}, 5(2):99--118, 1996.

\bibitem{oquab2014learning}
Maxime Oquab, Leon Bottou, Ivan Laptev, and Josef Sivic.
\newblock Learning and transferring mid-level image representations using
  convolutional neural networks.
\newblock In {\em Proceedings of the IEEE conference on computer vision and
  pattern recognition}, pages 1717--1724, 2014.

\bibitem{PapadakisCT18}
Mike Papadakis, Thierry~Titcheu Chekam, and Yves~Le Traon.
\newblock Mutant quality indicators.
\newblock In {\em 2018 {IEEE} International Conference on Software Testing,
  Verification and Validation Workshops, {ICST} Workshops, V{\"{a}}ster{\aa}s,
  Sweden, April 9-13, 2018}, pages 32--39. {IEEE} Computer Society, 2018.

\bibitem{PapadakisHHJT16}
Mike Papadakis, Christopher Henard, Mark Harman, Yue Jia, and Yves~Le Traon.
\newblock Threats to the validity of mutation-based test assessment.
\newblock In {\em Proceedings of the 25th International Symposium on Software
  Testing and Analysis, {ISSTA} 2016, Saarbr{\"{u}}cken, Germany, July 18-20,
  2016}, pages 354--365. {ACM}, 2016.

\bibitem{Papadakis+2019}
Mike Papadakis, Marinos Kintis, Jie Zhang, Yue Jia, Yves~Le Traon, and Mark
  Harman.
\newblock Chapter six - mutation testing advances: An analysis and survey.
\newblock {\em Adv. Comput.}, 112:275--378, 2019.

\bibitem{PapadakisM10b}
Mike Papadakis and Nicos Malevris.
\newblock Automatic mutation test case generation via dynamic symbolic
  execution.
\newblock In {\em {IEEE} 21st International Symposium on Software Reliability
  Engineering, {ISSRE} 2010, San Jose, CA, USA, 1-4 November 2010}, pages
  121--130. {IEEE} Computer Society, 2010.

\bibitem{PapadakisM10}
Mike Papadakis and Nicos Malevris.
\newblock An empirical evaluation of the first and second order mutation
  testing strategies.
\newblock In {\em Third International Conference on Software Testing,
  Verification and Validation, {ICST} 2010, Paris, France, April 7-9, 2010,
  Workshops Proceedings}, pages 90--99. {IEEE} Computer Society, 2010.

\bibitem{PapadakisSYB18}
Mike Papadakis, Donghwan Shin, Shin Yoo, and Doo{-}Hwan Bae.
\newblock Are mutation scores correlated with real fault detection?: a large
  scale empirical study on the relationship between mutants and real faults.
\newblock In Michel Chaudron, Ivica Crnkovic, Marsha Chechik, and Mark Harman,
  editors, {\em Proceedings of the 40th International Conference on Software
  Engineering, {ICSE} 2018, Gothenburg, Sweden, May 27 - June 03, 2018}, pages
  537--548. {ACM}, 2018.

\bibitem{Goran}
Goran Petrovic and Marko Ivankovic.
\newblock State of mutation testing at google.
\newblock In {\em 40th {IEEE/ACM} International Conference on Software
  Engineering: Software Engineering in Practice Track, {ICSE-SEIP} 2018, May 27
  - 3 June 2018, Gothenburg, Sweden}, 2018.

\bibitem{raffel2019exploring}
Colin Raffel, Noam Shazeer, Adam Roberts, Katherine Lee, Sharan Narang, Michael
  Matena, Yanqi Zhou, Wei Li, and Peter~J. Liu.
\newblock Exploring the limits of transfer learning with a unified text-to-text
  transformer.
\newblock {\em CoRR}, abs/1910.10683, 2019.

\bibitem{russakovsky2015imagenet}
Olga Russakovsky, Jia Deng, Hao Su, Jonathan Krause, Sanjeev Satheesh, Sean Ma,
  Zhiheng Huang, Andrej Karpathy, Aditya Khosla, Michael Bernstein, et~al.
\newblock Imagenet large scale visual recognition challenge.
\newblock {\em International journal of computer vision}, 115(3):211--252,
  2015.

\bibitem{shazeer2017outrageously}
Noam Shazeer, Azalia Mirhoseini, Krzysztof Maziarz, Andy Davis, Quoc~V. Le,
  Geoffrey~E. Hinton, and Jeff Dean.
\newblock Outrageously large neural networks: The sparsely-gated
  mixture-of-experts layer.
\newblock In {\em 5th International Conference on Learning Representations,
  {ICLR} 2017, Toulon, France, April 24-26, 2017, Conference Track
  Proceedings}. OpenReview.net, 2017.

\bibitem{6824804}
M.~{Shepperd}, D.~{Bowes}, and T.~{Hall}.
\newblock Researcher bias: The use of machine learning in software defect
  prediction.
\newblock {\em IEEE Transactions on Software Engineering}, 40(6):603--616,
  2014.

\bibitem{lstmcomparison2019}
Apeksha Shewalkar, Deepika Nyavanandi, and Simone Ludwig.
\newblock Performance evaluation of deep neural networks applied to speech
  recognition: Rnn, lstm and gru.
\newblock {\em Journal of Artificial Intelligence and Soft Computing Research},
  9:235--245, 10 2019.

\bibitem{SunXLZ17}
Chang{-}ai Sun, Feifei Xue, Huai Liu, and Xiangyu Zhang.
\newblock A path-aware approach to mutant reduction in mutation testing.
\newblock {\em Information {\&} Software Technology}, 81:65--81, 2017.

\bibitem{sutskever_arxiv_2014}
Ilya Sutskever, Oriol Vinyals, and Quoc~V. Le.
\newblock Sequence to sequence learning with neural networks, 2014.

\bibitem{TangMRS18}
Gongbo Tang, Mathias M{\"{u}}ller, Annette Rios, and Rico Sennrich.
\newblock Why self-attention? {A} targeted evaluation of neural machine
  translation architectures.
\newblock In Ellen Riloff, David Chiang, Julia Hockenmaier, and Jun'ichi
  Tsujii, editors, {\em Proceedings of the 2018 Conference on Empirical Methods
  in Natural Language Processing, Brussels, Belgium, October 31 - November 4,
  2018}, pages 4263--4272. Association for Computational Linguistics, 2018.

\bibitem{tufano_icsme_2019}
Michele Tufano, Cody Watson, Gabriele Bavota, Massimiliano Di~Penta, Martin
  White, and Denys Poshyvanyk.
\newblock Learning how to mutate source code from bug-fixes.
\newblock {\em 2019 IEEE International Conference on Software Maintenance and
  Evolution (ICSME)}, Sep 2019.

\bibitem{tufano_tosem_2019}
Michele Tufano, Cody Watson, Gabriele Bavota, Massimiliano~Di Penta, Martin
  White, and Denys Poshyvanyk.
\newblock An empirical study on learning bug-fixing patches in the wild via
  neural machine translation.
\newblock {\em {ACM} Trans. Softw. Eng. Methodol.}, 28(4):19:1--19:29, 2019.

\bibitem{VarghaDelaney2000}
András Vargha and Harold~D. Delaney.
\newblock A critique and improvement of the "cl" common language effect size
  statistics of mcgraw and wong.
\newblock {\em Journal of Educational and Behavioral Statistics},
  25(2):101--132, 2000.

\bibitem{WangXSZH17}
Bo~Wang, Yingfei Xiong, Yangqingwei Shi, Lu~Zhang, and Dan Hao.
\newblock Faster mutation analysis via equivalence modulo states.
\newblock In Tevfik Bultan and Koushik Sen, editors, {\em Proceedings of the
  26th {ACM} {SIGSOFT} International Symposium on Software Testing and
  Analysis, Santa Barbara, CA, USA, July 10 - 14, 2017}, pages 295--306. {ACM},
  2017.

\bibitem{yao2014study}
Xiangjuan Yao, Mark Harman, and Yue Jia.
\newblock A study of equivalent and stubborn mutation operators using human
  analysis of equivalence.
\newblock In {\em Proceedings of the 36th international conference on software
  engineering}, pages 919--930, 2014.

\bibitem{zero-day}
Zero-day vulnerability.
\newblock
  \url{https://www.trendmicro.com/vinfo/us/security/definition/zero-day-vulnerability},
  (accessed October 12, 2021).

\bibitem{0050ZHH0019}
Jie Zhang, Lingming Zhang, Mark Harman, Dan Hao, Yue Jia, and Lu~Zhang.
\newblock Predictive mutation testing.
\newblock {\em {IEEE} Trans. Software Eng.}, 45(9):898--918, 2019.

\bibitem{ZhangGMK13}
Lingming Zhang, Milos Gligoric, Darko Marinov, and Sarfraz Khurshid.
\newblock Operator-based and random mutant selection: Better together.
\newblock In Ewen Denney, Tevfik Bultan, and Andreas Zeller, editors, {\em 2013
  28th {IEEE/ACM} International Conference on Automated Software Engineering,
  {ASE} 2013, Silicon Valley, CA, USA, November 11-15, 2013}, pages 92--102.
  {IEEE}, 2013.

\bibitem{ZhangMK13}
Lingming Zhang, Darko Marinov, and Sarfraz Khurshid.
\newblock Faster mutation testing inspired by test prioritization and
  reduction.
\newblock In Mauro Pezz{\`{e}} and Mark Harman, editors, {\em International
  Symposium on Software Testing and Analysis, {ISSTA} '13, Lugano, Switzerland,
  July 15-20, 2013}, pages 235--245. {ACM}, 2013.

\end{thebibliography}

\begin{IEEEbiography}[{\includegraphics[width=2.8cm,height=3.1cm,keepaspectratio,clip]{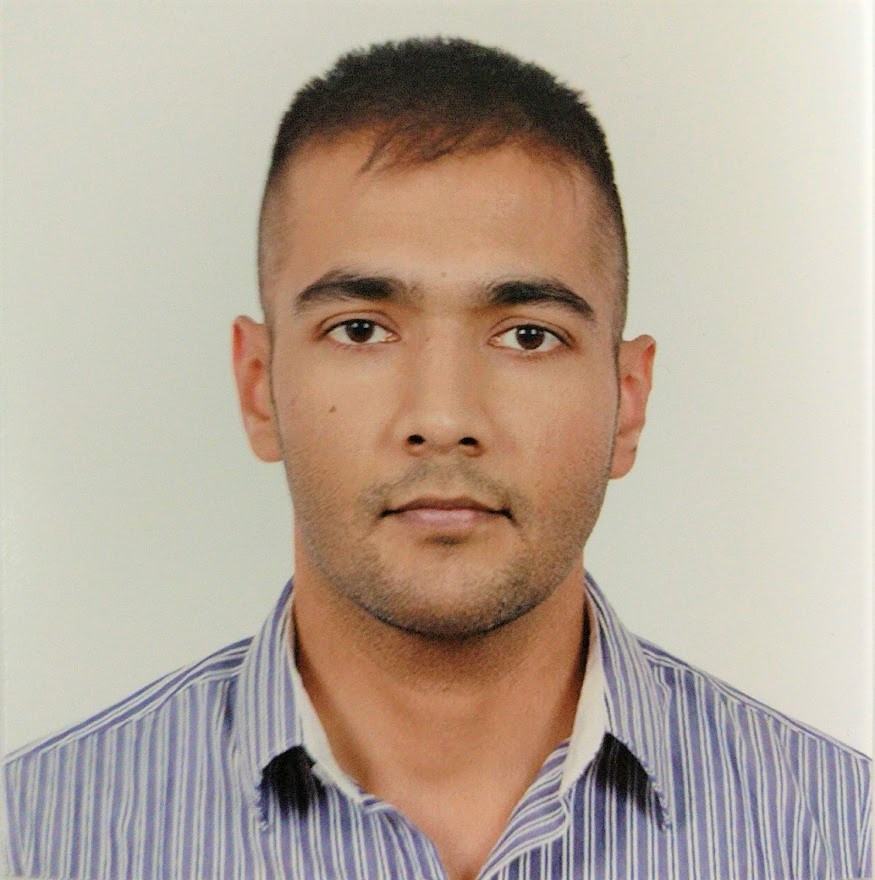}}]{Aayush~Garg}
is a doctoral researcher in the department of Computer Science at the Faculty of Science, Technology and Medicine~(FSTM), University of Luxembourg. He received his M.S. degree in Computer Science with a concentration in Security from Boston University, United States in 2019. He has several years of industrial experience as a Software Developer in Fintech organizations. His research areas comprise computer security, computational intelligence in software engineering, and mutation testing.
\end{IEEEbiography}

\begin{IEEEbiography}[{\includegraphics[trim=300 50 300 0,width=2.8cm,height=3.1cm,keepaspectratio,clip]{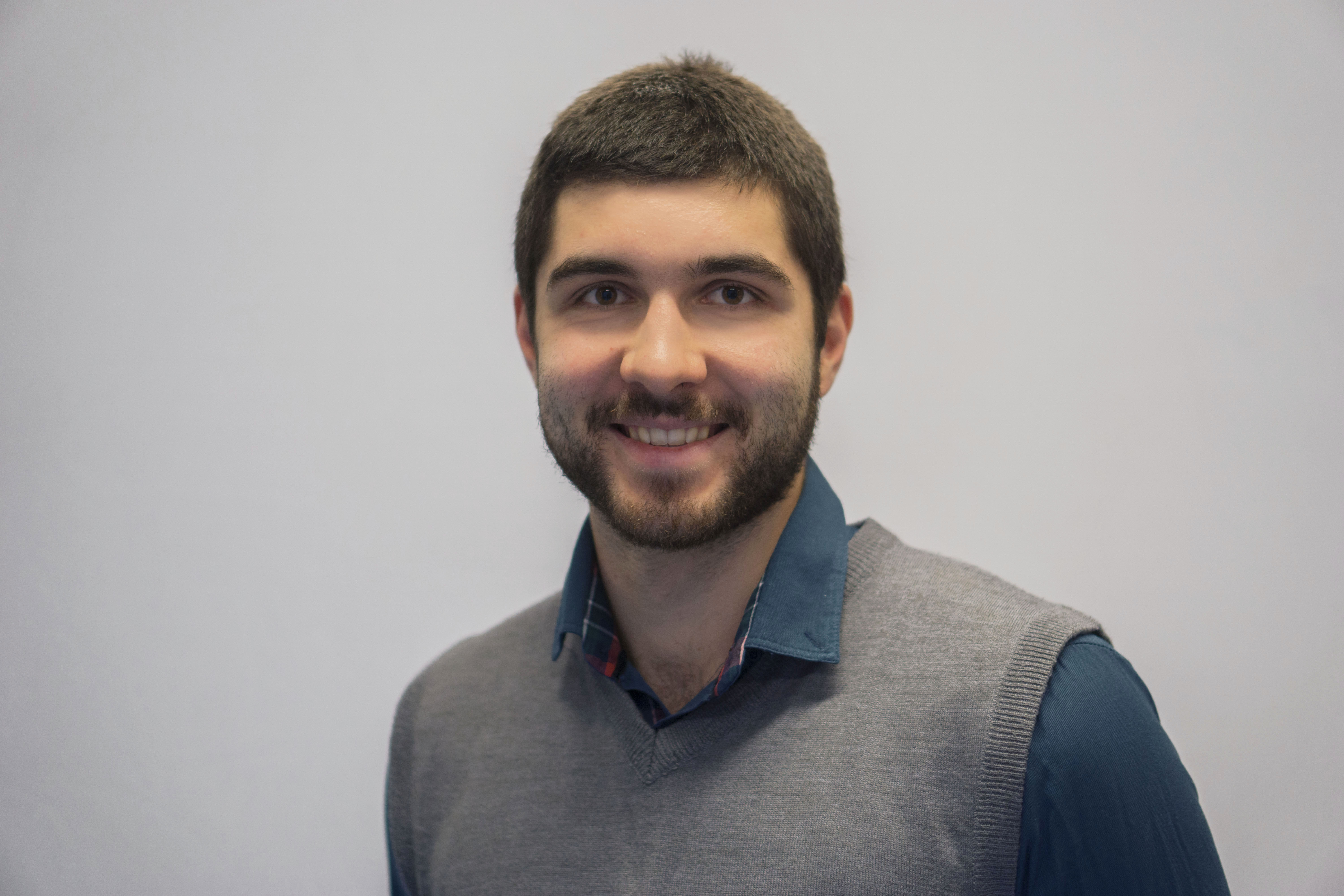}}]{Milos~Ojdanic}
is a doctoral researcher at the Interdisciplinary Center for Security, Reliability, and Trust (SnT) at the University of Luxembourg. He received his MSc degree from the Faculty of Innovation, Design, and Technology at the Mälardalen University (Sweden) in 2019. His research interests are in software development, testing, and evolution. In particular, he focuses on evolving systems, change-aware testing criteria, mutation testing, and prediction modeling.\end{IEEEbiography}




\begin{IEEEbiography}[{\includegraphics[width=2.8cm,height=3.1cm,keepaspectratio,clip]{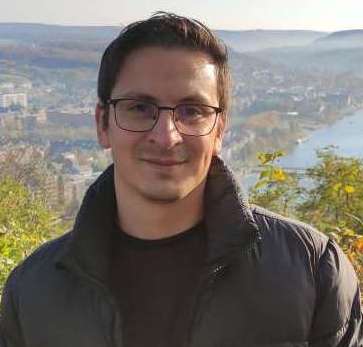}}]{Renzo~Degiovanni}
is a research associate at the Interdisciplinary Center for Security, Reliability and Trust (SnT) at the University of Luxembourg. He received a Ph.D. diploma in Computer Science from the National University of Cordoba, Argentina. His research interests  are in software engineering, specifically the validation and verification of software. His research has contributed to the automation of requirements engineering activities, software testing and formal software verification.
\end{IEEEbiography}

\begin{IEEEbiography}[{\includegraphics[width=2.8cm,height=3.1cm,keepaspectratio,clip]{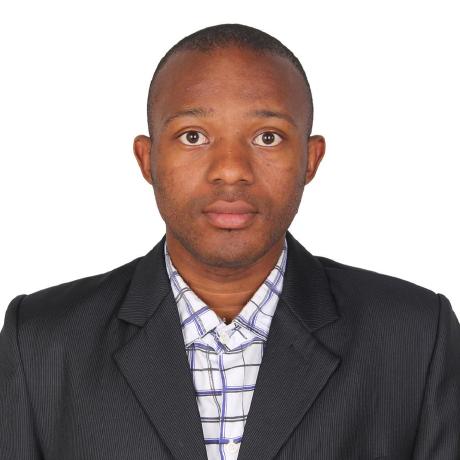}}]{Thierry~Titcheu~Chekam}
is a software engineer at SES, Luxembourg. He received his Ph.D. diploma in Computer Science from the University of Luxembourg. He received the B.Sc. degree in Computer Science and Technology from the University of Science and Technology of China in 2013, and the M.Eng. degree in Software Engineering from the School of Software, Tsinghua University in 2015. His research areas comprise software testing, mutation analysis, symbolic execution, and cloud computing/storage.
\end{IEEEbiography}

\begin{IEEEbiography}[{\includegraphics[width=2.8cm,height=3.1cm,keepaspectratio,clip]{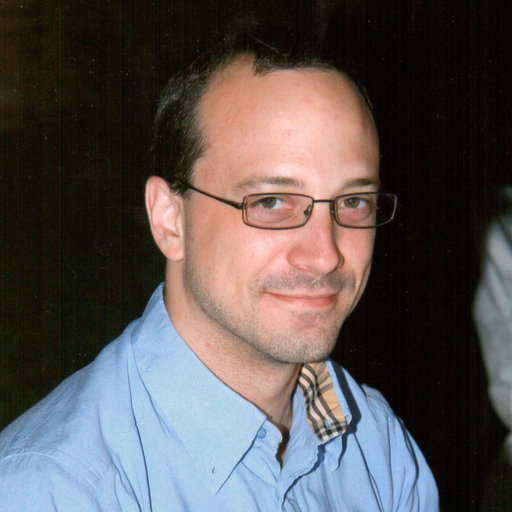}}]{Mike~Papadakis}
is a senior research scientist at the Interdisciplinary Center for Security, Reliability and Trust (SnT) at the University of Luxembourg. He received a Ph.D. diploma in Computer Science from the Athens University of Economics and Business. He is recognised for his work on software testing and in particular in the area of mutation testing. His research interests also include static analysis, prediction modelling and search-based software engineering.
\end{IEEEbiography}

\begin{IEEEbiography}[{\includegraphics[width=2.8cm,height=3.1cm,keepaspectratio,clip]{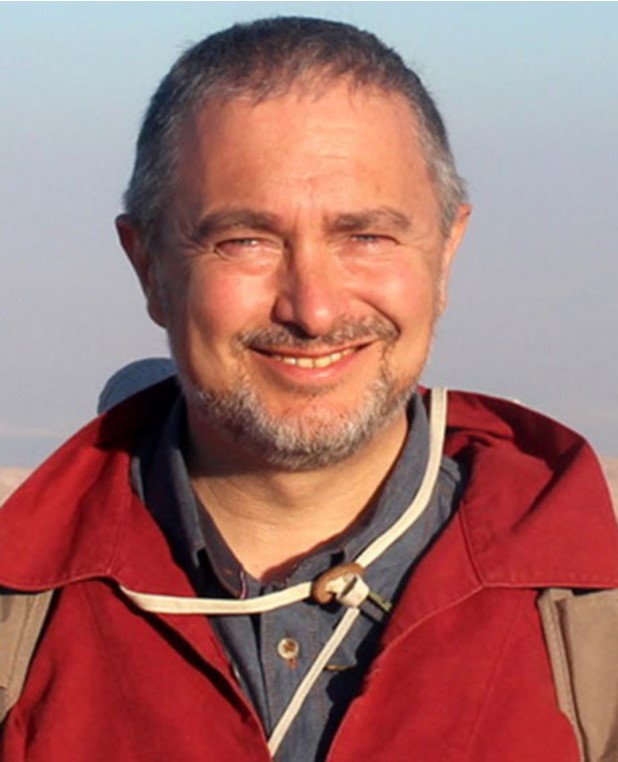}}]{Yves~Le~Traon}
is professor at the University of Luxembourg where he leads the SERVAL (SEcurity, Reasoning and VALidation) research team. His research interests within the group include (1) innovative testing and debugging techniques, (2) Android apps security and reliability using static code analysis, machine learning techniques and, (3) model-driven engineering with a focus on IoT and CPS. His reputation in the domain of software testing is acknowledged by the community. He has been General Chair of major conferences in the domain, such as the 2013 IEEE International Conference on Software Testing, Verification and Validation (ICST), and Program Chair of the 2016 IEEE International Conference on Software Quality, Reliability and Security (QRS). He serves at the editorial boards of several, internationally-known journals (STVR, SoSym, IEEE Transactions on Reliability) and is author of more than 150 publications in international peer-reviewed conferences and journals.
\end{IEEEbiography}

\end{document}